\PassOptionsToPackage{table}{xcolor}

\documentclass[10pt,journal,compsoc]{IEEEtran}

\ifCLASSOPTIONcompsoc
  \usepackage[nocompress]{cite}
\else
  \usepackage{cite}
\fi

\ifCLASSINFOpdf
\else
\fi

\usepackage{microtype}                 %
\PassOptionsToPackage{warn}{textcomp}  %
\usepackage{textcomp}                  %
\usepackage{mathptmx}                  %
\usepackage{times}                     %
\usepackage{cite}                      %
\usepackage{tabu}                      %
\usepackage{booktabs}                  %

\usepackage[hang]{subfigure}
\usepackage[export]{adjustbox}
\usepackage{hyperref}
\usepackage{multirow}

\usepackage[inline]{enumitem}
\usepackage{amssymb}
\usepackage{amsmath}

\newcommand{\etal}{\textit{et\penalty50\ al}.}
\newcommand{\ie}{\textit{i}.\textit{e}.,~}
\newcommand{\eg}{\textit{e}.\textit{g}.,~}

\newlist{inlinelist}{enumerate*}{1}
\setlist*[inlinelist,1]{%
  label=(\arabic*),
}

\usepackage{scrlayer-scrpage}

\usepackage[final]{changes}
\newcommand{\chA}[1]{{{#1}}}
\newcommand{\chB}[1]{{{#1}}}
\DeclareNewLayer[
  foreground, %
  headsep, %
  voffset=0.75cm, %
  contents={%
    \makebox[\layerwidth]{%
      \parbox{\dimexpr\layerwidth-1cm\relax}{%
        \centering%
        \scriptsize %
        \MakeLowercase{\MakeUppercase{T}his work has been submitted to the \MakeUppercase{IEEE} for possible publication. \MakeUppercase{C}opyright may be transferred without notice, after which this version may no longer be accessible.}%
      }%
    }%
  }
]{headerbox}

\AddLayersToPageStyle{@everystyle@}{headerbox, footerbox}

\definechangesauthor[name={RevA}, color=blue]{A}

\newcommand{\addA}[1]{\added[id=A]{#1}}
\newcommand{\delA}[1]{\deleted[id=A]{#1}}
\newcommand{\repA}[2]{\replaced[id=A]{#1}{#2}}

\usepackage{tabularx}

\definecolor{TRcolor}{RGB}{255, 151, 151} %
\definecolor{TLcolor}{RGB}{255, 203, 151} %

\definecolor{BRcolor}{RGB}{163, 151, 255} %
\definecolor{BLcolor}{RGB}{151, 176, 255} %
\definecolor{Bcolor}{RGB}{151, 245, 255} %

\definecolor{Rcolor}{RGB}{255, 151, 253} %
\definecolor{Lcolor}{RGB}{151, 255, 163} %
\definecolor{Tcolor}{RGB}{238, 255, 151} %

\definecolor{crimson}{RGB}{150, 20, 60}
\definecolor{darkerGreen}{RGB}{0,100,0}

\usepackage{filecontents}

\usepackage{xr}
\makeatletter
\newcommand*{\addFileDependency}[1]{
  \typeout{(#1)}
  \@addtofilelist{#1}
  \IfFileExists{#1}{}{\typeout{No file #1.}}
}
\makeatother

\newcommand*{\myexternaldocument}[1]{
    \externaldocument{#1}
    \addFileDependency{#1.tex}
    \addFileDependency{#1.aux}
}

\myexternaldocument{supplementary}

\newcommand{\Autoref}[1]{%
    \begingroup%
    \renewcommand\figureautorefname{Figure}%
    \renewcommand\tableautorefname{Table}%
    \autoref{#1}%
    \endgroup%
}
\def\tableautorefname{Tab.}
\def\figureautorefname{Fig.}

\begin{document}

\title{From Top-Right to User-Right: \\Perceptual Prioritization of Point-Feature\\Label Positions}

\author{
    Petr Bob\'{a}k, 
    Ladislav \v{C}mol\'{\i}k, 
    Martin \v{C}ad\'{\i}k
    \IEEEcompsocitemizethanks{
        \IEEEcompsocthanksitem P. Bob\'{a}k and M. \v{C}ad\'{\i}k are with the Faculty of Information Technology, Brno University of Technology. E-mail: \{ibobak\,$|$\,cadik\}@fit.vut.cz.
        \IEEEcompsocthanksitem L. \v{C}mol\'{\i}k 
        is with the Faculty of Electrical Engineering, Czech Technical University in Prague. E-mail: 
        cmolikl%
@fel.cvut.cz.
    }
}

\IEEEtitleabstractindextext{%
\begin{abstract}
\chB{In cartography, Geographic Information Systems (GIS), and visualization, the position of a label relative to its point feature is crucial for readability and user experience. Alongside other factors, the point-feature label placement (PFLP) is typically governed by the Position Priority Order (PPO), a systematic raking of potential label positions around a point feature according to predetermined priorities. While there is a broad consensus on factors such as avoiding label conflicts and ensuring clear label-to-feature associations, there is no agreement on PPO. Most PFLP techniques rely on traditional PPOs grounded in typographic and cartographic conventions established decades ago, which may no longer meet today’s user expectations. In contrast, commercial products like Google Maps and Mapbox use non-traditional PPOs for unreported reasons.
Our extensive user study introduces the Perceptual Position Priority Order (\texttt{PerceptPPO}), a user-validated PPO that significantly departs from traditional conventions. A key finding is that labels placed above point features are significantly preferred by users, contrary to the conventional top-right position. We also conducted a supplementary study on the preferred label density, an area scarcely explored in prior research. Finally, we performed a comparative user study assessing the perceived quality of \texttt{PerceptPPO} over existing PPOs, advocating its adoption in cartographic and GIS applications, as well as in other types of visualizations.
Our research, supported by nearly 800 participants from 48 countries and over 45,500 pairwise comparisons, offers practical guidance for designers and application developers aiming to optimize user engagement and comprehension, paving the way for more intuitive and accessible visualizations.}
\end{abstract} %

\begin{IEEEkeywords}
Point-feature Label Placement, Cartographic Visualization, Geographic Information Systems, User-centered Design
\end{IEEEkeywords}
}

\maketitle
\thispagestyle{plain}

\IEEEdisplaynontitleabstractindextext

\IEEEpeerreviewmaketitle

\IEEEraisesectionheading{\section{Introduction}\label{sec:introduction}}

\IEEEPARstart{A}{utomatic} label placement holds a crucial role across various domains such as cartography, data visualization, and geographic information systems (GIS). It  facilitates clear communication and enhances the user's ability to interpret the underlying data \chB{as the placement of short textual annotations or \textit{labels} in short, play an essential part in improving the comprehension of visualized data.} %
Labels convey essential information about distinct data points, ranging from cities and landmarks within cartographic and geographic information systems (GIS) contexts to pivotal elements in diagrams, charts in visualization. Recognized by the ACM Computational Geometry Impact Task Force~\cite{Chazelle99} as an area of significant research interest, the challenge of optimally positioning textual annotations remains \repA{an active}{a vibrant} field of study, especially concerning point-feature label placement (PFLP)\chA{~\cite{Lessani2024, Kittivorawong2024, Pavlovec2022}}.

Point-feature label placement primarily deals with the maximization problem -- aiming to position labels for the maximum number of point features, also called \textit{anchors}, possible within a given set \chB{without causing conflicts. A \textit{conflict} refers to a situation where a label overlaps or interferes with another label, an anchor, or other important map features, which can obscure critical information and reduce the map's readability or visual clarity.} This task is often constrained by the \textit{fixed-position} model, which restricts label placements to a limited number of predefined positions around a point feature. This limitation necessitates using a systematic order of preference for these positions, which we term as Position Priority Order (PPO). The PPO ranks potential label positions according to predetermined priorities, guiding the selection process.

However, a closer review of existing literature reveals a disconcerting lack of consensus in PPOs as described in detail in \autoref{sec:related-work}. Various authors have ascribed different priorities to the same label positions, often without a clear or unified rationale. Priorities have historically been based on typographic and cartographic conventions \addA{or printer capabilities}, with varying degrees of justification and consistency across the literature, leading to a fragmented understanding of optimal label placement practices. The identified inconsistency highlights a gap in the field and underlines the necessity for an empirically grounded methodology that reflects user perceptions and preferences.
\addA{Nevertheless, recent PFLP approaches~\cite{Lessani2024, Kittivorawong2024, Pavlovec2022} continue to rely on traditional PPOs. }

\addA{Interestingly, commercial products such as Google Maps, TomTom, and Mapbox tend to use non-traditional PPOs for reasons that have not been reported. This discrepancy between academic research and commercial practice further emphasizes the urgent need for updated, user-validated PPOs.}

Our research introduces Perceptual Position Priority Order (\texttt{PerceptPPO}), a user-centered methodology that seeks to redefine the prioritization of
point-feature label positions based on users' perceptual and cognitive preferences rather than traditional conventions. Through this effort, we aim to establish a new standard in automatic label placement that prioritizes user experience, paving the way for more intuitive and accessible map designs and setting a precedent for future research in the domain.
Our main contributions are summarized as follows:
\begin{enumerate}[label=(\arabic*)]
    \item We propose a comprehensive review of existing literature on Position Priority Orders (PPOs), analyzing existing PPOs and highlighting the missing consensus based on user preferences.
    \item We introduce Perceptual Position Priority Order (\texttt{PerceptPPO}), a novel, user-centered prioritization of point-feature label positions that prioritizes user perceptions and preferences over traditional conventions supported by a global user study involving nearly 800 participants from 48 countries.
    \item We uncover the optimal label density for maps, an aspect seldom explored in prior research. Our research shows that users prefer an overall label coverage of \repA{12.5\%}{17\%} on blind maps.
    \item We provide analysis demonstrating the superior user preference of \texttt{PerceptPPO} over existing PPOs, reinforcing its potential for improving map design, user experience in cartographic applications, and other types of visualizations.
\end{enumerate}

\section{Related Work}
\label{sec:related-work}
In the domain of point-feature label placement, the primary challenge involves the selection of the most appropriate position for each label from among $n$ potential candidates surrounding a point feature, with the objective of maximizing the number of labels that can be placed without label conflicts and with sufficient label-to-feature association. %
\chB{When multiple conflict-free position candidates are available for a point feature, the selection is determined by the Position Priority Order (PPO) of the label candidates to maintain consistency in the label layout (\ie ensuring that most labels have the same relative position to their anchors, particularly in sparse data scenarios). Therefore, each candidate is assigned a value within the $[1, n]$ range, where a lower value indicates a higher priority.}

\begin{figure}[t]
    \subfigure[] {
        \label{fig:fixed-4-model}
        \includegraphics[height=1.4cm]{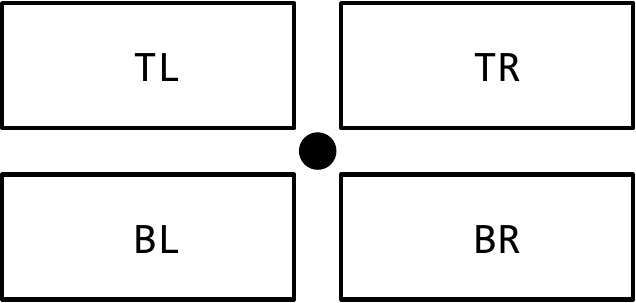}    
    }
    \hfill
    \subfigure[] {
        \label{fig:fixed-8-model}
        \includegraphics[height=1.4cm]{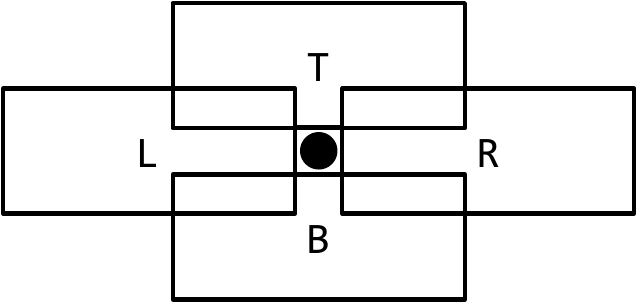}
    }
    \hfill
    \subfigure[] {
        \label{fig:fixed-10-model}
        \includegraphics[height=1.4cm]{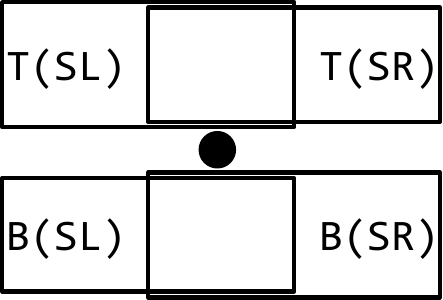}    
    }
    \caption{Common label positions for a point feature as presented in prior literature: \subref{fig:fixed-4-model} 4-position model, \subref{fig:fixed-8-model} additional positions of 8-position model, and \subref{fig:fixed-10-model} additional positions of 10-position model. \chA{Abbreviations: T -- top, B -- bottom, R -- right, L -- left, S -- slightly. Compounds, as T\textsubscript{SR} -- top slightly right, consists of the former trivial position names.}
    }
    \label{fig:position-models}
\end{figure}

Within the cartographic community and the domain of automated label placement, a variety of position models have been utilized, differing in the number of label candidates considered for each point feature. The 8-position model is the most prevalent, evaluating eight potential label placements per point feature, followed in popularity by the 6-position, 4-position, and 10-position models, respectively. \Autoref{fig:position-models} illustrates the label candidate positions for these models. Models with alternative configurations, such as the 5-position model, are used rarely.

Alongside the position models, authors frequently delineate the priorities assigned to label candidates. However, our review of the existing literature reveals inconsistencies and, at times, contradictions in these priorities. %
This section presents our analysis of point-feature labeling practices. \Autoref{tab:ppo-overview} shows the analyzed works detailing their chosen position models and the associated priorities. We have categorized the literature into four groups based on the similarity of their priority schemes. A fifth category encompasses works with distinct priority practices that do not align with those of any other group.
\begin{table}[t!]
\caption{
This overview outlines the PPO recommendations of various authors grouped by similarity. The top-right (TR) positioning emerges as a dominant preference, with authors consistently favoring it. While TR remains widely accepted across the board, there is noticeable diversity in selecting other positions.
The citation counts are derived from Google Scholar. $^*$The seminal work, \textit{Die Anordnung der Namen in der Karte}, was first introduced in German in 1962. Yoeli~\cite{Yoeli1972} and Zoraster~\cite{Zoraster1997} define more than 8 positions that are not included in our comparison as they are rarely used.
Abbreviations: A -- article, B -- book, SW -- software. \chA{Position names are abbreviated the same as in \autoref{fig:position-models}}.
}%
\centering
\setlength{\tabcolsep}{1.2pt}
\begin{adjustbox}{width=\columnwidth}
\begin{tabular}{|l|l|l|c|c|c|c|c|c|c|c|c|c|c|}
\hline
\multirow{2}{*}{Author} & \multirow{2}{*}{Year} & \multirow{2}{*}{Citations} & \multirow{2}{*}{Type} & \multicolumn{10}{c|}{Position Priority Order} \\ 
\cline{5-14}
& & & & 1 & 2 & 3 & 4 & 5 & 6 & 7 & 8 & 9 & 10\\ 
\hline
YoeliA~\cite{Yoeli1972} & 1972 & 231 & A & \cellcolor{TRcolor}TR & \cellcolor{TLcolor}TL & \cellcolor{BRcolor}BR & \cellcolor{BLcolor}BL & \cellcolor{Tcolor}T & \cellcolor{Bcolor}B & & & & \\
\hline
Robinson~\etal~\cite{Robinson1995} & 1995\small{ (6th)} & 3096\small{ (4th)} & B & \cellcolor{TRcolor}TR & \cellcolor{TLcolor}TL & \cellcolor{BRcolor}BR & \cellcolor{BLcolor}BL & \cellcolor{Tcolor}T & \cellcolor{Bcolor}B & T\textsubscript{SR} & B\textsubscript{SR} & T\textsubscript{SL} & B\textsubscript{SL} \\
\hline
\textbf{Brewer~\cite{Brewer2015}} & 2015 & 441 & B & \cellcolor{TRcolor}TR & \cellcolor{TLcolor}TL & \cellcolor{BRcolor}BR & \cellcolor{BLcolor}BL &  \cellcolor{Tcolor}T & \cellcolor{Bcolor}B & & & &\\
\hline
\hline

\textbf{YoeliB~\cite{Yoeli1972}} & 1972 & 231 & A & \cellcolor{TRcolor}TR & \cellcolor{TLcolor}TL & \cellcolor{BRcolor}BR & \cellcolor{BLcolor}BL & \cellcolor{Rcolor}R & \cellcolor{Lcolor}L & \cellcolor{Tcolor}T & \cellcolor{Bcolor}B & &\\
\hline
Dent~\cite{Dent2009} & 2009\small{ (6th)} & 1304\small{ (5th)} & B & \cellcolor{TRcolor}TR & \cellcolor{TLcolor}TL & \cellcolor{BRcolor}BR & \cellcolor{BLcolor}BL & \cellcolor{Rcolor}R & \cellcolor{Lcolor}L & \cellcolor{Tcolor}T\textsubscript{SL} & \cellcolor{Bcolor}B\textsubscript{SR} & &\\
\hline

Dobias (QGIS)\footnote{\url{https://github.com/qgis/QGIS/blob/f8edf729e24fdfa662aaccb8cad1d03968b4aeec/src/core/labeling/qgspallabeling.cpp}}  & 2009 & -- & SW & \cellcolor{TRcolor}TR & \cellcolor{TLcolor}TL & \cellcolor{BRcolor}BR & \cellcolor{BLcolor}BL & \cellcolor{Rcolor}R & \cellcolor{Lcolor}L & \cellcolor{Tcolor}T\textsubscript{SR} & \cellcolor{Bcolor}B\textsubscript{SR} &  &\\
\hline

Krygier~\etal~\cite{Krygier2016} & 2016\small{ (3rd)} & 361 & B & \cellcolor{TRcolor}TR & \cellcolor{TLcolor}TL & \cellcolor{BRcolor}BR & \cellcolor{BLcolor}BL & \cellcolor{Rcolor}R & \cellcolor{Lcolor}L & \cellcolor{Tcolor}T\textsubscript{SL} & \cellcolor{Bcolor}B\textsubscript{SR} & &\\
\hline
\hline

\textbf{Christensen and Marks}~\cite{Christensen1995} & 1995 & 520 & A & \cellcolor{TRcolor}TR & \cellcolor{TLcolor}TL & \cellcolor{BLcolor}BL & \cellcolor{BRcolor}BR & \cellcolor{Rcolor}R & \cellcolor{Tcolor}T & \cellcolor{Lcolor}L & \cellcolor{Bcolor}B & &\\
\hline
Yamamoto~\cite{Yamamoto2005} & 2005 & 40 & A & \cellcolor{TRcolor}TR & \cellcolor{TLcolor}TL & \cellcolor{BLcolor}BL & \cellcolor{BRcolor}BR & & & & & &\\
\hline
\hline

Ebinger and Goulette~\cite{Ebinger1989} & 1989 & -- & A & \cellcolor{TRcolor}TR & \cellcolor{BRcolor}BR & \cellcolor{TLcolor}TL & \cellcolor{BLcolor}BL & & & & & &\\
\hline
Wood~\cite{Wood2000} & 2000 & 28 & A & \cellcolor{TRcolor}TR & \cellcolor{BRcolor}BR & \cellcolor{TLcolor}TL & \cellcolor{BLcolor}BL & \cellcolor{Tcolor}T\textsubscript{SR} & \cellcolor{Bcolor}B\textsubscript{SL} & & & &\\
\hline
\textbf{Slocum}~\etal~\cite{Slocum2009} & 2022\small{ (4th)} & 962 & B & \cellcolor{TRcolor}TR & \cellcolor{BRcolor}BR & \cellcolor{TLcolor}TL & \cellcolor{BLcolor}BL & \cellcolor{Tcolor}T & \cellcolor{Bcolor}B & \cellcolor{Rcolor}R & \cellcolor{Lcolor}L & &\\
\hline
\hline

\textbf{Imhof~\cite{Imhof1975}} & 1975\small{ ($^*$1962)} & 439\small{ ($^*$88)} & A & \cellcolor{TRcolor}TR & \cellcolor{Rcolor}R & \cellcolor{Tcolor}T & \cellcolor{Bcolor}B & \cellcolor{Lcolor}L & & & & &\\
\hline
Zoraster~\cite{Zoraster1986} & 1986 & 108 & A & \cellcolor{TRcolor}TR & \cellcolor{Tcolor}T & \cellcolor{Rcolor}R & \cellcolor{TLcolor}TL & \cellcolor{BRcolor}BR & \cellcolor{Lcolor}L & \cellcolor{Bcolor}B & \cellcolor{BLcolor}BL & &\\
\hline
Jones~\cite{Jones1989} & 1989 & 79 & A & \cellcolor{TRcolor}TR & \cellcolor{Rcolor}R & \cellcolor{BRcolor}BR & \cellcolor{TLcolor}TL &  \cellcolor{Lcolor}L & \cellcolor{BLcolor}BL & & & &\\
\hline

Zoraster~\cite{Zoraster1990} & 1990 & 129 & A & \cellcolor{TRcolor}TR & \cellcolor{Tcolor}T & \cellcolor{TLcolor}TL & \cellcolor{Rcolor}R & \cellcolor{Lcolor}L & \cellcolor{BRcolor}BR & \cellcolor{Bcolor}B & \cellcolor{BLcolor}BL & &\\
\hline

\textbf{Zoraster~\cite{Zoraster1997}} & 1997 & 103 & A & \cellcolor{Tcolor}T & \cellcolor{TRcolor}TR & \cellcolor{TLcolor}TL & \cellcolor{Rcolor}R & \cellcolor{Lcolor}L & \cellcolor{BRcolor}BR & \cellcolor{Bcolor}B & \cellcolor{BLcolor}BL & &\\
\hline
\hline

\textbf{\texttt{PerceptPPO}} & 2024 & -- & A & \cellcolor{Tcolor}T & \cellcolor{Bcolor}B & \cellcolor{Rcolor}R & \cellcolor{TRcolor}TR & \cellcolor{BRcolor}BR & \cellcolor{Lcolor}L & \cellcolor{TLcolor}TL & \cellcolor{BLcolor}BL & &\\
\hline
\end{tabular}
\end{adjustbox}
\label{tab:ppo-overview}
\end{table}
The initial guidelines for selecting appropriate label candidates, among other labeling rules, were formulated by Imhof in 1962, published in German, and later translated into English~\cite{Imhof1975}. Imhof introduced a 5-position model tailored for left-to-right languages, leveraging his cartographic expertise to recommend the top-right position as the most favorable for label placement. His preference was rooted in typographic principles, \eg the top position (T) was favored over the bottom (B). This rationale was based on the observation that in the Latin alphabet, ascenders are more common than descenders, suggesting that labels placed at the top are likely to appear visually closer to their corresponding point features. This consideration is particularly relevant for city names, which typically begin with a capital letter.

Yoeli~\cite{Yoeli1972} introduced the first algorithms for the automated positioning of point-feature labels. He proposes two n-position models later adopted for point-feature label placement by other authors.
The first position model, denoted in \autoref{tab:ppo-overview} as \emph{YoeliA}, is a 10-position model where the label candidates are organized around the point feature as in \autoref{fig:position-models}. Notably, Yoeli proposed a grid system for typesetting the labels where the size of the grid cell is based on the size of the letters. Therefore, Yoeli distinguishes between labels with an odd and even number of letters. Since the latter cannot be centered above or below the point feature, Yoeli introduces additional top and bottom position modifiers that shift the label slightly left (SL) or slightly right (SR). In \autoref{tab:ppo-overview}, we report only the first six positions, as nowadays, even labels with an even number of letters can be easily centered above or below the point feature. 
Similarly, Brewer~\cite{Brewer2015} omits the four additional %
positions with shifted labels in his 6-position model. For Robinson~\etal~\cite{Robinson1995}, we report all positions in his 10-position model with priorities identical to Yoeli's, as he is not providing any information as to why the positions with shifted labels are included in his model.

The second position model, denoted in \autoref{tab:ppo-overview} as \emph{YoeliB}, is an 8-position model that Yoeli originally crafted for labeling small-area features. Nonetheless, other authors~\cite{Dent2009, Krygier2016} adopted or adapted Yoeli's 8-position model for point-feature label placement.
Similarly, as in the previous 10-position model, Yoeli~\cite{Yoeli1972} uses additional top and bottom positions slightly shifted to the left for labels with an even number of letters. However, we again ignore additional adjustments for even-lettered labels, as they can be precisely centered directly above or below the point feature nowadays, regardless of letter count.

Christensen and Marks~\cite{Christensen1995} introduced two algorithms for PFLP, employing gradient descent and simulated annealing techniques. The formulation of the proposed objective function draws upon Yoeli's~\cite{Yoeli1972} foundational work. Notably, they adopt an 8-position model similar to Yoeli's, but with swapped priorities of bottom-left (BL) and bottom-right (BR) positions and the top (T) and left (L) positions, which they describe as a standard PPO.

Ebinger and Goulette~\cite{Ebinger1989} proposed a 4-position model, %
see \autoref{fig:fixed-4-model}, which diverges in priority schemes from those proposed by Imhof~\cite{Imhof1962} and Yoeli~\cite{Yoeli1972}. While Yoeli prioritizes top positions over bottom positions, Ebinger and Goulette prioritize positions on the right side over the left.
Wood~\cite{Wood2000} proposed a 6-position model with the top position being shifted slightly right ($\mathrm{T_{SR}}$) and the bottom slightly left ($\mathrm{B_{SL}}$). The author remarks that the shifted positions should only be used in extreme cases. Moreover, he argues that the shifted positions help associate the label with the feature; unfortunately this claim is not supported by any justification. 
Similarly, Slocum~\etal~\cite{Slocum2009} proposed an 8-position model extending Wood's first for positions to include top (T), bottom (B), right (R), and left (L) positions, enhancing label placement flexibility.

The prior literature exhibits even more significant variety in priority schemes for label placement. Zoraster~\cite{Zoraster1986, Zoraster1990, Zoraster1997}, in his series of works, introduced three distinct 8-position models for oil well labeling, each with unique priorities. A notable trend across Zoraster's models is the elevated priority given to the top position, starkly contrasting with other authors' approaches.
Additionally, Jones~\cite{Jones1989}) proposed 8-position models that resemble the 4-position model by Ebinger and Goulette~\cite{Ebinger1989}, with a nuanced approach to prioritization: the right (R) position is placed between the top-right (TR) and bottom-right (BR) positions, while the left (L) position's priority is set between the top-left (TL) and bottom-left (BL) positions.

Upon examining \autoref{tab:ppo-overview}, it becomes clear that with the exception of Zoraster~\cite{Zoraster1997}, the top-right (TR) position %
emerges as the highest priority across multiple bodies of work. The designation of priorities to other positions shows even more significant variability, highlighting a lack of consensus.
Moreover, all reported priorities are based on the experience of the authors, and none of them were empirically verified by users. To our knowledge, the sole exception is a study by Scheuerman~\etal~\cite{Scheuerman2023}, which attempts to evaluate position priorities with user input but limits its scope to just three positions (TL, L, BL).  The study establishes the position preference order of $\mathrm{L} > \mathrm{TL} > \mathrm{BL}$, indicating L is favored over TL and BL, and TL is preferred over BL.

The disparities and notable gap in evaluations set the stage for our investigation. We ask the question whether specific Position Priority Orders (PPOs) influence the perceived label placement quality and we aim to establish a set of priorities validated by user-centered research. This will pave the way for more intuitive and accessible mapping solutions firmly grounded in empirical evidence. 

\section{Perceptual Position Priority Order}
\label{sec:percept-ppo}
Due to non-existing consensus on the ranking of label positions within cartography and GIS, we aim to create PPO rooted in user perception. Typographic and cartographic conventions used in previous works are valuable, but  originate from several decades-old practices that may not align with modern user needs. Herein, we detail the empirical user study that underpins \chB{the Perceptual Position Priority Order} (\texttt{PerceptPPO}) and lay the groundwork for a comparative analysis that underscores its efficacy.

\subsection{Data}
\label{sec:percept-ppo-data}
We randomly selected 30 locations worldwide. We excluded any locations on the sea or ocean and those with latitudes greater than -60 degrees to exclude Antarctica due to its sparse population. Each location served as the center of an area, defined by the location and a zoom level ranging from 5 (approximately the size of Europe) to 10 (roughly the size of Luxembourg), rendered as a vector SVG image at a size of $1305\times1025$ pixels. Settlements with more than 500 habitants, obtained from GeoNames, specifically Cities 500\footnote{\url{http://download.geonames.org/export/dump/}}, within these areas were used as anchors and were sorted by population size. Then we filtered only anchors such that all the 8-positions around are available in any configurations of labels without any conflict (we examined the occurrence of conflict for all anchors over bounding boxes containing all eight positions of labels). If an area contained fewer than 20 anchors,  we discarded it in favor of another area. Finally, we acquired 30 areas indexed from 0 to 29 at zoom levels 5 to 8, with 20 to 54 anchors. %

\begin{figure}[!ht]
    \subfigure[TR] {
        \includegraphics[width=0.4866\columnwidth, frame, trim={7.6cm 5cm 5.7cm 11.2cm}, clip]{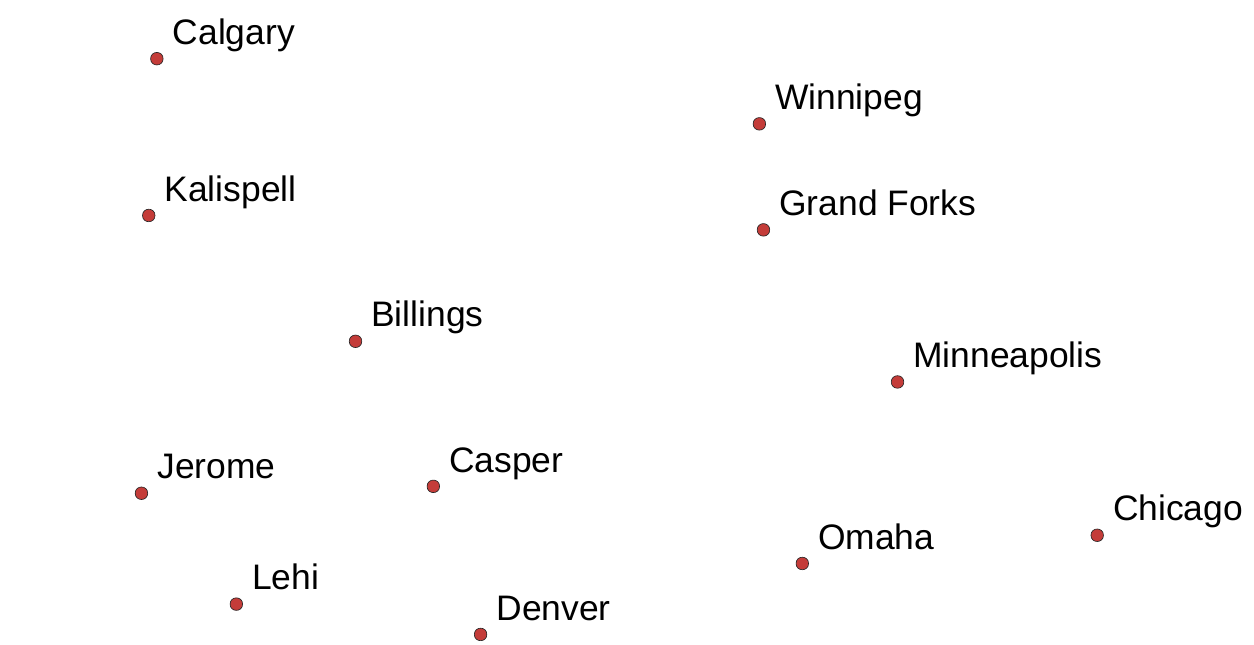}
    }
    \hspace{-1em}
    \subfigure[T] {
        \includegraphics[width=0.4866\columnwidth, frame, trim={7.6cm 5cm 5.7cm 11.2cm}, clip]{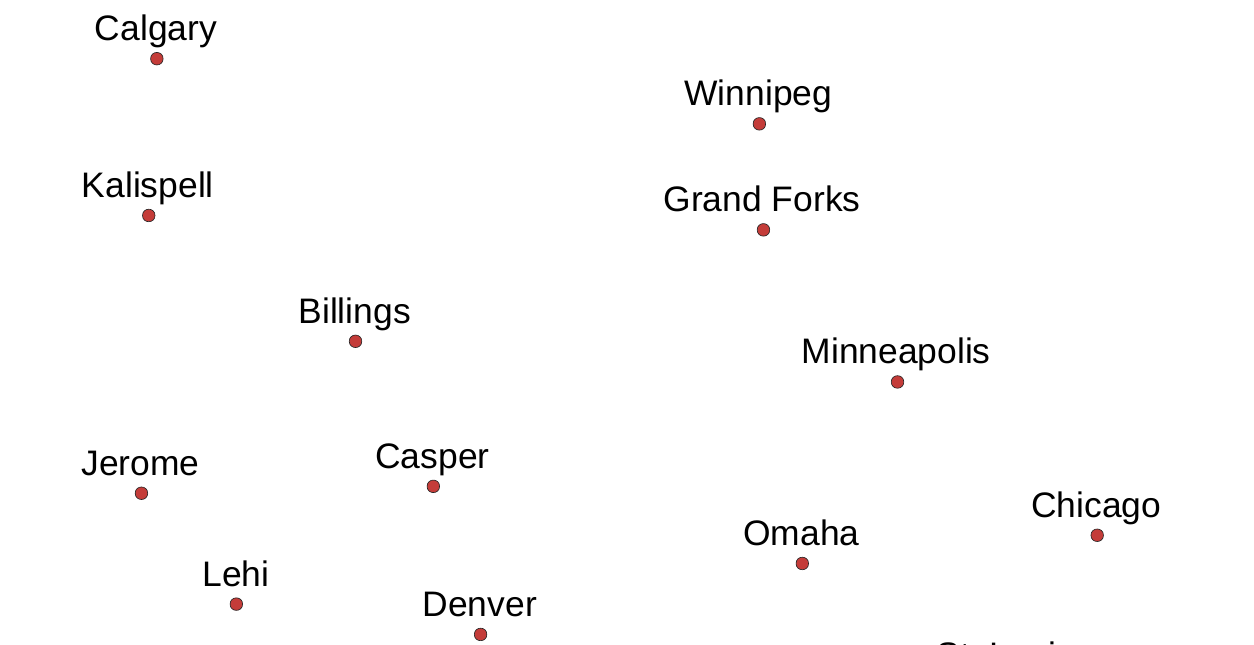}
    }
    \caption{
        Example of map area rendered consistently using TR and T position across all point features.
    }
    \label{fig:areas-ppo}
\end{figure}

Subsequently, we rendered each area eight times, placing all labels in one of the eight corresponding positions relative to the anchor: top-right (TR), top (T), top-left (TL), left (L), bottom-left (BL), bottom (B), bottom-right (BR), and right (R). This process yielded $30 \times 8 = 240$ blind maps featured with a white background, red anchors, and corresponding labels. See example in \autoref{fig:areas-ppo} and the suppl.\,material for \repA{additional}{all} renders. \addA{In order to understand user preferences for label positions in relation to the anchor, we opted for using blind maps to eliminate all other factors potentially influencing the judgment of the label placement, such as patterns and vivid colors in the map background. Each cartographic rule for label placement~\cite{Imhof1975, Robinson1995, Slocum2009, Brewer2015} focuses on a specific factor, such as avoiding overlaps, ensuring proper label alignment, or selecting appropriate font sizes and colors. Therefore, we believe that these factors can be analyzed separately. %
}

\subsection{Procedure}
\label{sec:ppo-procedure}
We employed the Two-Alternative Forced Choice (2AFC) approach to determine the position priority order based on the actual perception of users, which we call Perceptual Position Priority Order (\texttt{PerceptPPO}). Considering the eight label positions under examination, we have $\binom{8}{2} = 28$ pairwise comparisons on an area. Consequently, the entire \repA{evaluation}{experiment} consists of $30\times28 = 840$ pairwise comparisons to cover all of them once. In order to alleviate potential fatigue among the participants during the study, we allocated only three areas to each participant, resulting in a batch of $3\times28 = 84$ pairwise comparisons. Therefore, 10 participants were required to cover all pairwise comparisons once.

We engaged Mechanical Turk workers and university students to conduct the \repA{evaluation}{experiment}. \addA{Participation in the evaluation was voluntary, and the compensation for Mechanical Turk workers was set to match the average compensation rate of other requesters on Amazon Mechanical Turk.} Initially, the participants were introduced to the \chA{experiment}\addA{, informed about its duration, and that the provided data would be collected and used for research purposes}. Subsequently, we asked about their country of residence, age, gender, and education. Each participant was allowed to take only one batch of 84 pairwise comparisons to mitigate the carry-over effect. During the \repA{evaluation}{experiment}, the pairs of area and label position were distributed randomly but uniformly between the left and right sides of the shown comparison pair.
Participants were shown two maps sequentially, each depicting the same area but with different positions of labels relative to the anchors. Notably, the position of a label was kept consistent for all points within a single map, see \autoref{fig:areas-ppo}. Participants were asked to select the map they \repA{prefer}{found aesthetically more pleasing} from each presented pair. At the end of the \repA{evaluation}{experiment}, the participants were allowed to leave an additional note about anything regarding the \chA{experiment}.

\subsection{Statistical Analysis}
\label{sec:statistical-analysis}

To derive insights from the pairwise comparison data of $n$ objects, we transform the data into the $n \times n$ preference matrix $P$ for each participant individually. Each element $p_{ij}$ in the matrix indicates the number of times that method $i$ was selected over method $j$ by a participant. The conversion allows for a preference analysis, facilitating the identification of PPO patterns across participants in the study.

\subsubsection{Coefficient of Consistency \texorpdfstring{$\zeta$}{zeta}}
Assessing whether the particular participant can form a reliable judgment of the quality under examination is crucial when using paired comparison. Kendall and Babington~\cite{Kendall1947} proposed \textit{coeficient of consistency} $\zeta$ to measure the consistency based on the transitivity of participant's choices. For example, when evaluating three objects: $A$, $B$, and $C$, a participant might choose that $A > B$ ($A$ is preferred to $B$), $B > C$ and $C > A$. In this case, the \textit{triad} is called \textit{circular} and the pair comparison inconsistent. The $\zeta=1$ if there are no circular triads / no inconsistencies. On the other hand, when the number of circular triads/inconsistencies increases, the $\zeta$ decreases towards zero. Inconsistency can arise due to incompetence, participant's attention changes during the \repA{evaluation}{experiment}, or the examined objects being too alike. The definition of $\zeta$ can be simplified as $\zeta = 1-\frac{T}{T_{max}}$, where $T$ is the observed number of circular triads, $T_{max}$ is the maximum number of circular triads.
For more details, see Kendall and Babington~\cite{Kendall1947} and David~\cite{David1988}. We employ the implementation \delA{of the EBA package} proposed by Wickelmaier and Schmid~\cite{Wickelmaier2004}, which also provides the expected number of circular triads $E(T)$ when choices are made at random.

\subsubsection{Coefficient of Agreement \texorpdfstring{$u$}{u}}
While the coefficient of consistency provides a measure of consistency within participants, the \textit{coefficient of agreement} $u$ introduced by Kendall and Babington~\cite{Kendall1947} measures the variety of choices among $m$ participants. Complete agreement $u=1$ is achieved when all participants make identical choices for all pairs. In other words, the half $p_{ij}$ of the overall preference matrix $\mathbf{P} = \sum_{k \in (0, m)} P_k$ is equal to the number of participants $m$, while the other half is a zero. \chB{On the other hand, the minimum agreement occurs when the preference for each pair is equally divided among participants. Specifically, this happens} when $p_{ij} = \frac{m}{2}$ if $m$ is even or when $p_{ij} = \tfrac{m \pm 1}{2}$ otherwise. Correspondingly, minimum coefficient of agreement is $u_\mathrm{min} = \tfrac{-1}{m-1}$ or $u_\mathrm{min} = \tfrac{-1}{m}$. For cases in between the $u$ range, Kendall and Babington~\cite{Kendall1947} defines $u$ as
\begin{align}
	u = \frac{2
            \sum_{i \neq j}^{} \binom{p_{ij}}{2}
    }{ \binom{m}{2} \binom{n}{2} } - 1.
\end{align}

The statistical significance of $u$ with the null hypothesis that all participants choose the preferences randomly (or there is no agreement among participants) can be approximated by $\chi^2$ variate as described in David~\cite{David1988}. Again, we employ the implementation \delA{of the EBA package} proposed by Wickelmaier and Schmid~\cite{Wickelmaier2004}.

\subsubsection{Pairwise Comparison Model}
\label{sec:pwcmp}
To transform the preference matrices to quality scores, we employ Thurstone's statistical judgment model proposed by Thurstone \cite{Thurstone1927}, as recommended by Tsukida and Gupta\cite{Tsukida2011} and P\'{e}rez-Ortiz and Mantiuk~\cite{Perez2017}. The model assumes that the quality score of object $A$ is a Normal random variable $A \sim \mathcal{N}(q_A, \sigma_A^2)$ where \chB{mean} $q_A$ is assumed to be the true quality score\chB{, and $\sigma_A^2$ is the variance}. Similarly, for object $B \sim \mathcal{N}(q_B, \sigma_B^2)$.
Normal distribution captures the fact that different participants have various preferences regarding the quality of examined objects (inter-participant variance). Moreover, participants' preferences are also likely to change when they repeat the same \repA{evaluation}{experiment} (intra-participant variance). We apply Thurstone's Case V model, which assumes that a Normal distribution can explain inter- and intra-participant variance. At the same time, the variance $\sigma^2$ describes the uncertainty and is the same for all examined objects (in our example $\sigma_A^2 = \sigma_B^2$) while the correlation \chB{$\rho$} among objects is zero ($\rho_{AB} = 0$). The difference between the two Normal distributions is again Normal distribution $A-B \sim \mathcal{N}(q_{AB}, \sigma_{AB}^2)$. Without loss of generality, we can assume that variance $\sigma_A^2=\sigma_B^2=\tfrac{1}{2}$ so that $\sigma_{AB}^2= \sigma_A^2 + \sigma_B^2 - 2\rho_{AB}\sigma_A \sigma_B = 1$ which corresponds to standard Normal distribution.  
The quality difference estimation of two objects $A$ and $B$ $\hat{q}_{AB}$ is then defined as 
\begin{equation}
	\hat{q}_{AB} = \Phi^{-1}\left(\frac{\mathbf{P}_{A, B}}{\mathbf{P}_{A, B} + \mathbf{P}_{B, A}}\right),
\end{equation}
where \chB{$\mathbf{P}_{A, B}$ is the value from the preference matrix $\mathbf{P}$ for $A > B$, and }$\Phi^{-1}(x)$ is the inverse CDF of standard Normal distribution that can be interpreted as \textit{z-score} as it represents the distance of $x$ from the mean in units of the standard deviation.

To determine quality scores of $m$ objects, Tsukida and Gupta\cite{Tsukida2011} recommend using the maximum likelihood estimate (MLE). To this end, we employ the MLE implementation of P\'{e}rez-Ortiz and Mantiuk~\cite{Perez2017}, which also includes confidence interval estimation based on random sampling with replacement and methodology to perform a two-tailed test of the null hypothesis ``There is no difference among examined objects.'' at a significance level of $\alpha = 0.05$.

\subsection{Online Study Precautions}
\label{sec:online-precautions}
When dealing with online study, there is always a risk of ingenue responses and result fabrication. Therefore, to address these pitfalls, we conducted a pilot study with 50 participants who are qualified as Master Mechanical Turk Workers and consistently demonstrated high accuracy in performing various Human Intelligence Tasks (HITs). The pilot results serve as a calibration group to determine an \repA{evaluation}{experiment}'s statistics, \repA{evaluation}{experiment} duration, time spent on the introduction page, time spent filling out the survey, response time for individual pairs, participant's coefficient of consistency, and balance of choosing the left or right option.
Afterward, we made the study available to a broader range of Mechanical Turk Workers while following the general recommendations: HIT approval rate 95\%, number of approved HITs $> 2000$,  and restricted repetition of study by one worker. Additionally, we implemented Google reCAPTCHA to reduce the risk of bot fabrication, mitigating bot activity and ensuring that participants are genuine. Moreover, to assess and control the data quality, we used the statistics from the calibration group to eradicate workers of insufficient quality that significantly deviated from the standard deviation. We intentionally did not automate the elimination process to interpolate the measured statistics (deviations from the calibration group) while considering the expected number of circular triads $E(T)$ and workers' feedback, allowing a more nuanced understanding of worker performance and potential issues within the tasks. In particular, we identify participants as potentially inconsistent if $T < E(T)$ is observed in at least two of the three areas assigned for their evaluation. Once the HIT was approved, it became part of the calibration group, and all statistics for this group were recalculated.

To gain an even deeper understanding, we employed Smartlook\footnote{\url{www.smartlook.com}}\label{fn:smartlook} to analyze workers' behavior while working on the \repA{evaluation}{experiment} by manually reviewing activity recordings. By doing so, we identified three pitfalls:
\begin{enumerate*}[label=(\arabic*)]
    \item some workers were likely modifying JavaScript to alter the behavior of the study,
    \item workers often copied the text of the task/questions presumably because they did not understand the text and were translating it,
    \item some workers disregarded the instructions and tried to complete the task as quickly as possible.
\end{enumerate*}

To address these issues, we
\begin{enumerate*}[label=(\arabic*)]
    \item implemented obfuscation to prevent manipulation of JavaScript,
    \item integrated Google Translate into the study and slightly modified the instructions to accommodate non-native English speakers better, and
    \item introduced a mechanism where a participant must first review the pair and spend a minimum of 5 seconds selecting their preferred option.
\end{enumerate*}
 
\subsection{Results}
\label{sec:ppo-results}
We eliminated participants who exhibited inconsistencies in their responses as described in \autoref{sec:online-precautions}. Following this data refinement, we were left with a total of 225 participants with a dropout of 27\%. %
These participants generated 18,900 pairwise comparisons. Therefore, on average, each comparison pair was evaluated by approximately 23 different participants. %

We apply methodology as described in \autoref{sec:pwcmp} to compute the quality z-score and assess statistical significance at the significance level of $\alpha = 0.05$ to evaluate the null hypothesis $H_0^1$: ``There is no clear user preference among the label positions.''

Our initial findings of aggregated preferences suggested that label positions could be ordered by the perceptual preferences of participants as follows: $\mathrm{T} > \mathrm{B} > \mathrm{R} > \mathrm{TR} > \mathrm{BR} \geq \mathrm{L} > \mathrm{TL} > \mathrm{BL}$. A statistically significant difference was found between all pairs of label positions except for the $\mathrm{BR} \geq \mathrm{L}$ pair. Our finding also harmonizes with PPO proposed by Scheuerman~\etal~\cite{Scheuerman2023}, who claims that the $\mathrm{L} > \mathrm{TL} > \mathrm{BL}$ order is preferred among participants.

In order to determine statistical significance for the $\mathrm{BR} \geq \mathrm{L}$ pair, we engaged an additional 104 participants who were specifically asked to respond to the (BR, L) pair of positions. Each participant was presented with a single pair for each area, resulting in 30 pairwise comparisons per participant. This approach led to a total of 3,120 new pairwise comparisons specific to the pair of (BR, L) positions.

\begin{figure*}[!ht]
    \subfigure[] {
        \label{fig:ppo-quality-score}
        \includegraphics[width=0.5\textwidth]{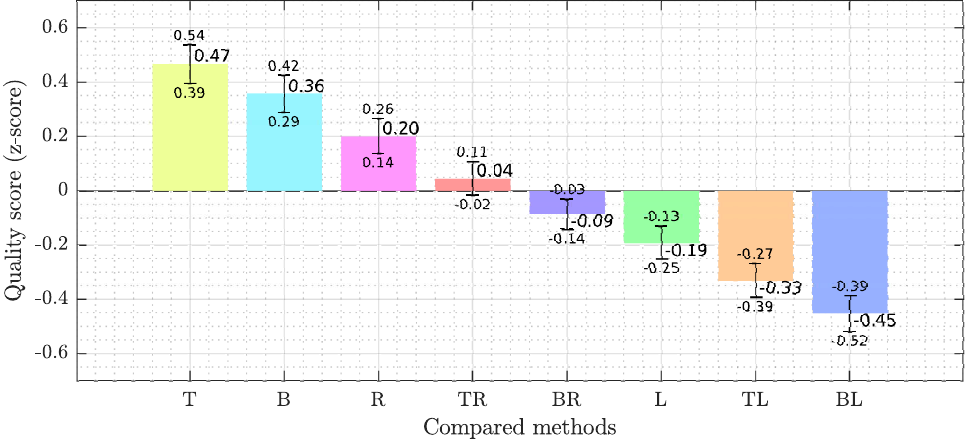}
    }
    \hspace{0em}
    \subfigure[] {
        \label{fig:ppo-significance}
        \includegraphics[width=0.5\textwidth]{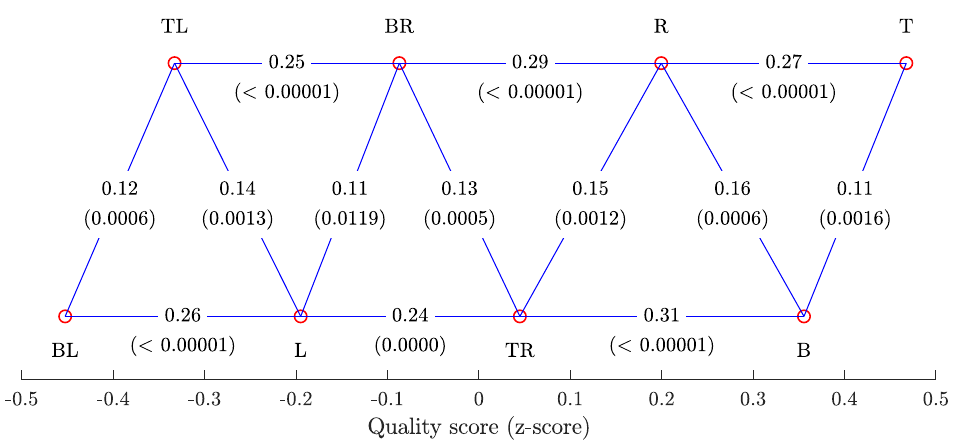}    
    }
    \caption{Results of the \texttt{PerceptPPO} user study. Chart \subref{fig:ppo-quality-score} depicts quality z-scores and 95\% confidence intervals. \chA{Chart \subref{fig:ppo-significance} shows a triangle plot visualizing significant differences between label position preferences as proposed by P\'{e}rez-Ortiz and Mantiuk~\cite{Perez2017}. Each red circle represents a label position, and the lines indicate significant differences between pairs. Solid blue lines represent statistically significant differences. The edge values show the absolute difference in z-scores between the compared label positions, with the $p$-values denoted in brackets. The label positions are plotted along the x-axis, with alternating y-axis offsets for clarity.}
    }
    \label{fig:ppo}
\end{figure*}

After the addition of the new comparisons, the final results, as depicted in \autoref{fig:ppo-quality-score}, show a statistically significant difference between all pairs of label positions, as illustrated by \autoref{fig:ppo-significance}. Therefore, we can reject the null hypothesis $H_0^1$ and claim that there is a clear preference of label positions in the order given by z-scores.

In total, we engaged 329 participants for this study, comprising of 217 males and 112 females. A majority of the participants hailed from the USA (155), followed by Czech Republic (88), India (35), and Slovakia (29). The most common age range among the participants was 20-30 years, with 159 individuals falling into this category. Regarding educational qualifications, the highest number of participants held bachelor's degrees  (119), followed by high school diplomas (117), and master's degrees (65).
On average, participants completed a batch of 84 pairwise comparisons in 5 minutes and 56 seconds, with a standard deviation of 3 minutes and 4 seconds. For the additional comparisons specifically acquired to determine statistical significance for the (BR, L) pair, the average completion time was 2 minutes and 4 seconds, with a standard deviation of 1 minute and 10 seconds.

The overall \repA{average}{median of} consistency $\zeta$ across participants is \repA{0.67}{0.75} ($SD = 0.29$, \repA{$MD = 0.75$}{$\bar{x}=0.67$}), which indicates that they were fairly consistent in their decisions \addA{and the consistency is reasonably leveled for each map area. For more details see \autoref{tab:ppo-results} in the suppl.\,material.}
\repA{The}{However, the} overall coefficient of agreement $u=0.12$ ($\mathrm{min}\ u=-0.001$) \delA{reveals relatively low agreement among participants, although} \addA{with} the $p$-value $=0$ clearly indicates that we can reject the null hypothesis $H_0^2$: ``There is no agreement among participants'' at $\alpha=0.05$ and conclude there is indeed \repA{statistically significant}{some} agreement among participants.

\begin{figure*}[!ht]
    \subfigure[\scriptsize{\textbf{C1} ($N=93$, $\zeta_1=0.833$, $u_1=0.335$, $p=0$)}] {
        \label{fig:ppo-cluster1}
        \includegraphics[height=4cm]{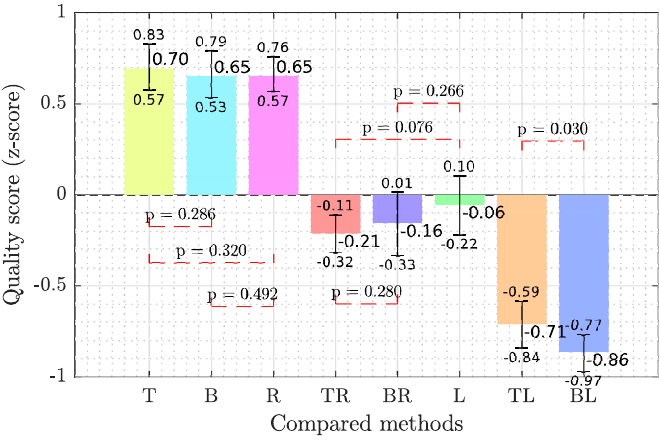}
    }
    \hspace{0em}
    \subfigure[\scriptsize{\textbf{C2} ($N=41$, $\zeta_2=0.844$, $u_2=0.37$, $p=1.5e^{-258}$)}] {
        \label{fig:ppo-cluster2}
        \includegraphics[height=4cm, trim={0.55cm 0cm 0cm 0cm}, clip]{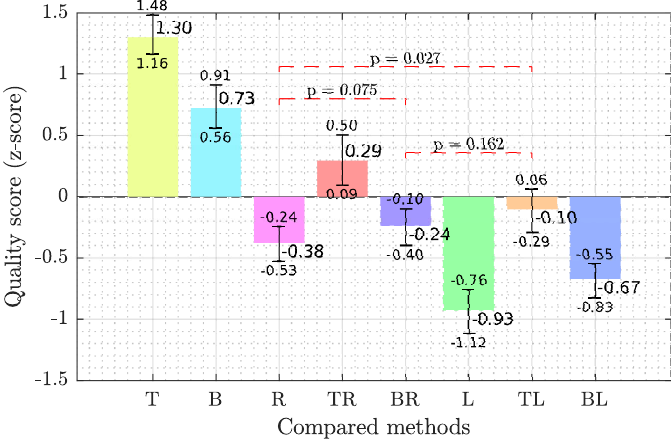}
    }
    \hspace{0em}
    \subfigure[\scriptsize{\textbf{C3} ($N=91$, $\zeta_3=0.435$, $u_3=0.021$, $p=2.7e^{-25}$)}] {
        \label{fig:ppo-cluster3}
        \includegraphics[height=4cm, trim={0.55cm 0cm 0cm 0cm}, clip]{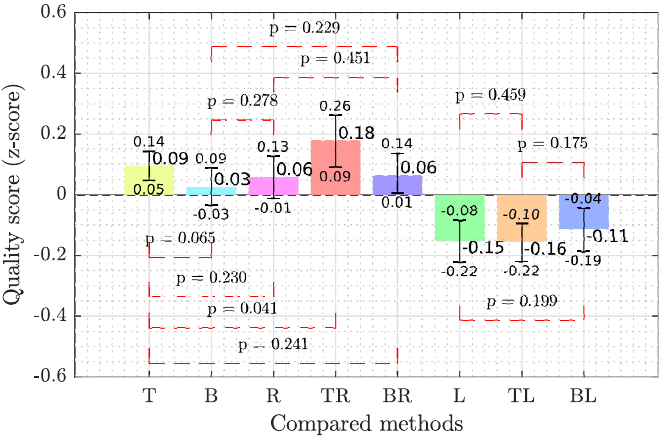}
    }
    \caption{
        Identified clusters of users in \texttt{PerceptPPO} study with depicted quality z-scores and 95\% confidence intervals. The red dashed brackets, with the corresponding $p$-values reported above, represent the PPOs pairs without evidence of statistically significant differences at $\alpha=0.05$ using the two-tail test. Conversely, between the PPOs pairs without brackets are detected statistically significant differences.
    }
    \label{fig:ppo-clusters}
\end{figure*}

\repA{However, the relatively low overall coefficient of agreement $u=0.12$ suggests that there might be underlying patterns or segments within the participant data that are not immediately apparent from the aggregated overall results. Therefore, we use hierarchical clustering applying Ward's minimum variance method to uncover these patterns and provide a more nuanced interpretation of the data~\cite{Jain1999, Hastie2009}.}{Using hierarchical clustering applying Ward's minimum variance method.} We identified three participant clusters as shown in \autoref{fig:ppo-clusters}. 
Even though the $p$-value for the coefficient of agreement $u$ \addA{within clusters} is sometimes greater than $\alpha=0.05$ for individual areas, \delA{which disallows us to reject the null hypothesis $H_0^2$ for several areas,} especially in Cluster 3, aggregation of the choices over all areas leads to $p$-values lower than $\alpha=0.05$ for all clusters. Therefore, among all clusters, there is indeed \repA{statistically significant}{some} agreement among participants.
Cluster 1 ($N=93$) shown in \autoref{fig:ppo-cluster1} with mean consistency $\zeta_1=0.833$ $(SD = 0.179,\ \repA{MD}{Q2} = 0.900)$ and fairly high agreement $u_1=0.335$ ($\mathrm{min}\ u_1=-0.004$), comprises participants \repA{that show strong preference in}{prefer} central positions T, B, R, and partly L over to corner positions BL, TL, TR, and BR. 
Cluster 2 ($N=41$) depicted in \autoref{fig:ppo-cluster2} with mean consistency $\zeta_2=0.844$ $(SD = 0.155,\ \repA{MD}{Q2} = 0.900)$ also shows considerable agreement $u_2=0.370$ ($\mathrm{min}\ u_3=-0.008$) and contains participants that \addA{strongly} favor label positions T, B, TR as opposed to L, BL, R, BR, and TL. 
Cluster 3 ($N=91$) presented in \autoref{fig:ppo-cluster3} with mean consistency $\zeta_3=0.435$ $(SD = 0.259, \repA{MD}{Q2} = 0.350)$ and relatively low agreement $u=0.021$ ($\mathrm{min}\ u=-0.004$) includes participants that are uncertain in their preferences but lean towards TR position.

\section{Evaluation of \texttt{PerceptPPO}}
In order to evaluate the established \texttt{PerceptPPO} and compare it with the other PPOs typesetted in bold in \autoref{tab:ppo-overview}, we have conducted a series of the following \repA{evaluations}{experiments}.
We do not include the position modifiers SL (Slightly Left) and SR (Slightly Right) in our study. Initially, the modifiers were introduced due to limitations in grid printing, specifically for labels with an even number of letters that could not be precisely centered. However, with current advancements in typesetting, the necessity for these auxiliary positions has become obsolete. Furthermore, we examine at most eight positions to reduce the complexity of a time-demanding pairwise comparison study, as the additional positions are relatively uncommon. The additional criteria for the selection were the number of citations (applied for Brewer~\cite{Brewer2015}, Christensen and Marks~\cite{Christensen1995}, Slocum~\cite{Slocum2009}, Imhof~\cite{Imhof1975}), the founder aspect (Imhof~\cite{Imhof1975}, Yoeli~\cite{Yoeli1972}), and similarity with \texttt{PerceptPPO} (Zoraster~\cite{Zoraster1997}). In the following text, we typeset these using \texttt{typewriter} font to denote the corresponding PPO and abbreviate the PPO proposed by Christensen and Marks only by the first author's name (\texttt{Christensen}).

\subsection{\texorpdfstring{\repA{Evaluation}{Experiment}}{Evaluation} 1: Label Density}
\label{sec:label-density}
We need map renders to evaluate and compare the \texttt{PerceptPPO} with existing PPOs. 
\addA{Again, we intend to create a blind map to eliminate factors potentially influencing the judgment of the label placement other than its position relative to the anchor.}
However, by doing so, we faced a question. How many labels should be presented in such a map area? We reviewed existing cartographic books and found that, surprisingly, just a handful of works studied this topic \cite{Liao19}. Therefore, we conducted a dedicated \chA{experiment} to see users' preferences on label density.

\subsubsection{Data}
\label{sec:label-density-data}
We selected ten populated areas (0, 4, 5, 6, 9, 12, 13, 17, 27, 29), providing a wide range of possible label density samples from areas employed in the \texttt{PerceptPPO} study. \addA{We limited our selection to ten areas because not all 30 locations described in \autoref{sec:percept-ppo-data} had sufficient populations to sample the varying levels of label density required for the experiment. Additionally, reducing the number of areas streamlined the experiment, making it more manageable and efficient for participants.}
For each area, we produced maps with varying density of labels which the participants can compare. We have several requirements on each map: 
\begin{enumerate*}[label=(\arabic*)]
\item The Global Label Density ($GLD$) of the whole map $M$ is lower than a given threshold $LD_\mathrm{thr}$ and
\item to prevent local dense clusters of labels in highly populated areas, the Local Label Density ($LLD$) of local neighborhood of each anchor has to be lower than the given threshold $LD_\mathrm{thr}$. 
\end{enumerate*}
\begin{figure*}[!ht]
    \subfigure[\fontsize{6.2pt}{2pt}\selectfont{$LD_\mathrm{thr} = 2.5\%$, \chB{$\widehat{LLDF} = 3.4\%$, $\overline{LLDF} = 3.9\%$}, $GLD = 2.5\%$}] {
        \includegraphics[width=0.332\textwidth, frame, trim={0cm 0cm 0cm 4cm}, clip]{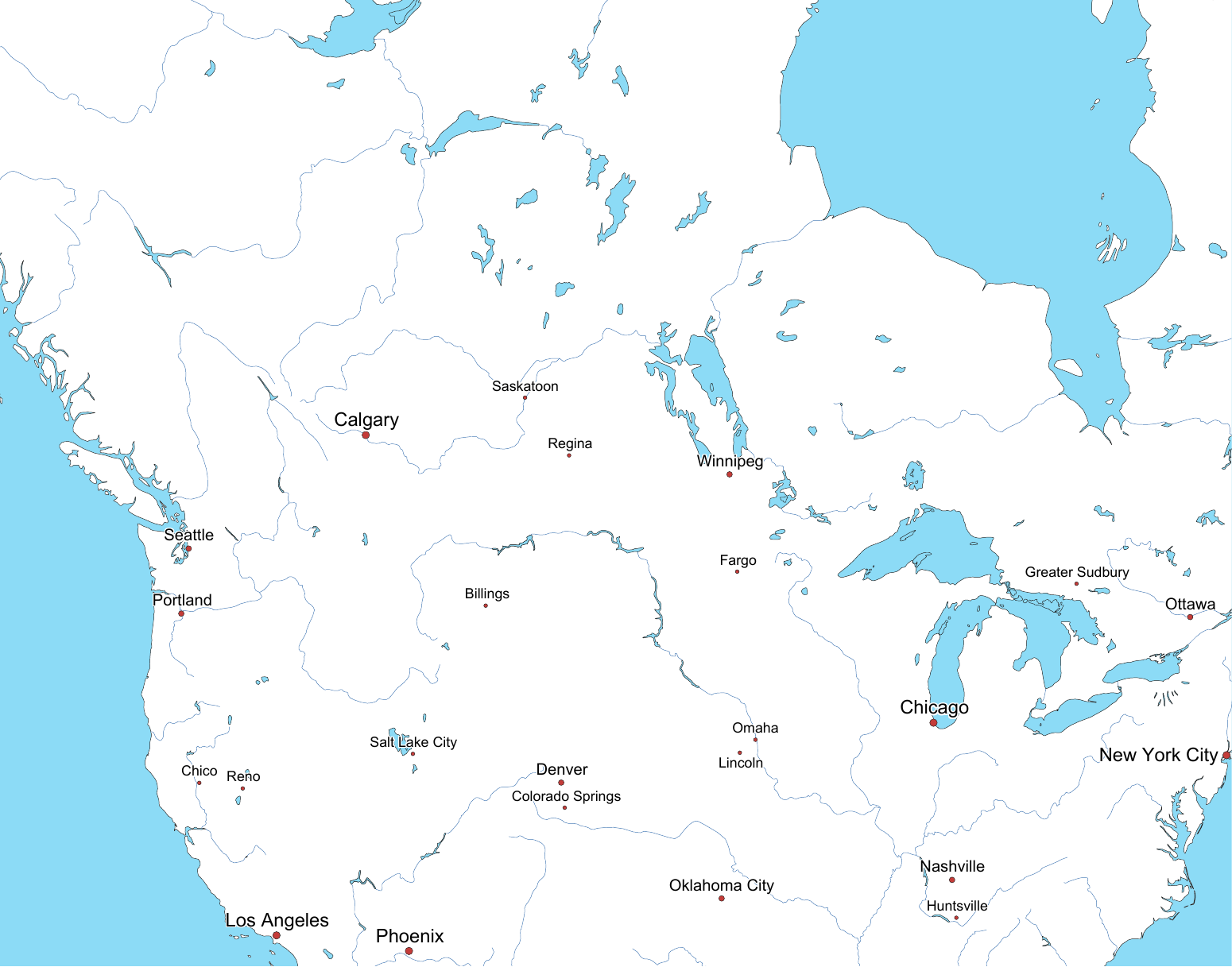}
    }
    \hspace{-1em}
    \subfigure[\fontsize{6.2pt}{2pt}\selectfont{$LD_\mathrm{thr} = 12.5\%$, \chB{$\widehat{LLDF} = 16\%$, $\overline{LLDF} = 15.8\%$}, $GLD = 12.5\%$}] {
        \includegraphics[width=0.332\textwidth, frame, trim={0cm 0cm 0cm 4cm}, clip]{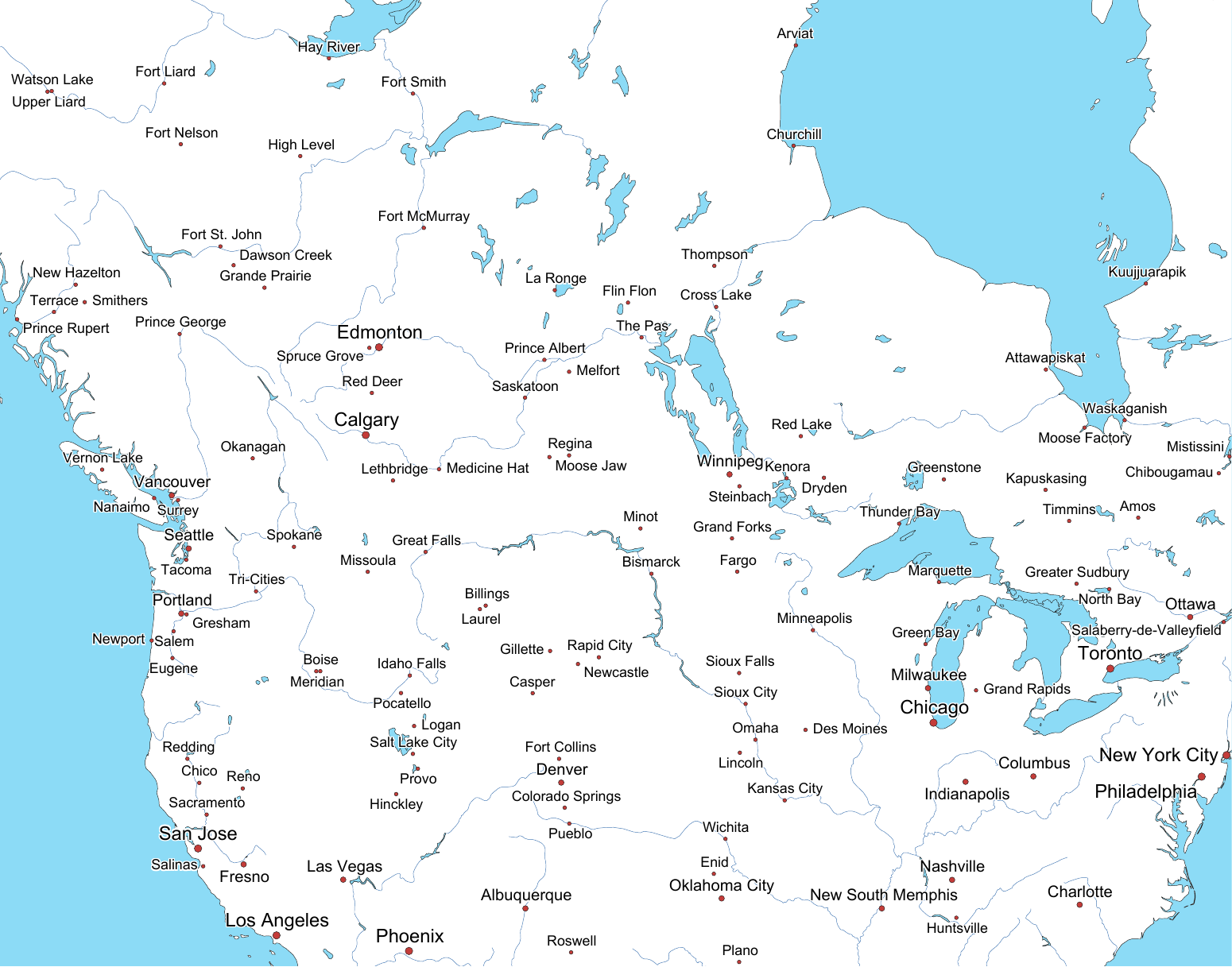}
    }
    \hspace{-1em}
    \subfigure[\fontsize{6.2pt}{2pt}\selectfont{$LD_\mathrm{thr} = 100\%$, \chB{$\widehat{LLDF} = 50.1\%$, $\overline{LLDF} = 46.7\%$}, $GLD = 31.3\%$}] {
        \includegraphics[width=0.332\textwidth, frame, trim={0cm 0cm 0cm 4cm}, clip]{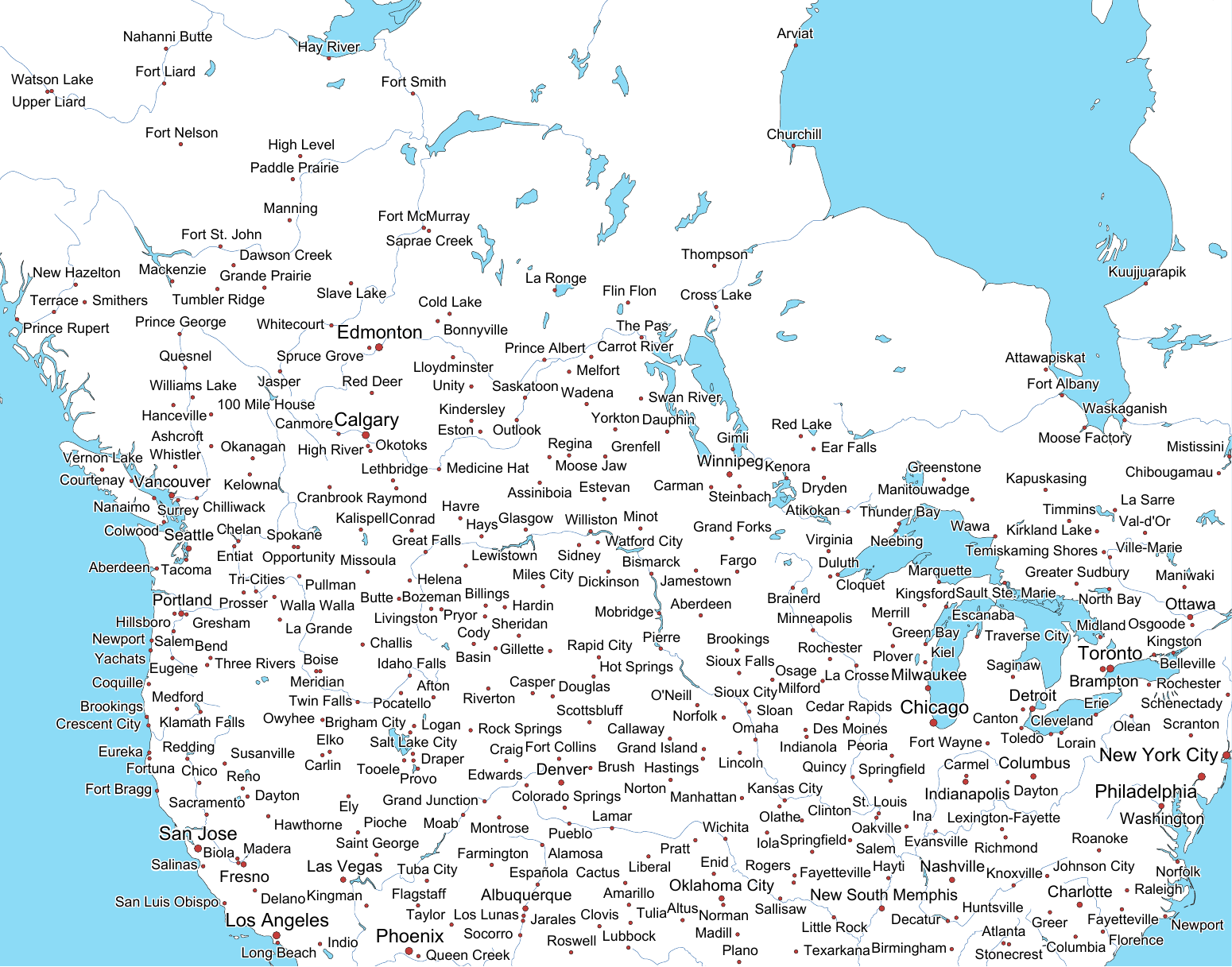}
    }
    \caption{
        Renders of map area 0 using \texttt{PercepPPO} at various values of $LD_\mathrm{thr}$ employed in \chA{the evaluation of} label density 
        described in \autoref{sec:label-density}.
    }
    \label{fig:areas-density}
\end{figure*}
Both these densities are expressed in percentages. 

We define $LLD$ for each anchor $\alpha$ in map $M$, such that the anchor is the center of a tile $T$ of size $256\times256$, which is the typical size of web-based raster maps. The $LLD$ is defined as:
\begin{equation}
    LLD(\alpha) = \frac{\sum_{a \in M} L(a) \cap T}{A(T)},
\end{equation}
where $a$ is an anchor, $T$ is the square tile, $L(a)$ is the rectangle enclosing the label text corresponding to an anchor $a$, and $A(T)$ is the area of the tile. If the anchor is in the map $M$ positioned such that \addA{$T$ is not entirely in the map $M$, we perform a minimal shift of the tile $T$ to be entirely in the map $M$.}

Similarly, we define GLD for each map $M$ as
\begin{equation}
    GLD(M) = \frac{\sum_{a \in M} L(a) \cap M}{A(M)}.
\end{equation}

\repA{We order the cities in the given map area by population size, from most populated to least populated, and process them iteratively. We sequentially try to add anchors and labels to the map, starting with the largest population and proceeding to the smallest.}{We order the cities in the given map area by population size and process them iteratively. We sequentially try to add anchors and labels to the map, from the largest to the smallest.} The anchor $\alpha$ is added to map $M$ at position $p$ from PPO if all of the following conditions are met: 
\begin{enumerate*}[label=(\arabic*)]
    \item \repA{Anchor $\alpha$ is in the map $M$,}{$\alpha \in M$,}
    \item \repA{the label $L$ of anchor $\alpha$ is entirely in the map $M$,}{$L(\alpha) \cap M = L(\alpha)$,}
    \item \repA{the label $L$ of anchor $\alpha$ is not overlapping any anchor,}{$\forall a \in M, \, L(\alpha) \cap a = \emptyset$,}
    \item \repA{the label $L$ of anchor $\alpha$ is not overlapping any already placed label,}{$\exists p \, \forall a \in M, \text{ such that } \alpha \neq a \, \land \, L(\alpha) \cap L(a) = \emptyset$,}
    \item \repA{the local label density $LLD$ of anchor $\alpha$ is lower than $LD_\mathrm{thr}$,}{$LLD(\alpha) \leq LD_\mathrm{thr}$,}
    \item \repA{the global label density $GLD$ of the map $M$ is lower than $LD_\mathrm{thr}$.}{$GLD(M) \leq LD_\mathrm{thr}$.}
\end{enumerate*}

After no other anchors can be labeled 
\repA{according to the above rules,}{without conflict} we recompute the final $LLD$ ($LLDF$) for each placed anchor as we do not update $LLD(a_1)$, when label $L(a_2)$ protrudes from $T_2$ into $T_1$ and label $L(a_2)$ is placed after $L(a_1)$. Therefore, $LLDF$ can be slightly above $LD_\mathrm{thr}$.

\chB{We aggregate $LLDF$ within each map $M$ by calculating the median and mean values across all anchors $\alpha \in M$, capturing overall statistics and allowing for consistent comparison of local label density across different maps. Specifically, these aggregated values are calculated as follows:}
\begin{align}
    \chB{\widehat{LLDF}(M)} &= \text{median}_{\alpha \in M}(LLDF(\alpha)) \\
    \chB{\overline{LLDF}(M)} &= \text{mean}_{\alpha \in M}(LLDF(\alpha))
\end{align}

To prepare various samples of selected areas with \repA{various}{different} levels of label densities, we repeatedly rendered the area with $LD_\mathrm{thr} = (2.5\% - 40\%, \mathrm{step\ size}\ 2.5\%) \cup (45\%, 50\%, 75\%, 100\%)$. The upper part is more sparse because, with increasing levels of label density, the renders became perceptually similar. For example, only a few labels create differences between 40\% and 42.5\% or 50\% and 55\%. Therefore, we increased the step size to 5\% from 40\% and to 25\% from 50\%. If any rendered maps of a given area were the same, we kept only unique map renders.

Following this procedure, we rendered ten selected areas (see \autoref{fig:areas-density} and for more examples, refer to the suppl.\,material) for selected PPOs from \autoref{tab:ppo-overview}. \delA{Again, }We used the same data source for cities as in \autoref{sec:percept-ppo-data}. 

We \repA{employed}{rendered} the city labels in three different sizes, as suggested by \addA{various guidelines}~\cite{Robinson1995, Slocum2009, Brewer2015, Krygier2016} \repA{, because}{as} more than three categories are perceived with difficulty according to Robinson~\etal~\cite{Robinson1995}. As suggested by Tyner~\cite{Tyner2014}, we split the population into three intervals: \delA{(}$<500,000$\delA{)}; \delA{(}500,001--1,000,000\delA{)}; and \delA{(}$>1,000,000$\delA{)}; and apply three different font sizes of 11pt, 13pt, and 15pt, differing by 2pt as recommended by Robinson~\etal~\cite{Robinson1995}.

\addA{For the evaluation of label density, we are not using a blind map, but we included the borders of continents and water bodies in terms of rivers, lakes, seas, and oceans (see \autoref{fig:areas-density} and the suppl.\,material). By doing so, we aimed to provide the participants with information about where the cities may occur and where they might not (\eg a lake or ocean) %
while minimizing the factors affecting the participants' decision process. We used the data provided by Natural Earth}\footnote{\url{https://www.naturalearthdata.com/downloads/}}, \addA{specifically land, lakes and reservoirs, rivers, and lake central lines. We selected data from available scales at 1:50,000,000 due to the trade between precision and image size as we employ SVG vector format at a size of $1305\times1025$ pixels.}

\subsubsection{Data Verification}
\begin{figure*}[!ht]
    \subfigure[All $LD_\mathrm{thr}$ aggregated at $1305\times1025$] {
        \label{fig:density-placed-labels-count-all_1305}
        \includegraphics[width=0.335\textwidth]{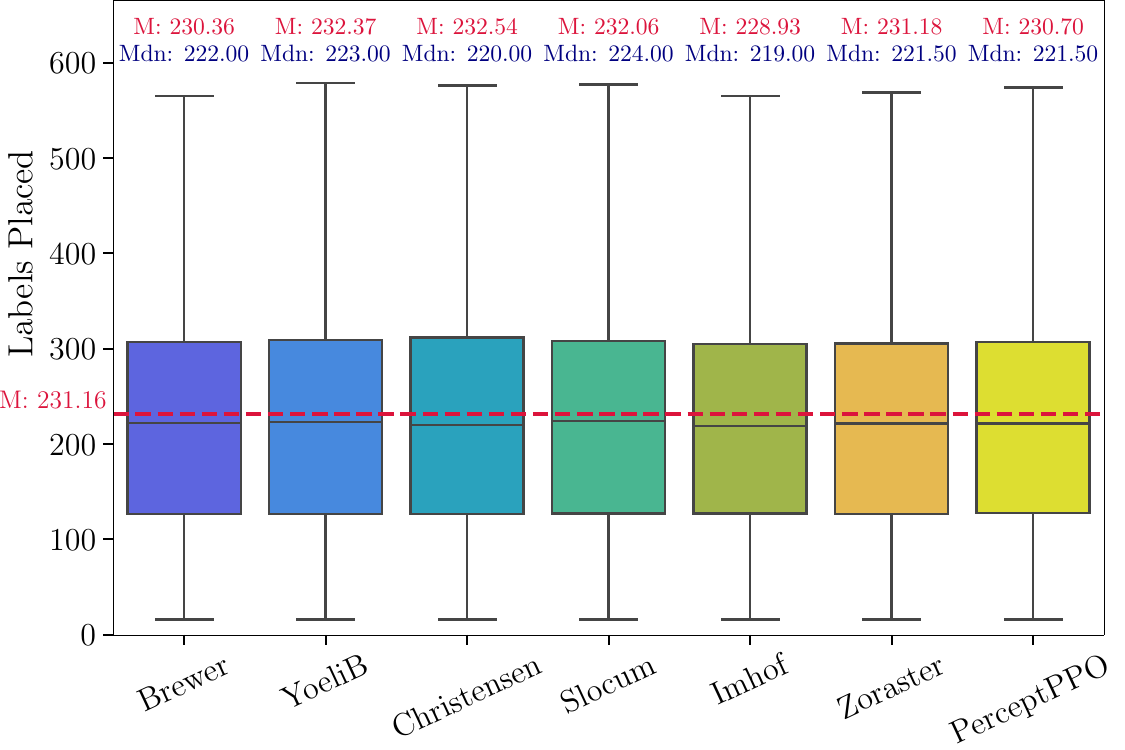}
    }
    \hspace{-1em}
    \subfigure[$LD_\mathrm{thr} = 12.5\%$ at $1305\times1025$] {
        \label{fig:density-placed-labels-count-12-5_1305}
        \includegraphics[width=0.335\textwidth]{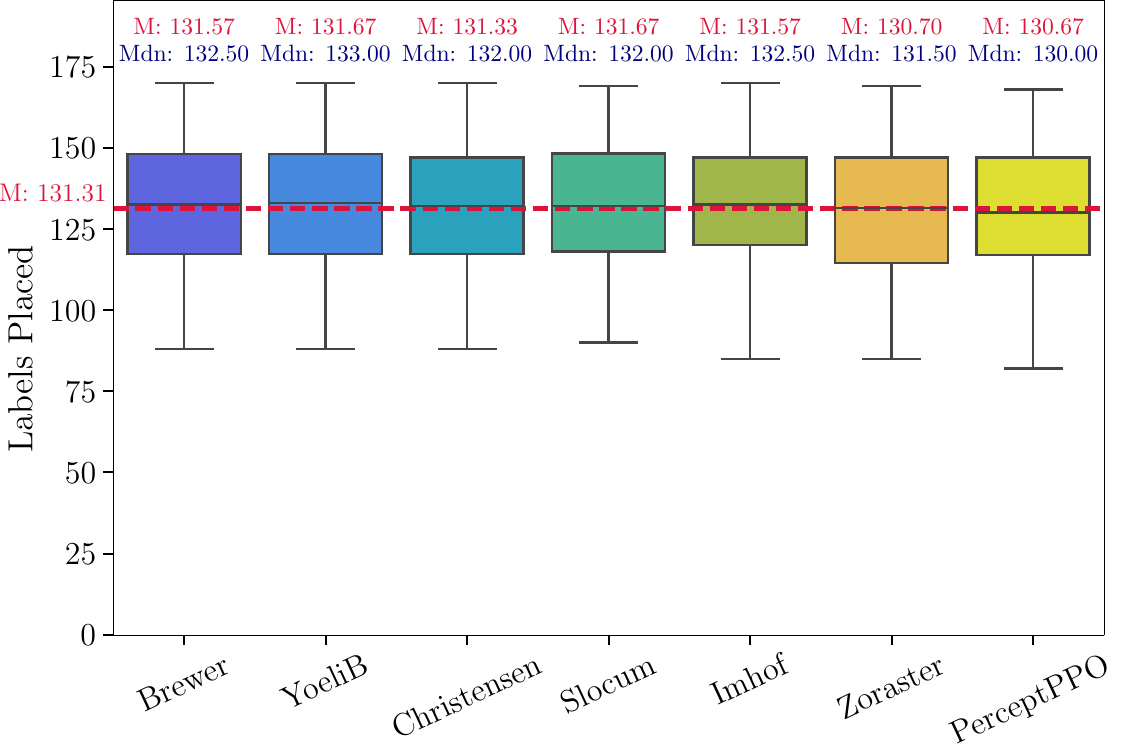}
    }
    \hspace{-1em}
    \subfigure[$LD_\mathrm{thr} = 12.5\%$ at $652\times512$] {
        \label{fig:ppo-eval-placed-labels-count12-5_652}
        \includegraphics[width=0.335\textwidth]{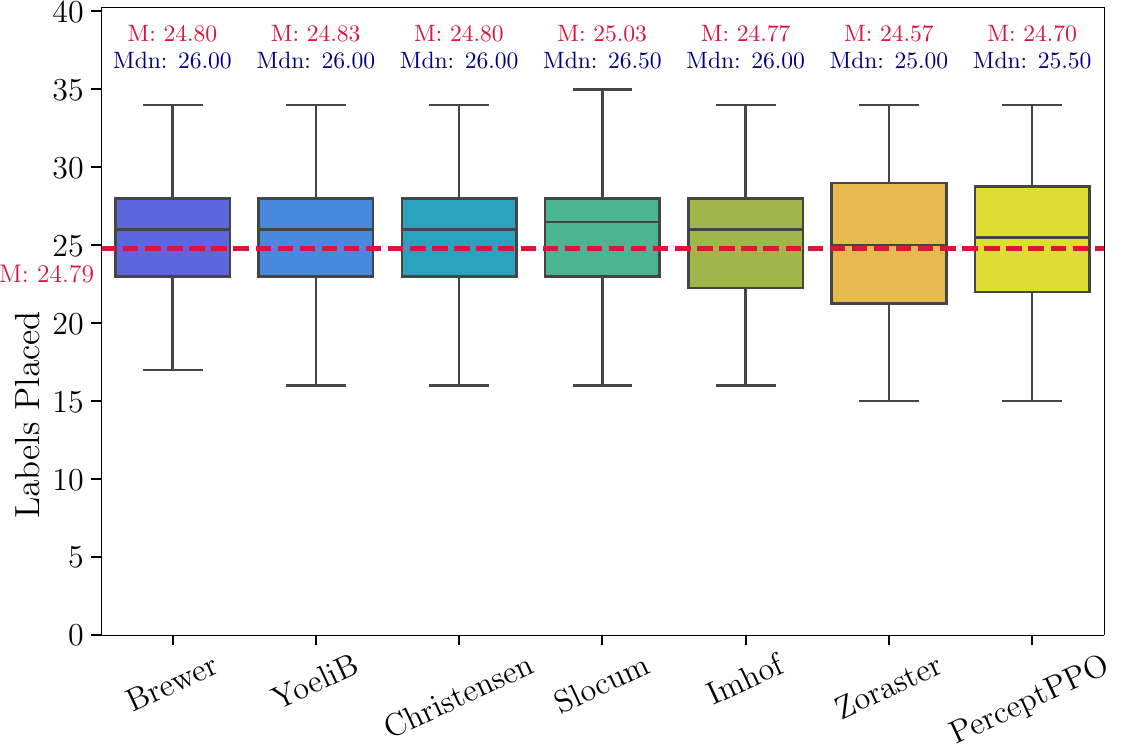}
    }
    \caption{
            \subref{fig:density-placed-labels-count-all_1305} The box plot of placed labels for each examined PPO aggregated for all values of $LD_\mathrm{thr}$ at $1305\times1025$.
            \subref{fig:density-placed-labels-count-12-5_1305} The box plot of placed labels for label density $LD_\mathrm{thr} = 12.5\%$ at $1305\times1025$. \subref{fig:density-placed-labels-count-12-5_1305} The box plot of placed labels for label density $LD_\mathrm{thr} = 12.5\%$ at $652\times512$.
        }
    \label{fig:labels-placed-data-compare}
\end{figure*}

We measured the number of positioned labels for all PPOs to ensure that anchors' spatial configurations within selected areas do not discriminate against or favor a particular PPO. The \autoref{fig:density-placed-labels-count-all_1305} and \autoref{fig:density-placed-labels-count-12-5_1305} show that the number of placed labels is very similar for all PPOs.

\addA{Additionally,} we explored the adherence of label placements to their designated priorities and examined how increasing label densities influence the label placement. For each selected PPO outlined in \autoref{tab:ppo-overview}, we have measured the probability of labels occupying specific positions and observed how these probabilities shift with escalating label densities.

\begin{figure*}[!ht]
    \includegraphics[width=\textwidth]{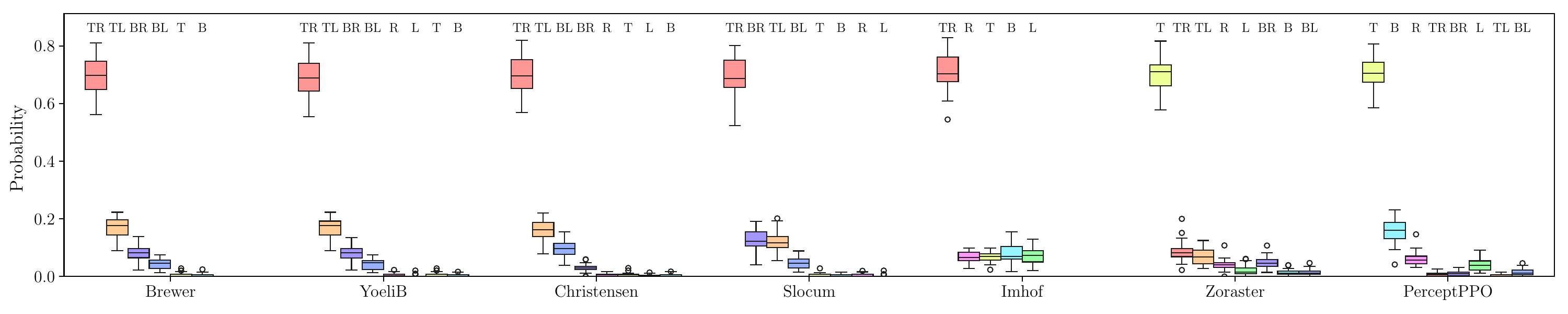}
    \caption{
        The probability of a label being placed at a given position over all 30 locations rendered as described in \autoref{sec:label-density-data} with label density $LD_\mathrm{thr}=12.5\%$.
    }
    \label{fig:probability_all_methods}
\end{figure*}

As expected, labels are most likely to be positioned in their highest-priority position. The probability of the label being placed at the second highest priority was significantly lower, decreasing progressively for lower-priority positions (refer to \autoref{fig:probability_all_methods} and \autoref{fig:probabilities_density_all_methods} in the suppl.\,material for additional insights into label density impacts). Nonetheless, some patterns deviated from this trend.

We observed a striking deviation with priorities of Imhof~\cite{Imhof1975}, where labels at lower densities were equally likely to occupy the second to fifth priority positions. However, as label density increased, this likelihood inverted, favoring lower-priority positions, realigning the priority sequence from the intended TR, R, T, B, L to TR, L, B, T, R for maps with greater label density. This density-dependent realignment in positioning probabilities is detailed in \autoref{fig:probabilities_density_all_methods} in the suppl.\,material.

We also noted exceptions in the PPOs set by Slocum~\cite{Slocum2009}, Zoraster~\cite{Zoraster1997}, and our \texttt{PerceptPPO}. For the priorities of \texttt{Slocum}, labels at higher densities are more likely to be placed at TL than at BR, contrary to the intended order BR, TL. Similarly, for the priorities of \texttt{Zoraster}, the likelihood of position BR is higher than for L or R, contrary to intended order R, L, BR. Likewise, \texttt{PerceptPPO} demonstrated a higher placement likelihood at L over TR or BR. The suppl.\,material further illustrates these individual probability patterns across different PPOs and their relation to label density in \autoref{fig:probabilities_density_all_methods}.

\subsubsection{Procedure}
Participants could access the study directly via a web application or indirectly via the Mechanical Turk interface, which embedded the same web app within its environment. \addA{Participation in the evaluation was voluntary, and the compensation for Mechanical Turk workers was set to match the average compensation rate of other requesters on Amazon Mechanical Turk.} The web application included an introduction, a survey, an \repA{evaluation}{experiment}, and a feedback section. The introduction defined the participant's task and how to use the application. Within the \repA{evaluation}{experiment} we presented rendered maps of selected areas in randomized order. Each participant was assigned a single PPO for all ten areas. Each participant was allowed to participate only once to mitigate the carry-over effect. For each area, we preloaded all renders with various label density levels. Using the slider, the participants were asked to \repA{``}{"}choose a label density they found comfortable without being overwhelmed by the amount of information.\repA{''}{"} The leftmost position was the lowest label density, and the rightmost position was the highest density. The participant must spend at least 5 seconds carefully selecting the preferred label density and explore the full range of label densities. When they complete all ten areas, we provide them the option to provide any feedback.

\subsubsection{Results}
We eliminated participants who exhibited inconsistencies within their responses across different areas and ones that deviated more than two standard deviations from the \chB{mean of} $\overline{LLDF}$ and \repA{evaluation}{experiment}
duration, as described in \autoref{sec:online-precautions}. After this data refinement, we engaged 110 %
participants from Mechanical Turk with a relatively high dropout of 45\%, which we attribute to a spike of inattentive workers according to the review of activities recorded by Smartlook. Eliminated workers consistently set the slider in the right or leftmost position, which is highly unlikely to be the preferred position according to the control group. 

\chB{We found the following overall statistics on the preference of label density threshold $LD_\mathrm{thr}$ from participants: $\mathrm{median}(LD_\mathrm{thr}) = 12.5\%$ ($\mu = 14.85,\ SD=7.22\%$). We determined the preferred $LD_\mathrm{thr}$ using the median, as it provides a robust measure that minimizes the impact of potential outliers. The preferred $LD_\mathrm{thr} = 12.5\%$ corresponds to the $\mathrm{median}(GLD) = 12.5\%$ ($\mu = 14.54\%$,\ $SD = 6.59\%$), and a median of \chA{150} labeled anchors ($\mu = 159$,\ $SD = 82$).
To accurately reflect the trends in participants' opinions on the local label density, we calculated the overall preference statistics from the aggregated median $\widehat{LLDF}$ and mean $\overline{LLDF}$, respectively. 
Specifically, the preferred $LD_\mathrm{thr} = 12.5\%$ corresponds to $\mathrm{median}(\widehat{LLDF}) = 16.3\%$ ($\mu = 17\%$,\ $SD= 7.35\%$).
}

Upon conducting an Analysis of Variance (ANOVA) to investigate the effect of different PPOs on the $LD_\mathrm{thr}$, our results indicated no significant differences across the various PPOs. Specifically, the ANOVA test, utilizing a Type II sum of squares approach, yielded an F-statistic of $1.660$ with a corresponding $p$-value of $0.127$. The $p$-value suggests that we found no statistically significant differences in $LD_\mathrm{thr}$ values among PPOs at significance level $\alpha = 0.05$.

The finding implies that the examined PPOs do not significantly affect the $LD_\mathrm{thr}$ value, reinforcing the idea that the variations observed in $LD_\mathrm{thr}$ across different PPOs might be attributed to random chance rather than inherent differences in PPOs. Therefore, our analysis supports the conclusion that the selection of PPOs does not significantly influence the \chA{preferred} $LD_\mathrm{thr}$. 

\subsection{\texorpdfstring{\repA{Evaluation}{Experiment}}{Evaluation} 2: Comparison of PPOs}
\label{sec:ppo_eval}
To compare the proposed \texttt{PerceptPPO} with existing PPOs, we conducted an \repA{evaluation}{experiment} that follows findings from the \addA{evaluation of} label density \delA{experiment} 
described in \autoref{sec:label-density}. We aim to validate that the \texttt{PerceptPPO} is preferred when the label cannot always be placed in a single position (see \autoref{fig:areas-ppo_eval}), unlike the \texttt{PerceptPPO} \chA{experiment} described in \autoref{sec:percept-ppo}.

\subsubsection{Data}
\begin{figure*}[!ht]
    \subfigure[Brewer] {
        \includegraphics[width=0.246\textwidth, frame, trim={0cm 0cm 0cm 2cm}, clip]{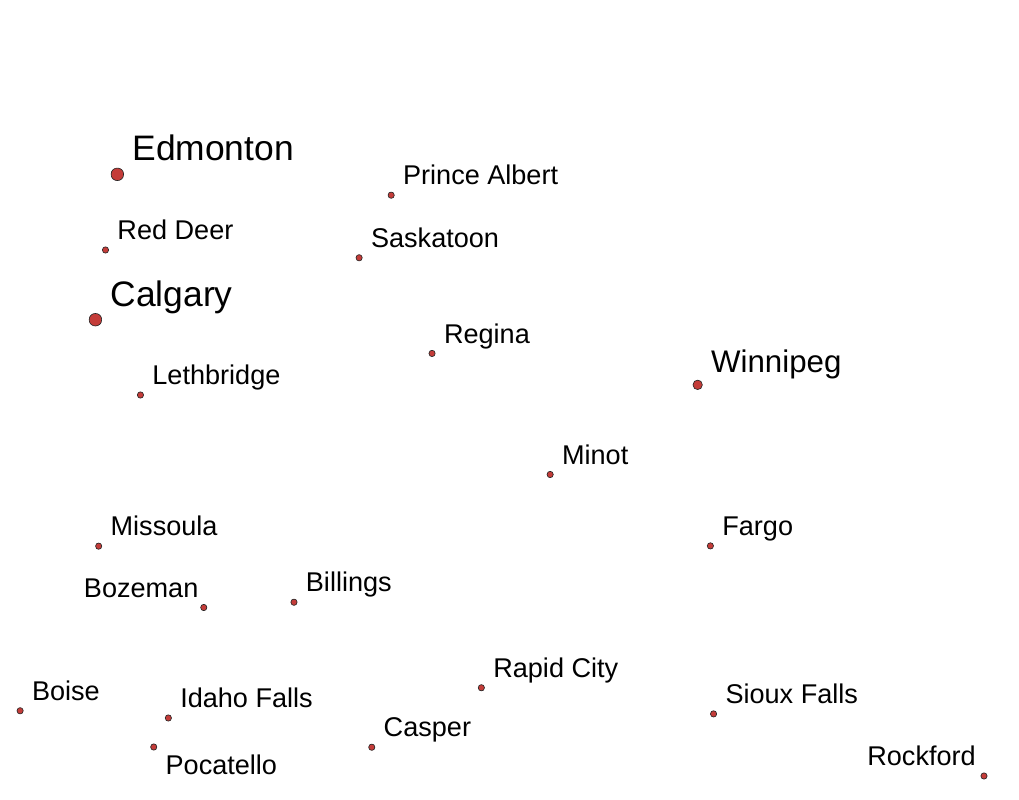}
    }
    \hspace{-1em}
    \subfigure[Christensen] {
        \includegraphics[width=0.246\textwidth, frame, trim={0cm 0cm 0cm 2cm}, clip]{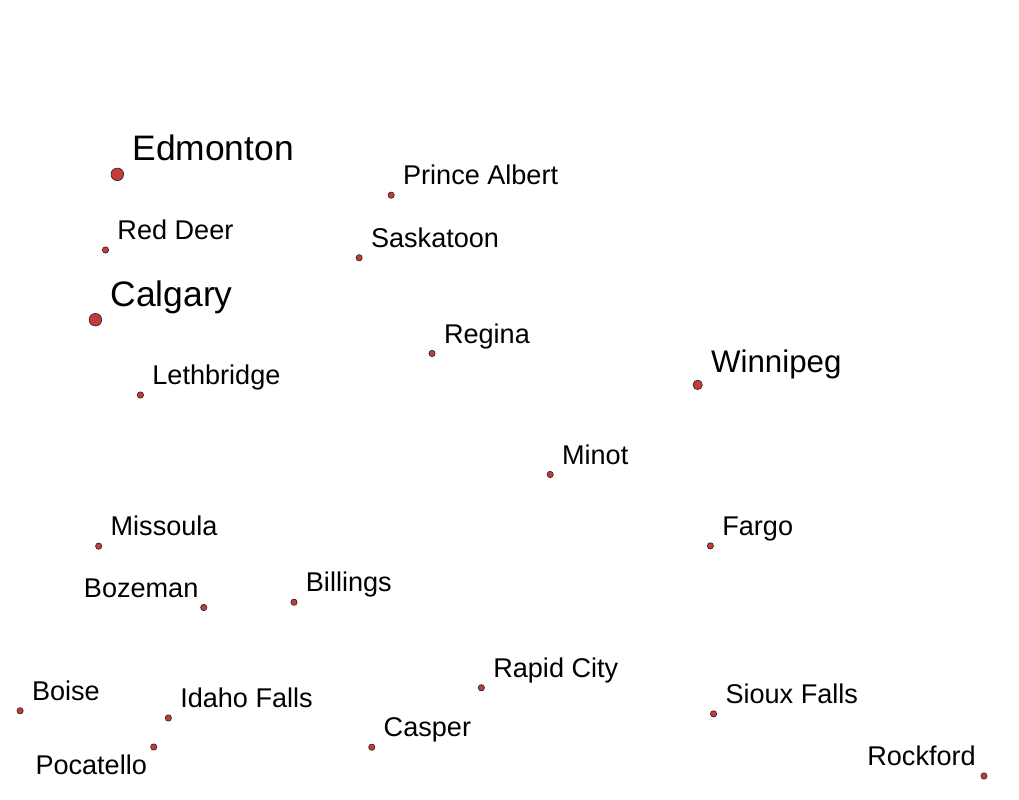}
    }
    \hspace{-1em}
    \subfigure[Zoraster] {
        \includegraphics[width=0.246\textwidth, frame, trim={0cm 0cm 0cm 2cm}, clip]{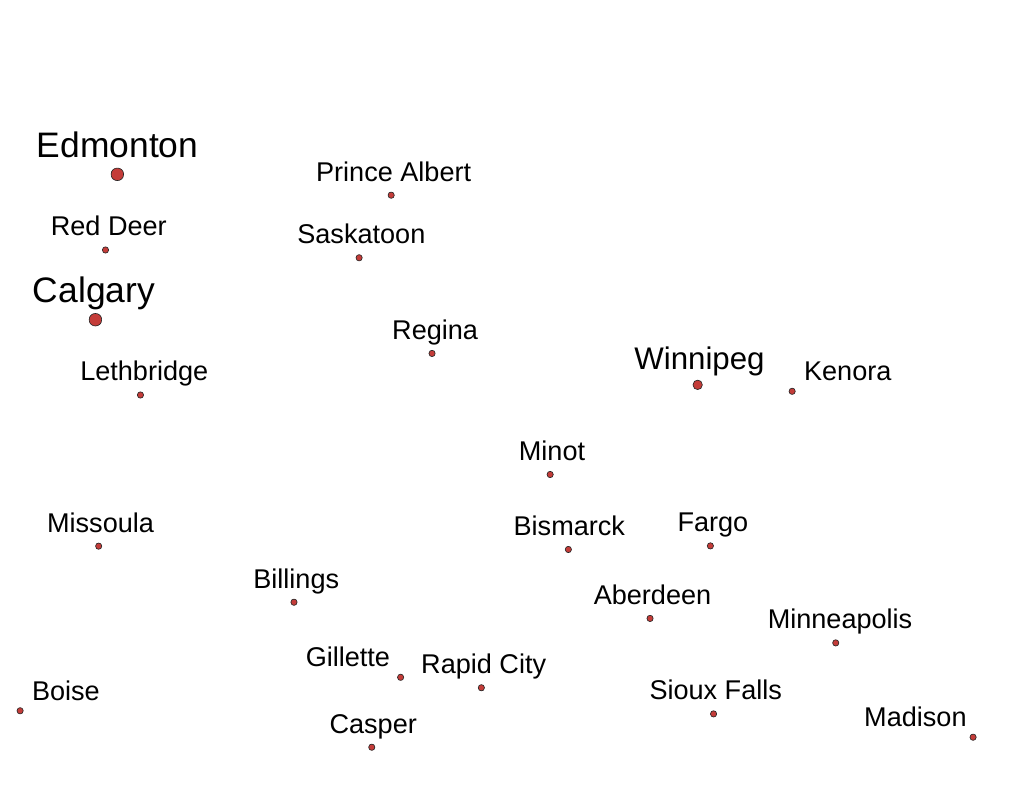}
     }
     \hspace{-1em}
     \subfigure[PerceptPPO] {
        \includegraphics[width=0.246\textwidth, frame, trim={0cm 0cm 0cm 2cm}, clip]{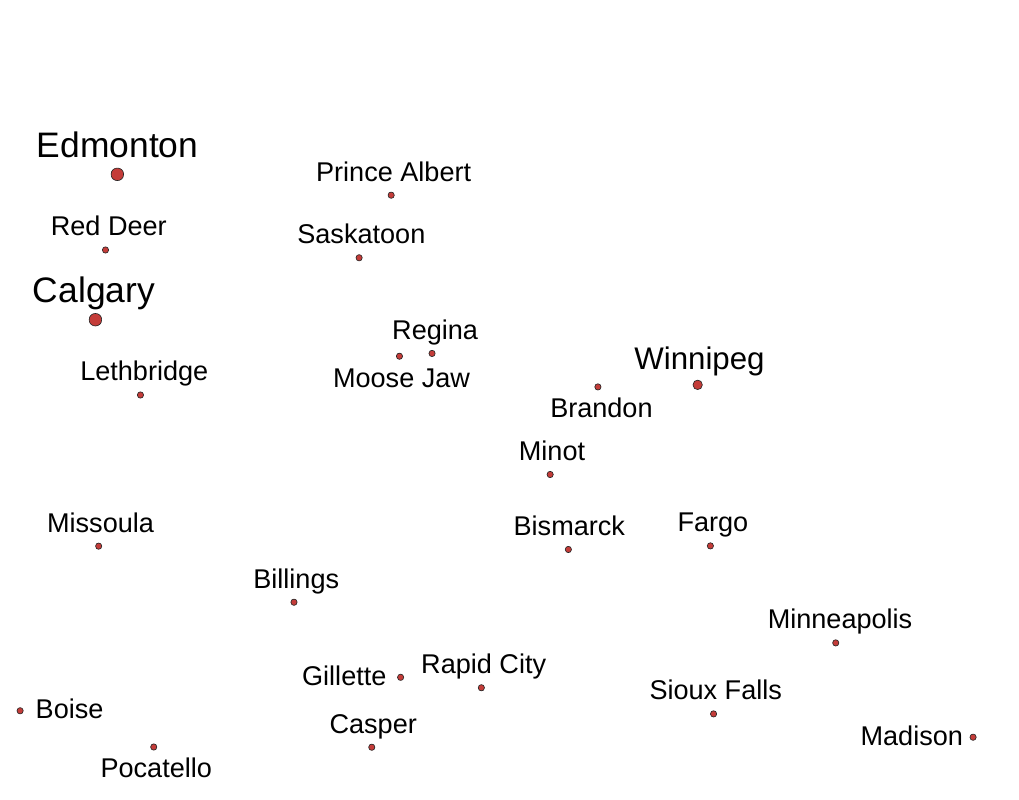}
     }
    \caption{
        Renders of map area 0 using various PPOs (at $LD_\mathrm{thr}=12.5\%$) employed in comparison of PPOs described in \autoref{sec:ppo_eval}.
    }
    \label{fig:areas-ppo_eval}
\end{figure*}

In this \repA{evaluation}{experiment}, we employ the same approach as described in \autoref{sec:label-density-data} with $LD_\mathrm{thr}=12.5\%$. However, this time, we \addA{again use a blind map to eliminate factors potentially influencing the judgment of the label placement other than its position relative to the anchor.}  We have also reduced the size of renders to half $652\times512$, which means that the %
\chA{average} number of shown anchors and labels is \chA{24.79} (%
$SD = 5.91$), contrary to \chA{131.31} (%
$SD = 21.01$) for $1305\times1025$, see \autoref{fig:labels-placed-data-compare} and the suppl.\,material \addA{for more details}. We found out that comparing map pairs with 131.31 labels on average is overly demanding and cannot be completed in a reasonable time. Therefore, we chose to reduce the size of the map area while still maintaining the $LD_\mathrm{thr}=12.5\%$ but with 24.79 presented labels on average, see \autoref{fig:areas-ppo_eval}. In this \repA{evaluation}{experiment}, we also use all 30 locations worldwide as described in \autoref{sec:percept-ppo-data}. Other data characteristics hold with \autoref{sec:label-density-data}, except the background is blank as in \autoref{sec:percept-ppo-data}.

\subsubsection{Data Verification}
We measured the number of placed labels for all PPOs again to ensure that anchors' spatial configurations within selected areas do not discriminate against or favor a particular PPO. The \autoref{fig:ppo-eval-placed-labels-count12-5_652} show that the number of placed labels is very similar for all PPOs.

Finally, we explored the adherence of label placements to their designated priorities for the data used in the \repA{evaluation}{experiment} ($652\times512$ size of renders at $LD_\mathrm{thr}=12.5\%$). Refer to \autoref{fig:probabilities_density_all_methods_2} in the suppl.\,material \addA{for more details}. We found the same deviations for the priority orders of \texttt{Imhof}, \texttt{Zoraster}, and \texttt{PerceptPPO} as for the data with $1305\times1025$ size of renders at $LD_\mathrm{thr}=12.5\%$ (see \autoref{sec:label-density-data}).

\subsubsection{Procedure}
We applied the 2AFC paradigm to determine the preferred PPO based on the perception of users. Considering the seven PPOs under examination, we have $\binom{7}{2} = 21$ pairwise comparisons on an area. Consequently, the entire \repA{evaluation}{experiment} consists of $30\times21 = 630$ pairwise comparisons to cover all of them once. In order to alleviate potential fatigue among the participants during the study, we allocated only three areas to each participant, resulting in a batch of $3\times21 = 63$ pairwise comparisons. Therefore, 10 participants were required to cover all pairwise comparisons once.
We engaged Mechanical Turk workers and university students to perform the \repA{evaluation}{experiment}. The rest of the procedure is the same as for \autoref{sec:ppo-procedure} except, participants were shown two maps sequentially, each depicting the same area but with different PPOs.

\subsubsection{Results}
We eliminated participants with inconsistent responses as described in \autoref{sec:online-precautions}. Following this data refinement, we were left with a total of 352 participants \chA{(179 females and 173 males)} with a dropout of 24\%.

\chA{A majority of the participants hailed from the USA (221), followed by India (54), Czech Republic (32), Brazil (7), and United Kingdom (5). The most common age range among the participants was 20-30 years, with 98 individuals falling into this category, followed by 31-40 with 93 participants, 41-50 with 77 participants, 51-60 with 47 participants, $>60$ with 35 participants and 2 participants bellow 20. Regarding educational qualifications, the highest number of participants held bachelor’s degrees (163), master’s degrees (76), followed by high school diplomas (67), community college education (38), doctoral’s degrees (7), and elementary education (1). On average, participants completed a batch of 63 pairwise comparisons in 10 minutes and 46 seconds ($SD=$ 3 minutes and 13 seconds).}

\chA{The} %
participants generated 22,236 pairwise comparisons. Therefore, on average, each comparison pair was evaluated by approximately 35 different participants.

\chA{We apply methodology as described in \autoref{sec:pwcmp} to compute the quality z-score and assess statistical significance at the significance level of $\alpha = 0.05$ to evaluate the null hypothesis $H_0^3$: ``There is no clear user preference among the examined PPOs.''}

\begin{figure*}[!ht]
    \subfigure[] {
        \label{fig:ppo_eval-quality-score}
        \includegraphics[width=0.4798\textwidth]{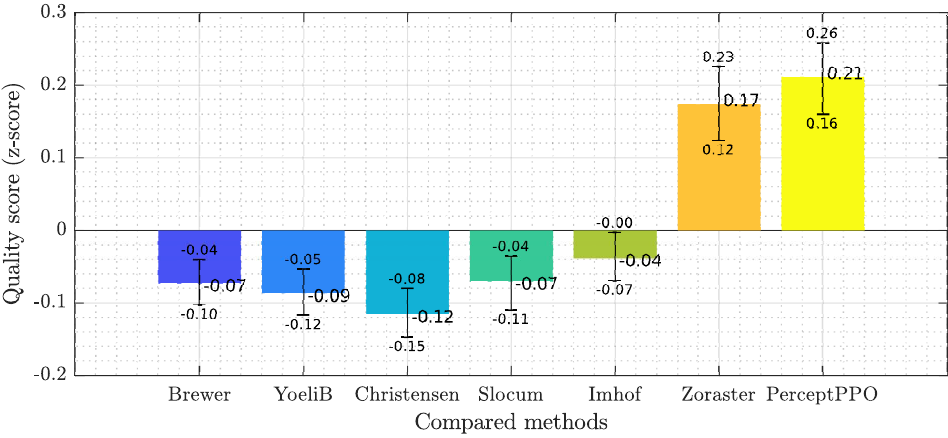}
    }
    \hspace{1em}
    \subfigure[] {
        \label{fig:ppo_eval-significance}
        \includegraphics[width=0.4798\textwidth]{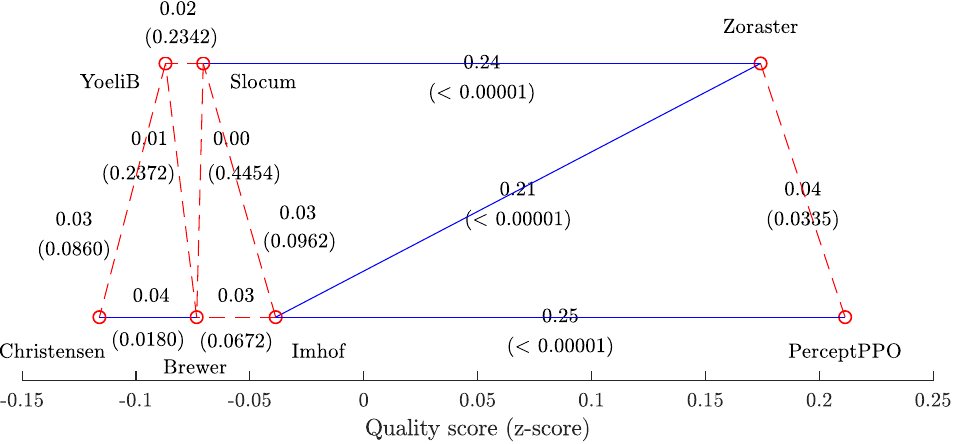}    
    }
    \caption{Overall comparison of PPOs. Chart \subref{fig:ppo_eval-quality-score} depicts quality score and 95\% confidence intervals. \chA{Chart \subref{fig:ppo_eval-significance} shows a triangle plot visualizing significant differences between PPO preferences as proposed by P\'{e}rez-Ortiz and Mantiuk~\cite{Perez2017}. Each red circle represents a PPO, and the lines indicate significant differences between pairs. Solid blue lines represent statistically significant differences, while the dashed red lines indicate non-significant ones. The edge values show the absolute difference in z-scores between the compared PPOs, with the $p$-values denoted in brackets. The PPOs are plotted along the x-axis, with alternating y-axis offsets for clarity.}
    }
    \label{fig:ppo_eval}
\end{figure*}

\chA{The overall results, as depicted in \autoref{fig:ppo_eval}, show a statistically significant difference between PPO groups $G_\mathrm{T}=$\{\texttt{PerceptPPO},\ \texttt{Zoraster}\} and $G_\mathrm{TR}=$\{\texttt{Brewer},\ \texttt{YoeliB},\ \texttt{Christensen},\ \texttt{Slocum},\ \texttt{Imhof}\}.
Specifically, the statistical analysis indicates that there is no significant difference within the groups $G_\mathrm{T}$, and $G_\mathrm{TR}$, but there is a significant difference between these two groups. \Autoref{fig:ppo_eval-significance} demonstrates that \texttt{PerceptPPO} and \texttt{Zoraster} consistently outperform other PPOs, justifying their grouping. Therefore, we can reject the null hypothesis $H_0^3$ and claim that there is a clear preference of PPOs only between groups $G_\mathrm{T}$ and $G_\mathrm{TR}$. In other words, participants perceived PPOs in $G_\mathrm{T}$ significantly better than in $G_\mathrm{TR}$.} 
\chA{The outcome validates the findings of our \texttt{PercepPPO} study described in \autoref{sec:percept-ppo} that participants prefer label position $T$ at the first place of PPO as proposed by \texttt{Zoraster} and \texttt{PerceptPPO}. Interestingly, the following label positions do not seem essential for perceiving the quality of PPOs.}

The overall \repA{average}{median of} consistency $\zeta$ across participants is \repA{0.59}{0.57} ($SD = 0.23$, \repA{$MD=0.57$}{$\bar{x}=0.59$}), which indicates that they were fairly consistent in their choices. The consistency remained reasonably uniform across each map area; details are provided in \autoref{tab:ppo-eval-results} in the suppl.\,material.
\repA{The}{However, the} overall coefficient of agreement $u=0.013$ ($\mathrm{min}\ u=-0.0009$) \repA{with}{reveals relatively low agreement among participants,  although} the $p$-value $= 7.362e^{-55}$ clearly shows that we can reject the null hypothesis $H_0^4$: ``There is no agreement among participants'' at $\alpha=0.05$ and conclude there is indeed \repA{statistically significant}{some} agreement\addA{ among participants}.

\begin{figure*}[!ht]
    \subfigure[\scriptsize{\textbf{C1} ($N=134$, $\zeta_1=0.659$, $u_1=0.218$, $p=0$)}] {
        \label{fig:ppo-eval-cluster1}
        \includegraphics[height=4cm]{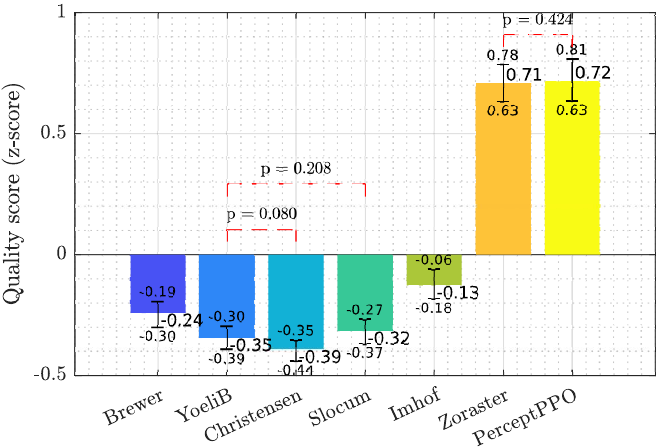}
    }
    \hspace{0em}
    \subfigure[\scriptsize{\textbf{C2} ($N=82$, $\zeta_2=0.612$, $u_2=0.131$, $p=1.2e^{-134}$)}] {
        \label{fig:ppo-eval-cluster2}
        \includegraphics[height=4cm, trim={0.55cm 0cm 0cm 0cm}, clip]{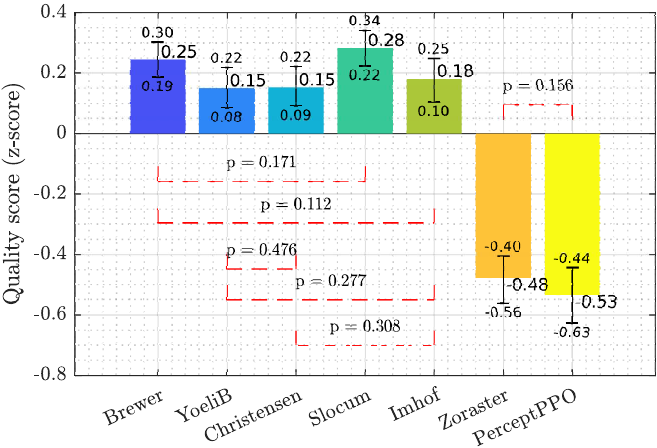}
    }
    \hspace{0em}
    \subfigure[\scriptsize{\textbf{C3} ($N=136$, $\zeta_3=0.502$, $u_3=0.005, p=7.6e^{-6}$})] {
        \label{fig:ppo-eval-cluster3}
        \includegraphics[height=4cm, trim={0.55cm 0cm 0cm 0cm}, clip]{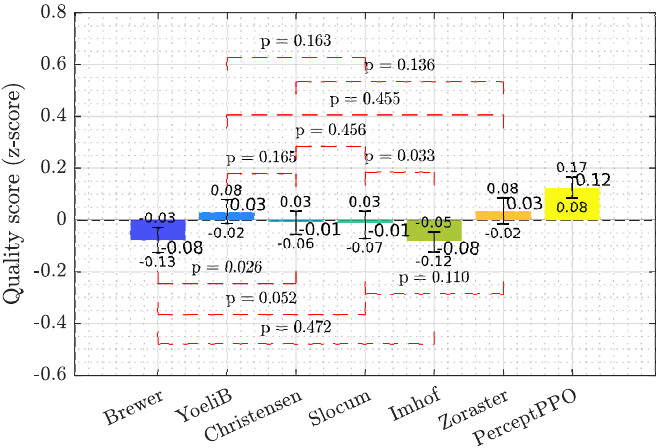}
    }
    \caption{
        Identified clusters of users in a comparison study of PPOs with depicted quality z-scores and 95\% confidence intervals. The red dashed brackets, with the corresponding $p$-values reported above, represent the PPOs pairs without evidence of statistically significant differences at $\alpha=0.05$ using the two-tail test. Conversely, between the PPOs pairs without brackets are detected statistically significant differences.
    }
    \label{fig:ppo-eval-clusters}
\end{figure*}

\repA{However, the relatively low overall coefficient of agreement $u=0.013$ suggests that there might be underlying patterns or segments within the participant data that are not immediately apparent from the aggregated overall results. Therefore, we apply the same hierarchical clustering technique as in \autoref{sec:ppo-results} to uncover these patterns and provide a more nuanced interpretation of the data.}{Using hierarchical clustering applying Ward's minimum variance method.} We identified three participant clusters as shown in \autoref{fig:ppo-eval-clusters}. 
Even though the $p$-value for the coefficient of agreement $u$ \addA{within clusters} is sometimes greater than $\alpha=0.05$ for individual areas, \delA{which disallows us to reject the null hypothesis $H_0^4$ for several areas,} aggregation of the choices over all areas leads to $p$-values lower than $\alpha=0.05$ for all clusters. Therefore, among all clusters, there is indeed \repA{statistically significant}{some} agreement among participants.
Detailed results supporting found clustering, including statistics and cluster compositions, can be found in \autoref{tab:ppo-eval-results} in the suppl.\,material. 

Cluster 1 ($N=134$) depicted in \autoref{fig:ppo-eval-cluster1} with mean consistency $\zeta_1=0.659$ $(SD = 0.232,\ \repA{MD}{Q2} = 0.714)$ shows fairly high agreement $u_1=0.218$ ($\mathrm{min}\ u_1=-0.002$, $p$-value$=0$) and contains participants that prefer PPOs in group $G_\mathrm{T}=\{\texttt{PerceptPPO},\ \texttt{Zoraster}\}$. 
Cluster 2 ($N=82$) shown in \autoref{fig:ppo-eval-cluster2} with mean consistency $\zeta_2=0.612$ $(SD = 0.223,\ \repA{MD}{Q2} = 0.643)$ and slightly lower agreement $u_2=0.131$ ($\mathrm{min}\ u_2=-0.004$, $p$-value$=1.25e^{-134}$), comprises participants that prefer PPOs in group $G_{\mathrm{TR}}=\{\texttt{Brewer},\ \texttt{YoeliB},\ \texttt{Christensen},\ \texttt{Slocum},\ \texttt{Imhof}\}$ contrary to \chA{PPOs in group} $G_\mathrm{T}$.
Cluster 3 ($N=136$) presented in \autoref{fig:ppo-eval-cluster3} with mean consistency $\zeta_3=0.502$ $(SD = 0.210,\ \repA{MD}{Q2} = 0.500)$ and relatively low agreement $u_3=0.005$ ($\mathrm{min}\ u_3=-0.002$, $p$-value$=7.63e^{-6}$) includes participants that are uncertain in their preferences but slightly incline towards \chA{PPOs in group} $G_\mathrm{T}$ over \chA{PPOs in group} $G_\mathrm{TR}$.

\section{Discussion}
The \texttt{PerceptPPO} study showcases the potential of a user-centered approach to enhancing the principles of label placement in cartography and GIS. The high coeficient of consistency $\zeta$ observed among the participants, with an overall mean of 0.67 and a median of 0.75, underscores the reliability of user judgments in determining the perceptual preference order of label positions around point features. Notably, the study established a clear overall preference order: top (T) $>$ bottom (B) $>$ right (R) $>$ top-right (TR) $>$ bottom-right (BR) $>$ left (L) $>$ top-left (TL) $>$ bottom-left (BL). 
\chB{Moreover, the analysis of the participants' feedback revealed that their choices were driven primarily by the functionality of the label placement rather than aesthetics. Many participants highlighted that their preferences were based on the ease of associating labels with their anchors. Specifically, labels positioned directly on top or bottom of the anchor were most effective in facilitating the association, with the top position being particularly favored.} 
This finding challenges traditional conventions and suggests a shift towards prioritizing labels above or bellow of point features for improved user experience.

The \repA{evaluation of}{experiment on} label density offers a brief view into how users perceive and prefer the density of labels on maps -- the median of \chB{the final} preferred local label density at 16.3\% reflects the preferred label density in the local region of a map. The median of global label density of 12.5\% suggests that users favor a moderate overall label presence, enough to inform without being too crowded.
\chB{Analysis of participants' feedback further supports our findings, as they consistently highlighted preference selection driven by functionality over aesthetics. Participants reported that when label density was too low, they struggled to orient themselves in the area due to a lack of contextual information. Conversely, when label density was too high, they found the maps cluttered, making it challenging to associate labels with the corresponding cities. Therefore, achieving the right balance is crucial for creating user-friendly maps that facilitate easy navigation and understanding.}

Finally, the comparison between \texttt{PerceptPPO} and existing PPOs reveals that \texttt{PerceptPPO}'s perceived quality surpasses the traditional PPOs and slightly outperforms \texttt{Zoraster}. Moreover, our comparison of \texttt{PerceptPPO} with \texttt{Zoraster} suggests that the initial position within a PPO plays a crucial role, while subsequent positions have less impact on overall user perception.

\section{Limitations}
While the study presents a statistically significant ordering of label positions, it also uncovers distinct clusters of user preferences, revealing the complexity of perceptual prioritization. We used generic blind maps that, on the one hand, minimize the degree of freedom and enable precise evaluation of preferences, but on the other hand, it is essential to acknowledge that other elements of a map could also influence user perception. Additionally, this study did not account for semantic considerations -- such as the placement of city labels across state borders or the positioning of coastal city labels towards water bodies, as highlighted by Imhof~\cite{Imhof1975}. Preferences might also vary based on the map's intended use and audience, from military to recreational purposes. The demographic aspects may similarly affect the preferences for languages with right-to-left or top-to-bottom scripts.
\addA{Additionally, participants' familiarity with the presented area may influence their preferences.}
Notably, the study did not delve into the functional aspect of label placement, including ease of information search and overall readability. Despite stringent participant screening and measures to ensure data integrity, the inherent variables associated with the uncontrolled online environment could introduce unseen biases.

The outcomes of our study lay a solid foundation, yet further research is needed to explore the aforementioned factors. Future work should investigate how semantic rules, map purpose, target audience preferences, and functional aspects such as readability interplay with the perceptual prioritization of label positions to develop more nuanced guidelines for map design.

\section{Conclusion}
In this work, we introduced Perceptual Position Priority Order (\texttt{PerceptPPO}), fundamentally reviewing the point-feature label placement by prioritizing user preferences over traditional conventions. Engaging nearly 800 participants globally, we have established a user-preferred ordering of label positions along the feature point that challenges and diverges from the conventional top-right towards the top position. Moreover, we performed an additional study to find users' preferred label density, a domain narrowly studied in prior literature. According to the results, users prefer %
\repA{12.5\%}{17\%} of the generic map to be \addA{overall} covered by labels. Finally, the comparative analysis underscores \texttt{PerceptPPO}'s superiority in perceived quality against traditional PPOs, particularly highlighting the significance of the first label position's role in user perception. Our empirical study marks a significant step toward more intuitive and user-centered map designs, emphasizing the importance of aligning label placement visualization practices with actual user expectations.

\ifCLASSOPTIONcompsoc
  \section*{Acknowledgments}
\else
  \section*{Acknowledgment}
\fi

This work was supported by project \emph{LTAIZ19004 Deep-Learning Approach to Topographical Image Analysis}; 
by the Ministry of Education, Youth and Sports of the Czech Republic within the activity INTER-EXCELENCE (LT), subactivity
INTER-ACTION (LTA), ID: SMSM2019LTAIZ and by Grant Agency of CTU in Prague grant No. SGS22/173/OHK3/3T/13 - Research of Modern Computer Graphics Methods 2022-2024.
Computational resources were mainly supplied by the e-INFRA CZ project (ID:90254), supported by the Ministry of Education, Youth and Sports of the Czech Republic.

\ifCLASSOPTIONcaptionsoff
  \newpage
\fi

\bibliographystyle{IEEEtran}

\bibliography{references}

\newpage
\begin{IEEEbiography}
[{\includegraphics[width=1in,height=1.25in,clip,keepaspectratio]{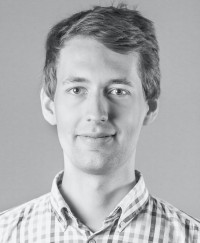}}]{Petr Bob\'{a}k}
is a PhD candidate at the Faculty of Information Technology at Brno University of Technology, Czechia. He received his Master's degree in Computer Science from the same institution in 2017. His research interests include information visualization, machine learning, computer vision, and mathematical optimization.
\end{IEEEbiography}

\begin{IEEEbiography}[{\includegraphics[width=1in,height=1.25in,clip,keepaspectratio]{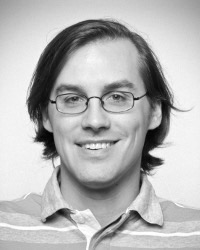}}]{Ladislav \v{C}mol\'{\i}k}
is an assistant professor at the Department of Computer Graphics and Interaction of the Czech Technical University in Prague, Czechia. He received his PhD from the same institution in 2011. His research interests include illustrative visualization, non-photorealistic rendering, and HCI.
\end{IEEEbiography}

\begin{IEEEbiography}[{\includegraphics[width=1in,height=1.25in,clip,keepaspectratio]{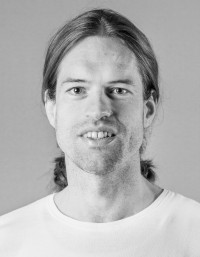}}]{Martin \v{C}ad\'{\i}k} is an associate professor of computer science at Brno University of Technology, 
where he heads his Computational Photography group. He received his PhD from the Czech Technical University in Prague in 2002. His research includes high dynamic range imaging, image processing, computer vision, human visual perception and image, and video quality assessment, among others.\end{IEEEbiography}

\vfill

\end{document}



\firstsection{Introduction}
\maketitle

In this supplementary material, we  provide a collection of additional resources, including various images, detailed tables, comprehensive instructions for conducting the experiments, and further results not included in the primary publication. These elements were omitted from the main document in an effort to preserve its conciseness and readability. Our aim is to ensure that readers have access to all necessary supplementary information to fully understand the scope and depth of our research findings.

\section{\texttt{PerceptPPO} Study}

Addressing the lack of uniformity in label position prioritization in cartography and GIS, our objective is to establish a PPO that is deeply anchored in user perceptions rather than relying on typographic and cartographic traditions. These traditional practices, though historically significant, stem from methodologies developed decades ago and might not reflect the requirements of today's users. With \texttt{PerceptPPO}, our aim is to transcend traditional label placement strategies by leveraging empirical user data to dictate label positioning, thereby enhancing the perceived clarity, usability, and overall user experience of maps. 

\subsection{User Study Interface}
\begin{figure}[H]
    \subfigure[Introduction] {
        
        \includegraphics[width=\columnwidth]{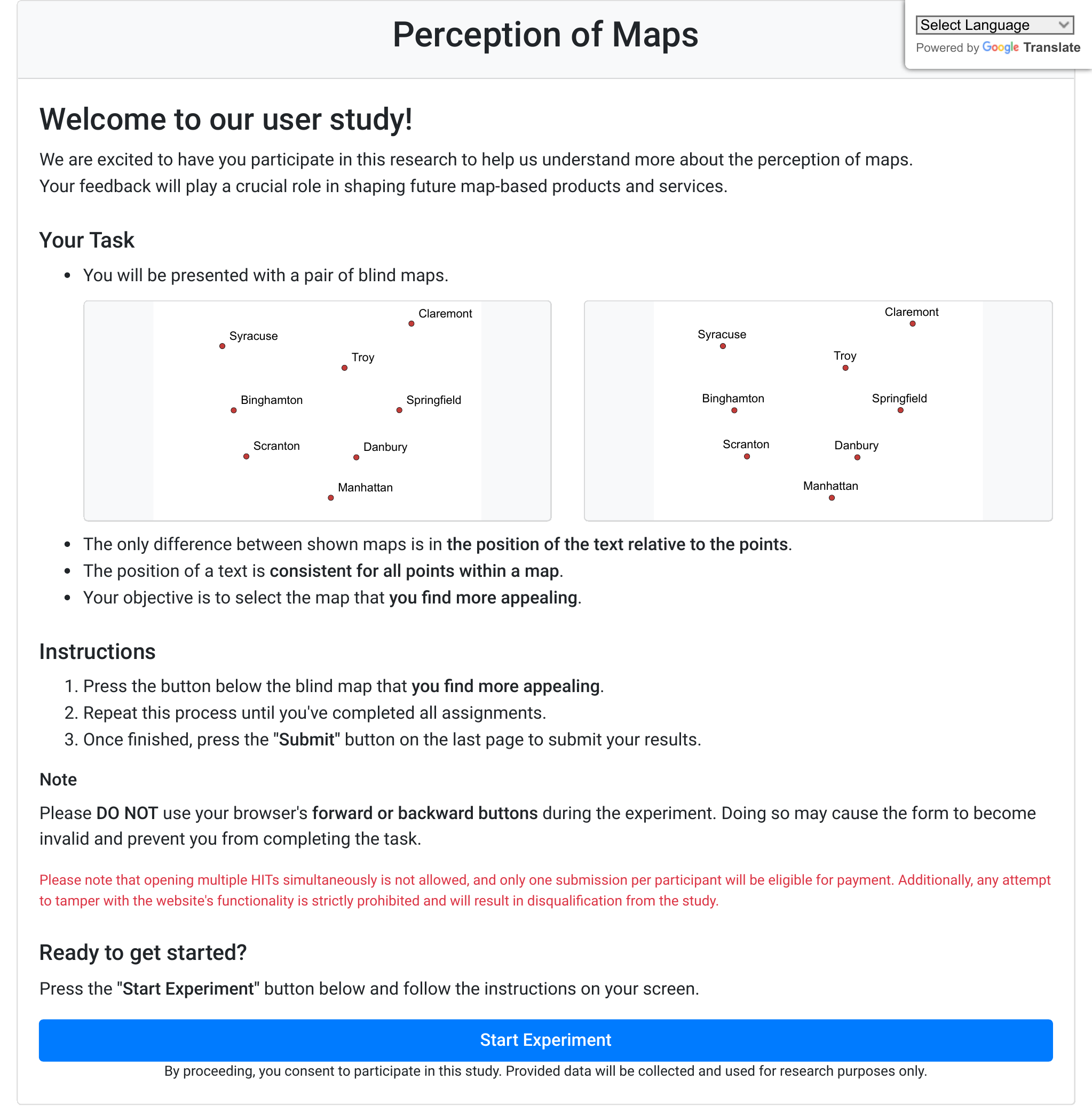}
    }
    \hspace{0em}
    \subfigure[Survey] {
        
        \includegraphics[width=\columnwidth]{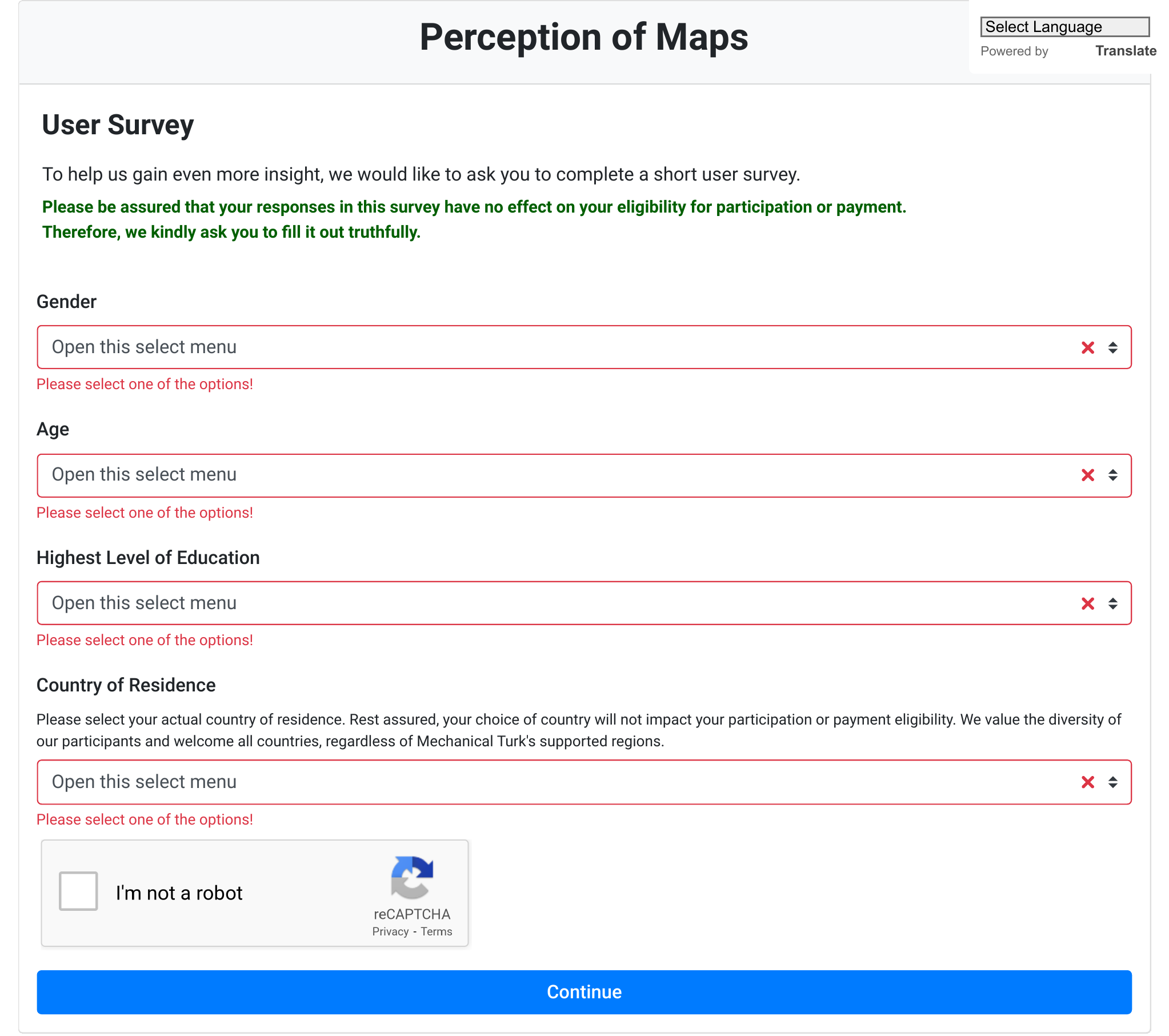}
    }
    \caption{
        \textbf{User Study -- \texttt{PerceptPPO}:} Web interface for introduction and survey.
    }
    \label{fig:web-ppo1}
\end{figure}

\begin{figure}[H]
    \subfigure[Experiment] {
        
        \includegraphics[width=\columnwidth]{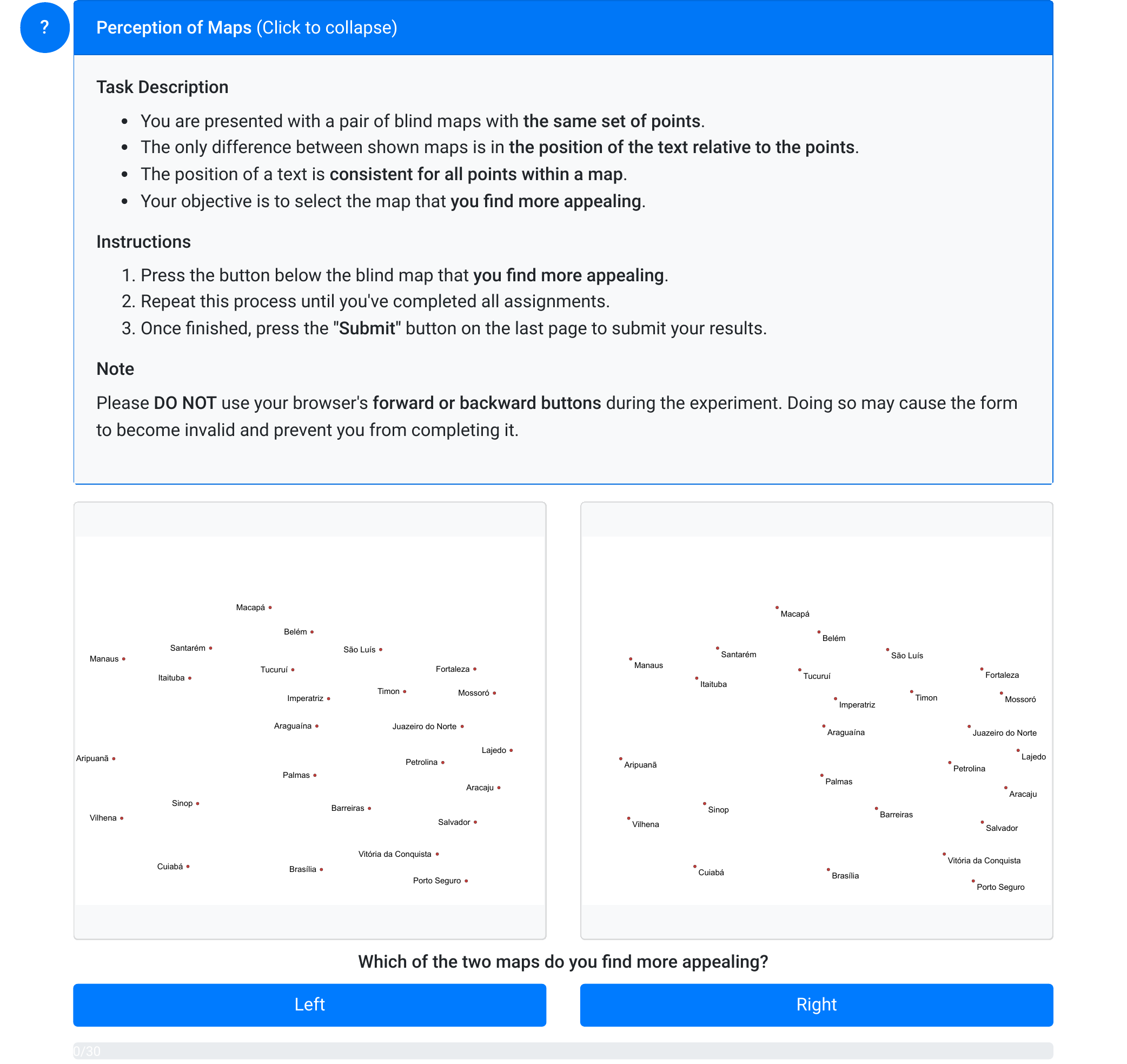}
    }
    \hspace{0em}
    \subfigure[Feedback] {
        
        \includegraphics[width=\columnwidth]{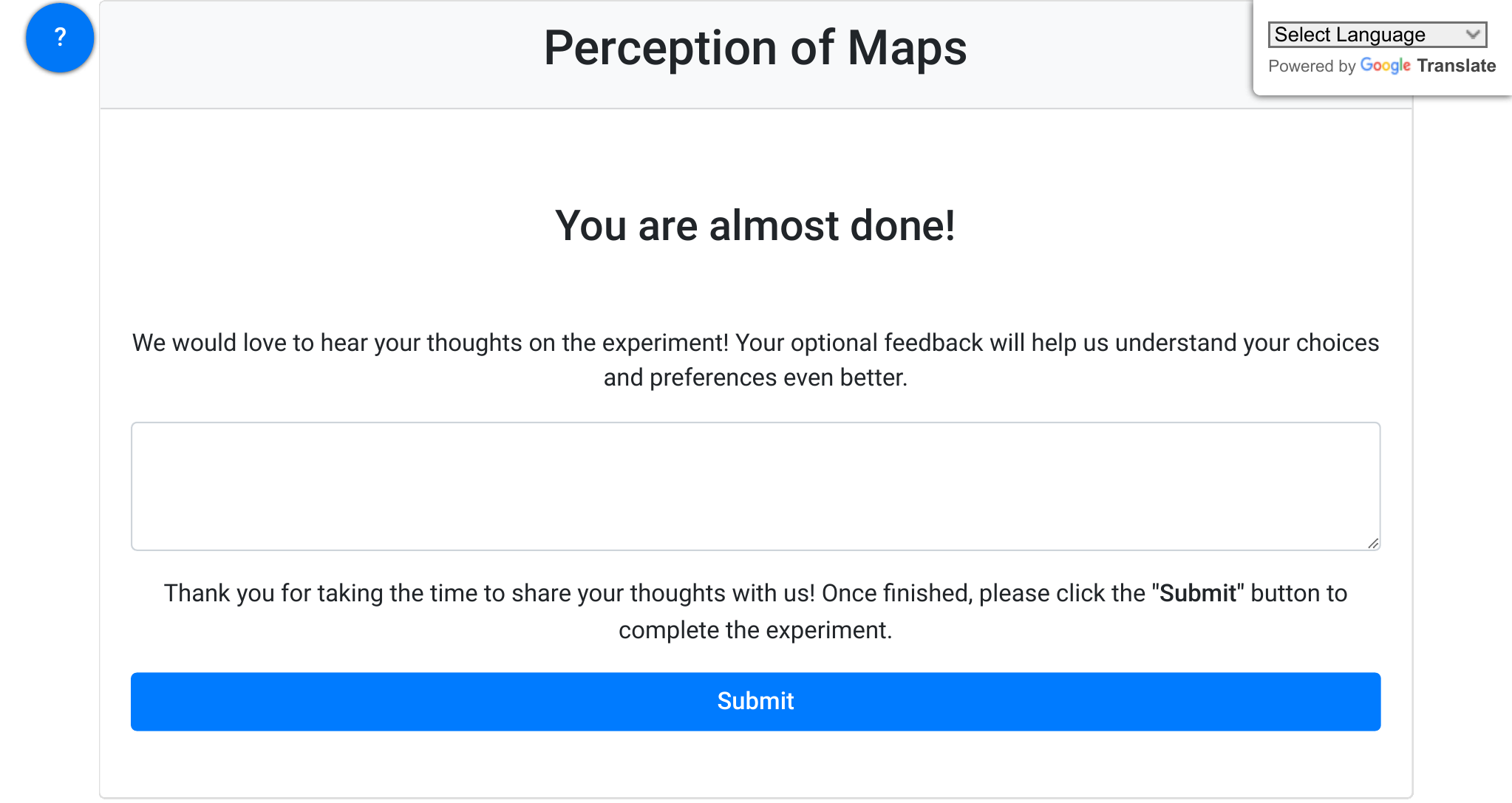}
    }
    \caption{
        \textbf{User Study -- \texttt{PerceptPPO}:} Web interface for experiment and feedback.
    }
    \label{fig:web-ppo2}
\end{figure}

This section provides examples of user interface used to establish \texttt{PerceptPPO}. The following paragraphs present textual information available to participants as seen in \autoref{fig:web-ppo1} and \autoref{fig:web-ppo2}.

\textbf{Introduction} Welcome to our user study! We are excited to have you participate in this research to help us understand more about the perception of maps. Your feedback will play a crucial role in shaping future map-based products and services. 

\textbf{Task} You will be presented with a pair of blind maps. The only difference between shown maps is in the position of the text relative to the points. The position of a text is consistent for all points within a map. Your objective is to select the map that you find more appealing.

\textbf{Instructions} Press the button below the blind map that you find more appealing. Repeat this process until you've completed all assignments. Once finished, press the "Submit" button on the last page to submit your results.

\textbf{Note} Please note that opening multiple HITs simultaneously is not allowed, and only one submission per participant will be eligible for payment. Additionally, any attempt to tamper with the website's functionality is strictly prohibited and will result in disqualification from the study.

\textbf{Start Experiment} Ready to get started? Press the "Start Experiment" button below and follow the instructions on your screen. By proceeding, you consent to participate in this study. Provided data will be collected and used for research purposes only.

\textbf{Call For Action} Which of the two maps do you find more appealing?

\subsection{Data}
\begin{figure*}[h!]
 
    \includegraphics[width=\textwidth]{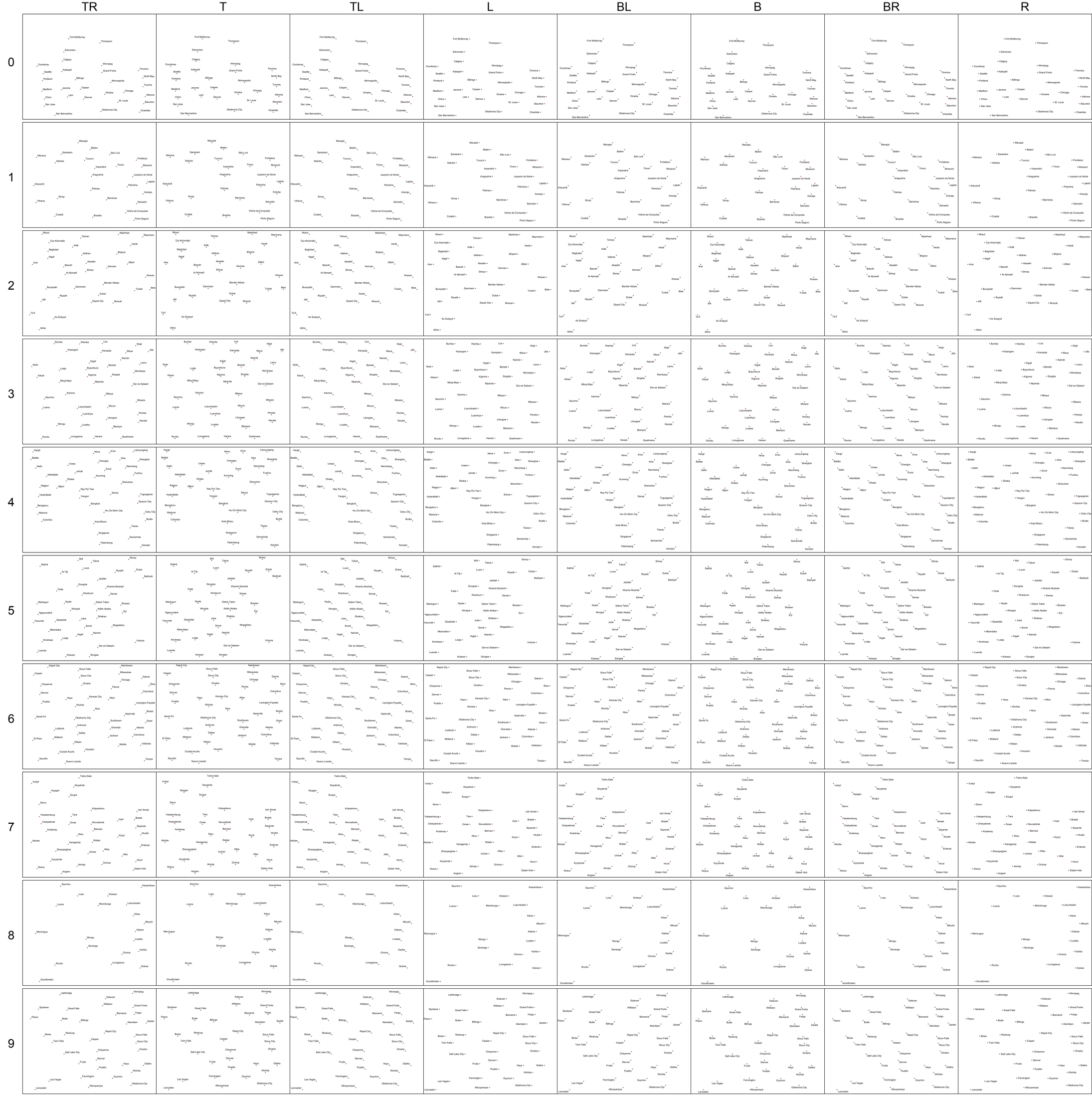}
    \caption{\textbf{\texttt{PerceptPPO} Study:} Visualizations of areas 0 through 9 used in the user study.}
    \label{fig:ppo_renders1}
\end{figure*}

\begin{figure*}[h!]
 
    \includegraphics[width=\textwidth]{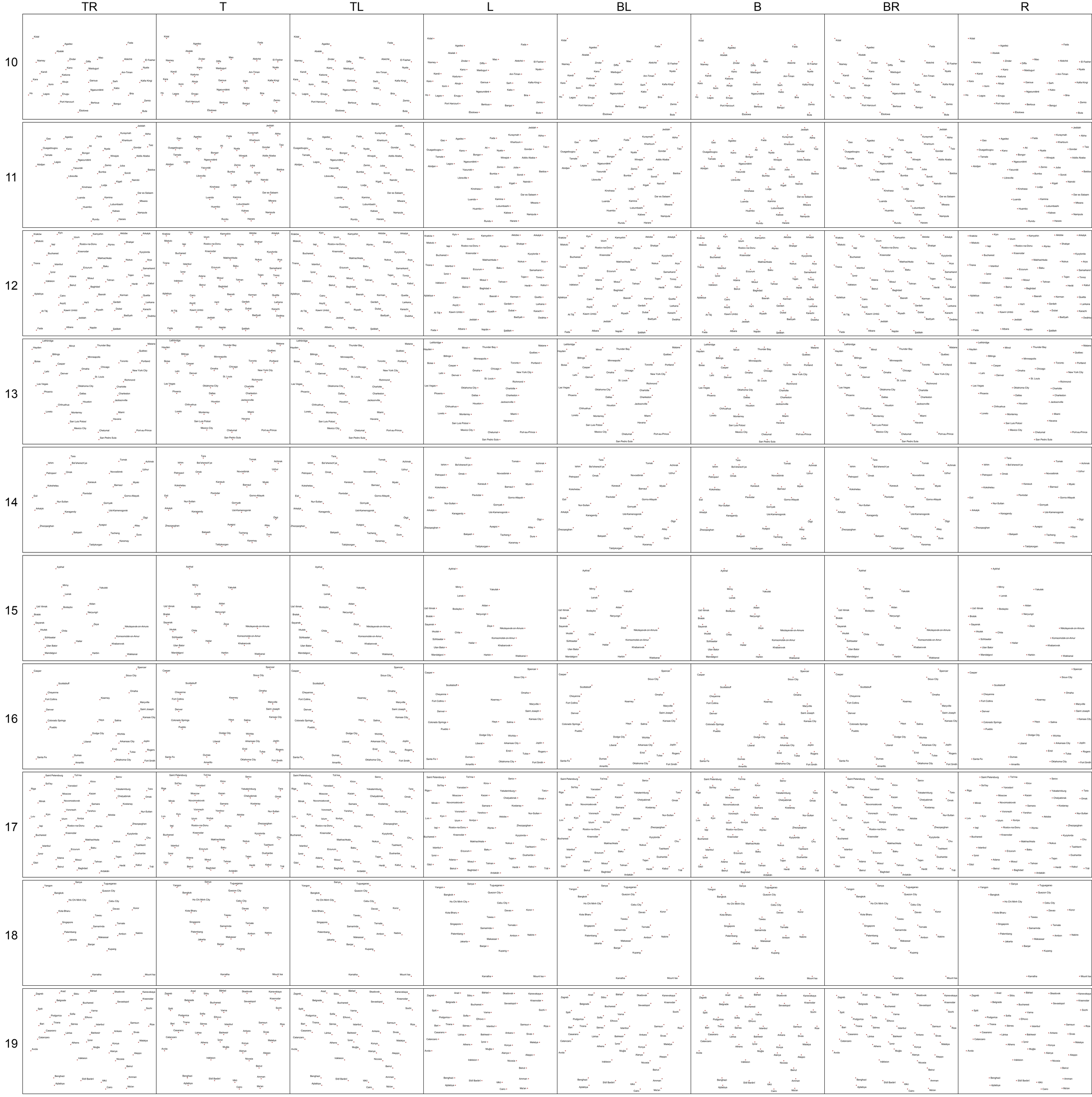}
    \caption{\textbf{\texttt{PerceptPPO} Study:} Visualizations of areas 10 through 19 used in the user study.}
    \label{fig:ppo_renders2}
\end{figure*}

\begin{figure*}[h!]
 
    \includegraphics[width=\textwidth]{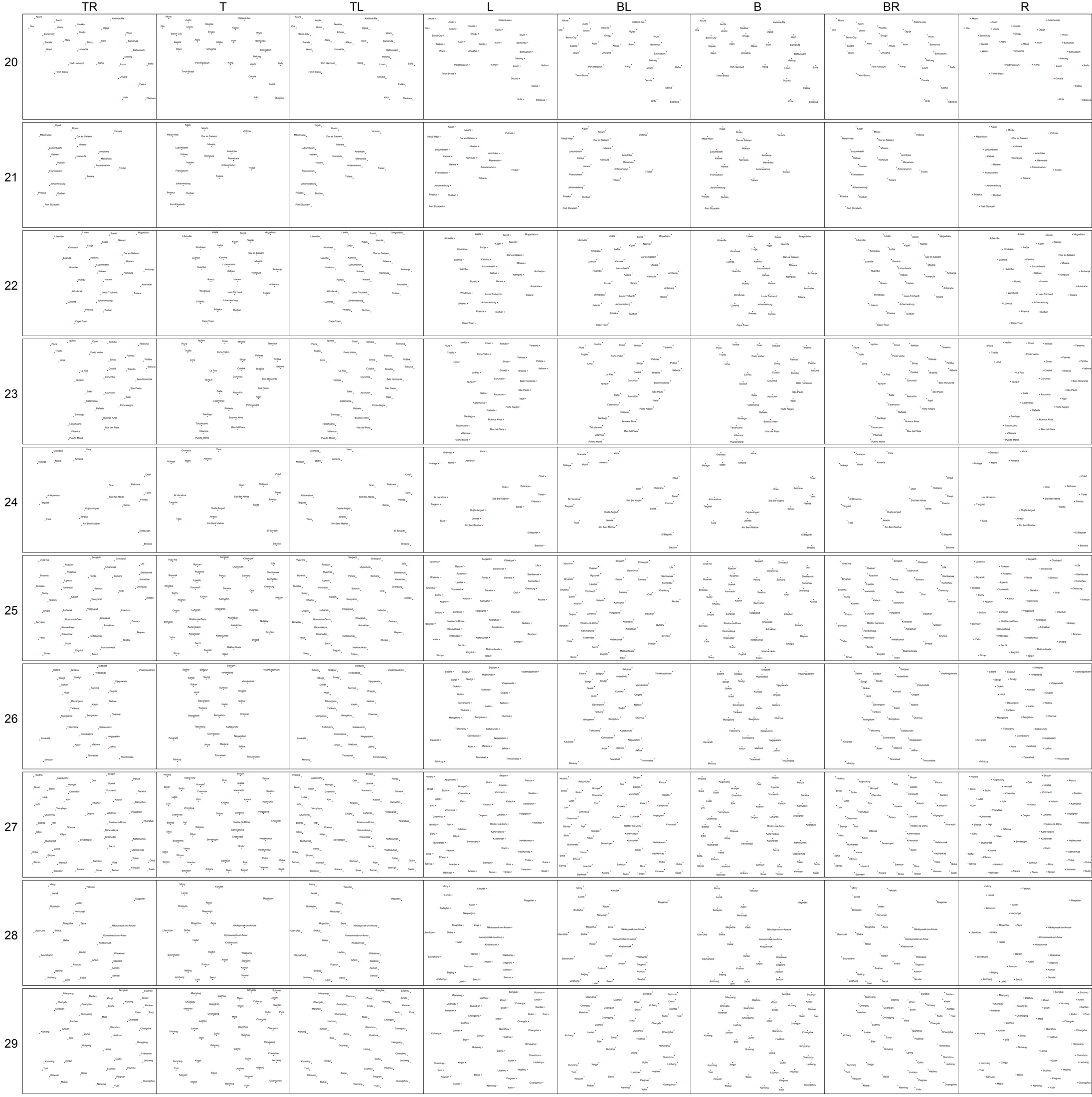}
    \caption{\textbf{\texttt{PerceptPPO} Study:} Visualizations of areas 20 through 29 used in the user study.}
    \label{fig:ppo_renders3}
\end{figure*}

\autoref{fig:ppo_renders1}, \autoref{fig:ppo_renders2}, and \autoref{fig:ppo_renders3} display the map renders utilized in the user study, showcasing all examined label positions within the 8-position model. We randomly selected 30 global locations, excluding maritime regions and areas south of -60 degrees latitude to omit Antarctica, given its minimal population. Each site was centered within a zone defined by specific coordinates and a zoom level between 5 (comparable to the size of Europe) and 10 (similar in area to Luxembourg), rendered into vector SVG images measuring $1305\times1025$ pixels. We identified settlements with populations exceeding 500, using data from GeoNames, specifically the Cities 500 dataset\footnote{\url{http://download.geonames.org/export/dump/}}, as anchor points. These anchors were ranked by population size, and only those with all eight potential label positions conflict-free were included. Areas with less than 20 suitable anchors were replaced. This process yielded 30 areas, numbered 0 to 29, at zoom levels 5 to 8, featuring between 20 to 54 anchors each.

\subsection{Results}

\begin{figure*}[!ht]
    \subfigure[] {
        
        \includegraphics[width=0.33\textwidth]{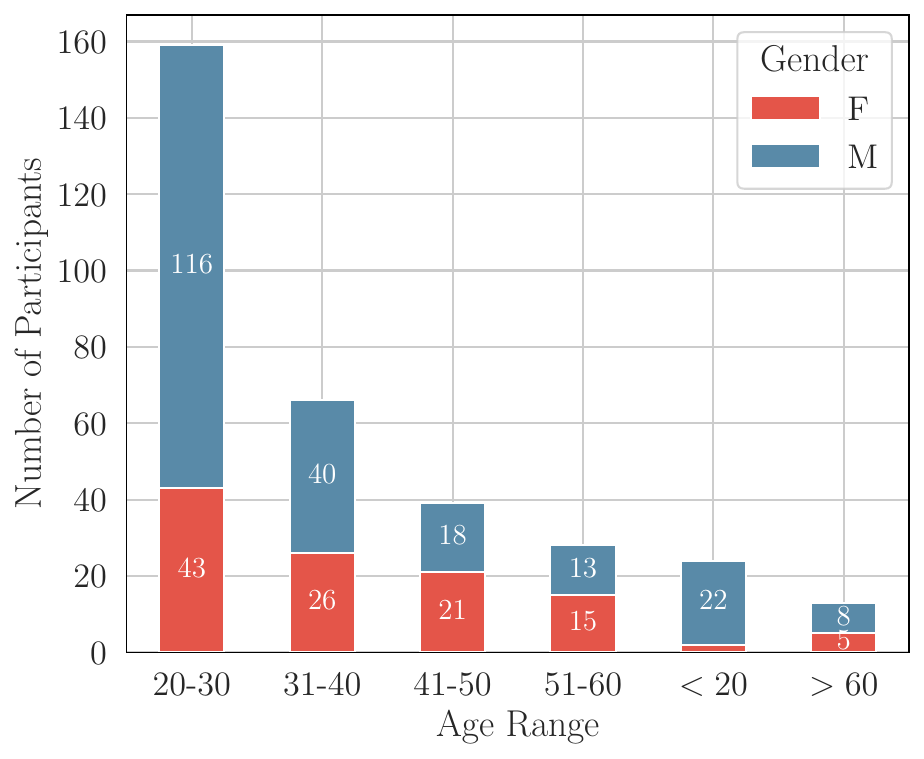}
    }
    \hspace{0em}
    \subfigure[] {
        
        \includegraphics[width=0.33\textwidth]{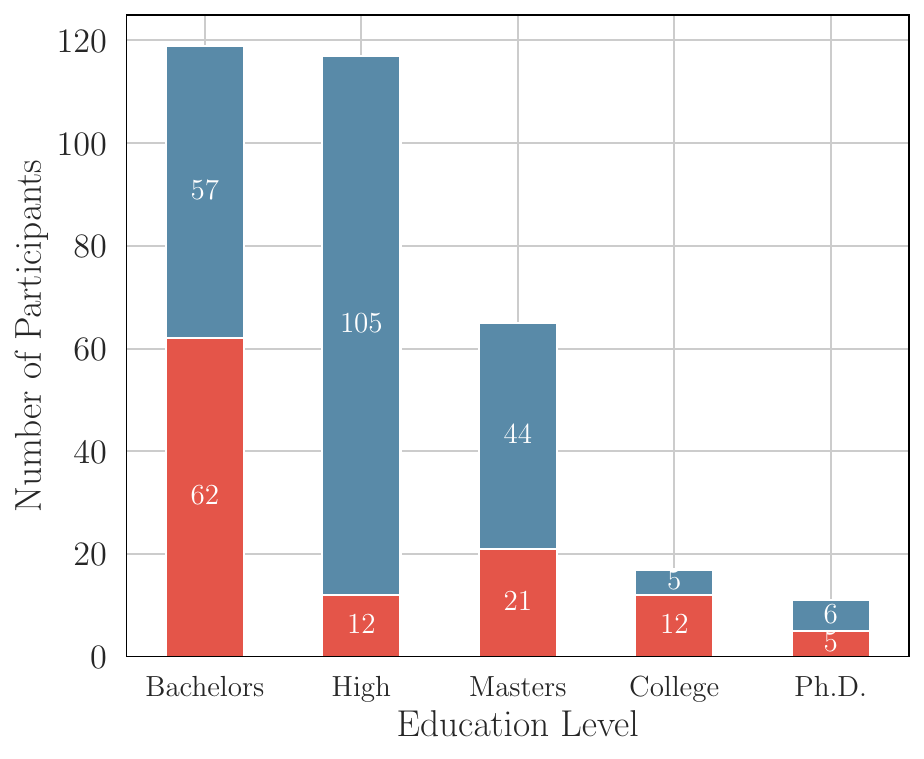}
    }
    \hspace{0em}
    \subfigure[] {
        
        \includegraphics[width=0.33\textwidth]{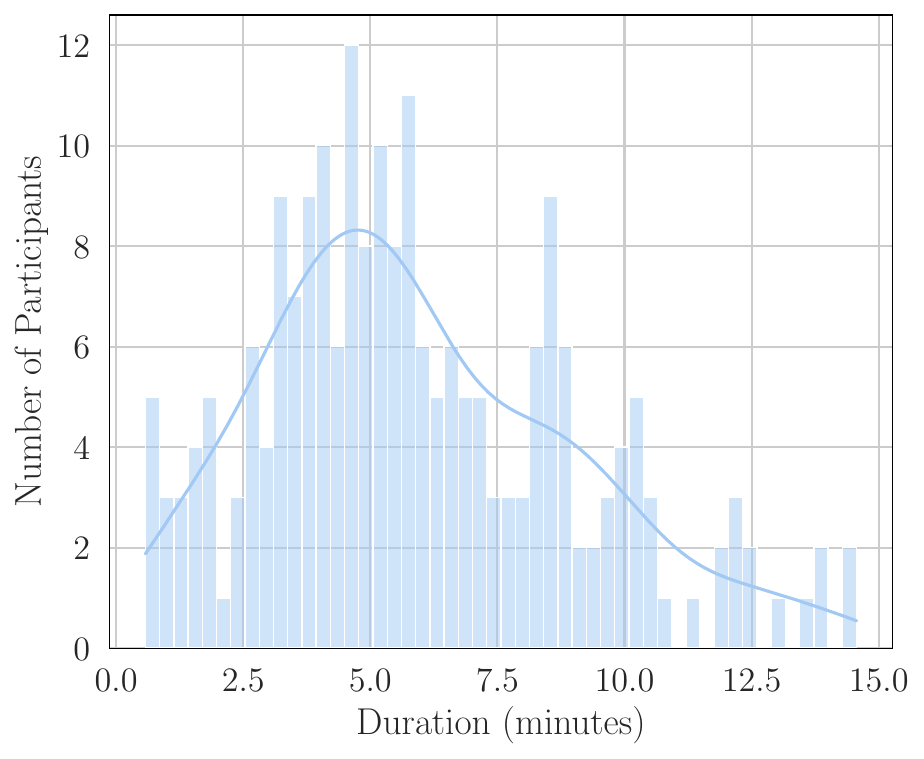}
    }
    \caption{
        \textbf{\texttt{PerceptPPO} Study:} User statistics of age, education, and and duration of study for all participants.
    }
\label{fig:ppo-results_stats}
\end{figure*}

\begin{figure*}[!ht]
    \subfigure[] {
        
        \includegraphics[width=0.5\textwidth]{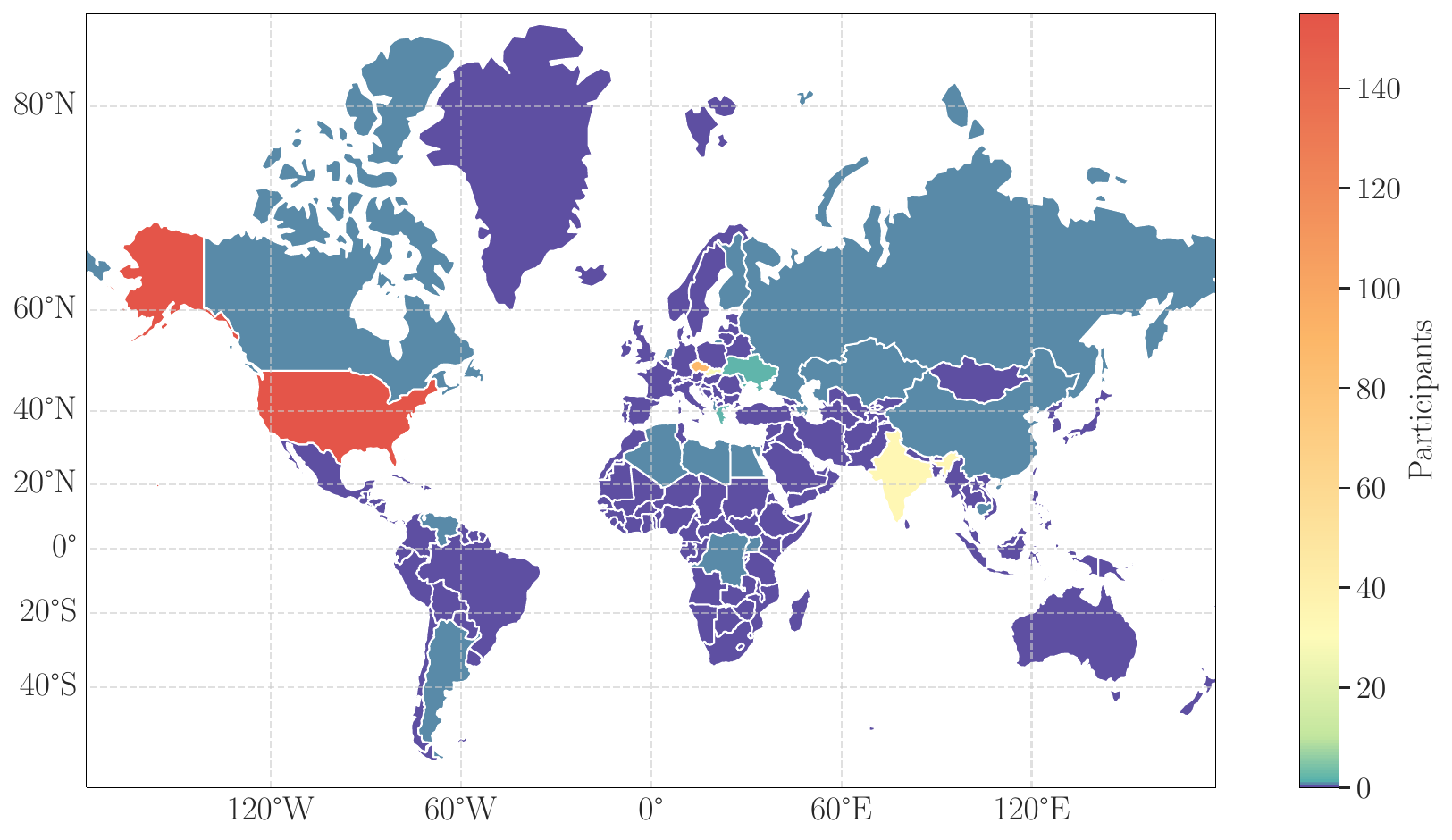}
    }
    \hspace{0em}
    \subfigure[] {
        
        \includegraphics[width=0.5\textwidth]{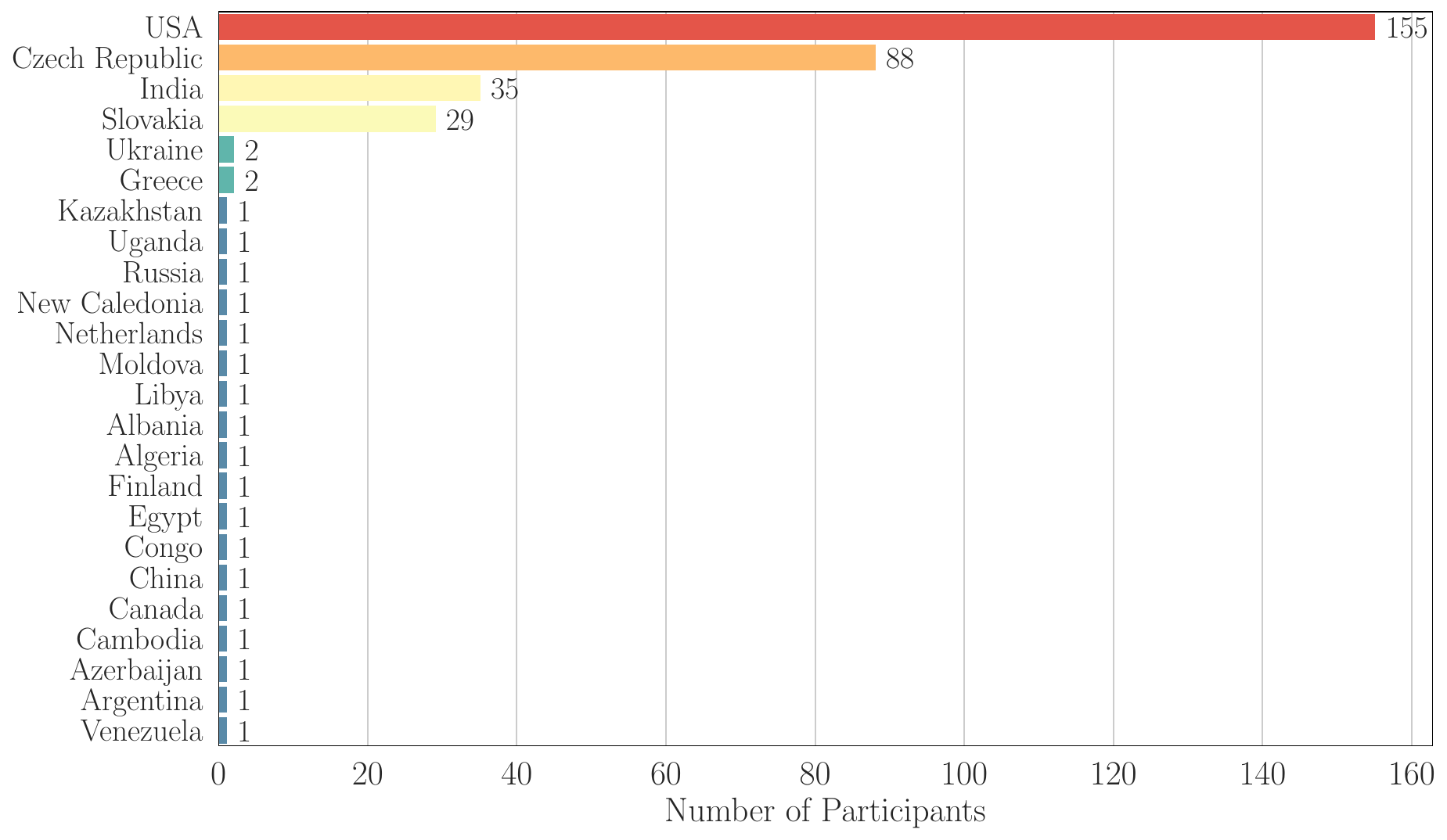}
    }
    \caption{
        \textbf{\texttt{PerceptPPO} Study:} Country distribution for all participants.
    }
    \label{fig:ppo-results_countries}
\end{figure*}

\autoref{tab:ppo-results_pvals} presents comprehensive $p$-values from pairwise comparisons of label positions. Additionally, \autoref{tab:ppo-results_pvals_c1}, \autoref{tab:ppo-results_pvals_c2}, and \autoref{tab:ppo-results_pvals_c3} detail the $p$-values for the pairwise comparisons within the identified clusters. \autoref{tab:ppo-results} shows the coefficient of consistency $\zeta$ and coefficient of agreement $u$ along with its corresponding values of \repA{$u_\mathrm{min}$}{min $u$}, $\chi^2$ and $p$-value for positions of \texttt{PerceptPPO} for each map area reported over all responses and clusters obtained by hierarchical clustering. \autoref{fig:ppo-results_stats} and \autoref{fig:ppo-results_countries} show the detailed statistics of engaged participants.  

\begin{table}[H]
\centering
\setlength{\tabcolsep}{3pt}
\begin{adjustbox}{width=\columnwidth}
\begin{tabular}{c|cccccccc}
\toprule
{} &                                  T &                                  B &                                  R &                                 TR &                                 BR &                                  L &                                 TL &                                 BL \\
\midrule
T  &                                 -- &  \textcolor{darkerGreen}{1.64e-03} &  \textcolor{darkerGreen}{3.81e-07} &  \textcolor{darkerGreen}{8.59e-17} &  \textcolor{darkerGreen}{3.31e-22} &  \textcolor{darkerGreen}{3.49e-35} &  \textcolor{darkerGreen}{1.28e-55} &  \textcolor{darkerGreen}{4.38e-50} \\
B  &  \textcolor{darkerGreen}{1.64e-03} &                                 -- &  \textcolor{darkerGreen}{6.40e-04} &  \textcolor{darkerGreen}{3.22e-08} &  \textcolor{darkerGreen}{3.58e-20} &  \textcolor{darkerGreen}{1.09e-29} &  \textcolor{darkerGreen}{2.93e-34} &  \textcolor{darkerGreen}{8.54e-48} \\
R  &  \textcolor{darkerGreen}{3.81e-07} &  \textcolor{darkerGreen}{6.40e-04} &                                 -- &  \textcolor{darkerGreen}{1.25e-03} &  \textcolor{darkerGreen}{4.46e-12} &  \textcolor{darkerGreen}{1.30e-21} &  \textcolor{darkerGreen}{2.42e-19} &  \textcolor{darkerGreen}{3.39e-31} \\
TR &  \textcolor{darkerGreen}{8.59e-17} &  \textcolor{darkerGreen}{3.22e-08} &  \textcolor{darkerGreen}{1.25e-03} &                                 -- &  \textcolor{darkerGreen}{4.57e-04} &  \textcolor{darkerGreen}{2.06e-05} &  \textcolor{darkerGreen}{6.10e-20} &  \textcolor{darkerGreen}{4.82e-24} \\
BR &  \textcolor{darkerGreen}{3.31e-22} &  \textcolor{darkerGreen}{3.58e-20} &  \textcolor{darkerGreen}{4.46e-12} &  \textcolor{darkerGreen}{4.57e-04} &                                 -- &  \textcolor{darkerGreen}{1.19e-02} &  \textcolor{darkerGreen}{1.25e-07} &  \textcolor{darkerGreen}{2.43e-20} \\
L  &  \textcolor{darkerGreen}{3.49e-35} &  \textcolor{darkerGreen}{1.09e-29} &  \textcolor{darkerGreen}{1.30e-21} &  \textcolor{darkerGreen}{2.06e-05} &  \textcolor{darkerGreen}{1.19e-02} &                                 -- &  \textcolor{darkerGreen}{1.30e-03} &  \textcolor{darkerGreen}{2.08e-09} \\
TL &  \textcolor{darkerGreen}{1.28e-55} &  \textcolor{darkerGreen}{2.93e-34} &  \textcolor{darkerGreen}{2.42e-19} &  \textcolor{darkerGreen}{6.10e-20} &  \textcolor{darkerGreen}{1.25e-07} &  \textcolor{darkerGreen}{1.30e-03} &                                 -- &  \textcolor{darkerGreen}{5.79e-04} \\
BL &  \textcolor{darkerGreen}{4.38e-50} &  \textcolor{darkerGreen}{8.54e-48} &  \textcolor{darkerGreen}{3.39e-31} &  \textcolor{darkerGreen}{4.82e-24} &  \textcolor{darkerGreen}{2.43e-20} &  \textcolor{darkerGreen}{2.08e-09} &  \textcolor{darkerGreen}{5.79e-04} &                                 -- \\
\bottomrule
\end{tabular}
\end{adjustbox}
\caption{\textbf{User Study -- \texttt{PerceptPPO}:} Detailed overall $p$ values disregard identified clusters for all position pairs. Green cells indicate statistically significant differences, or red otherwise. We use a Two-Tailed test with $\alpha=0.05$; therefore, $p$ has to be less than $0.025$ to be significant.}
\label{tab:ppo-results_pvals}
\end{table}

\begin{table}[H]
\centering
\setlength{\tabcolsep}{3pt}
\begin{adjustbox}{width=\columnwidth}
\begin{tabular}{c|cccccccc}
\toprule
{} &                                  T &                                  B &                                  R &                                 TR &                                 BR &                                  L &                                 TL &                                 BL \\
\midrule
T  &                                 -- &      \textcolor{crimson}{2.86e-01} &      \textcolor{crimson}{3.20e-01} &  \textcolor{darkerGreen}{6.10e-25} &  \textcolor{darkerGreen}{1.90e-10} &  \textcolor{darkerGreen}{9.61e-15} &  \textcolor{darkerGreen}{2.66e-48} &  \textcolor{darkerGreen}{3.87e-50} \\
B  &      \textcolor{crimson}{2.86e-01} &                                 -- &      \textcolor{crimson}{4.92e-01} &  \textcolor{darkerGreen}{5.93e-19} &  \textcolor{darkerGreen}{3.71e-14} &  \textcolor{darkerGreen}{3.05e-10} &  \textcolor{darkerGreen}{3.42e-31} &  \textcolor{darkerGreen}{2.54e-61} \\
R  &      \textcolor{crimson}{3.20e-01} &      \textcolor{crimson}{4.92e-01} &                                 -- &  \textcolor{darkerGreen}{3.01e-29} &  \textcolor{darkerGreen}{1.77e-19} &  \textcolor{darkerGreen}{8.66e-13} &  \textcolor{darkerGreen}{3.40e-37} &  \textcolor{darkerGreen}{1.39e-83} \\
TR &  \textcolor{darkerGreen}{6.10e-25} &  \textcolor{darkerGreen}{5.93e-19} &  \textcolor{darkerGreen}{3.01e-29} &                                 -- &      \textcolor{crimson}{2.80e-01} &      \textcolor{crimson}{7.63e-02} &  \textcolor{darkerGreen}{2.01e-11} &  \textcolor{darkerGreen}{1.70e-16} \\
BR &  \textcolor{darkerGreen}{1.90e-10} &  \textcolor{darkerGreen}{3.71e-14} &  \textcolor{darkerGreen}{1.77e-19} &      \textcolor{crimson}{2.80e-01} &                                 -- &      \textcolor{crimson}{2.66e-01} &  \textcolor{darkerGreen}{3.72e-06} &  \textcolor{darkerGreen}{1.01e-13} \\
L  &  \textcolor{darkerGreen}{9.61e-15} &  \textcolor{darkerGreen}{3.05e-10} &  \textcolor{darkerGreen}{8.66e-13} &      \textcolor{crimson}{7.63e-02} &      \textcolor{crimson}{2.66e-01} &                                 -- &  \textcolor{darkerGreen}{3.06e-11} &  \textcolor{darkerGreen}{3.84e-16} \\
TL &  \textcolor{darkerGreen}{2.66e-48} &  \textcolor{darkerGreen}{3.42e-31} &  \textcolor{darkerGreen}{3.40e-37} &  \textcolor{darkerGreen}{2.01e-11} &  \textcolor{darkerGreen}{3.72e-06} &  \textcolor{darkerGreen}{3.06e-11} &                                 -- &      \textcolor{crimson}{2.97e-02} \\
BL &  \textcolor{darkerGreen}{3.87e-50} &  \textcolor{darkerGreen}{2.54e-61} &  \textcolor{darkerGreen}{1.39e-83} &  \textcolor{darkerGreen}{1.70e-16} &  \textcolor{darkerGreen}{1.01e-13} &  \textcolor{darkerGreen}{3.84e-16} &      \textcolor{crimson}{2.97e-02} &                                 -- \\
\bottomrule
\end{tabular}
\end{adjustbox}
\caption{\textbf{User Study -- \texttt{PerceptPPO}:} Detailed $p$ values of Cluster 1 for all position pairs. Green cells indicate statistically significant differences, or red otherwise. We use a Two-Tailed test with $\alpha=0.05$; therefore, $p$ has to be less than $0.025$ to be significant.}
\label{tab:ppo-results_pvals_c1}
\end{table}

\begin{table}[H]
\centering
\setlength{\tabcolsep}{3pt}
\begin{adjustbox}{width=\columnwidth}
\begin{tabular}{c|cccccccc}
\toprule
{} &                                  T &                                  B &                                  R &                                 TR &                                 BR &                                  L &                                 TL &                                 BL \\
\midrule
T  &                                 -- &  \textcolor{darkerGreen}{3.76e-10} &  \textcolor{darkerGreen}{1.80e-45} &  \textcolor{darkerGreen}{8.22e-14} &  \textcolor{darkerGreen}{1.15e-36} &  \textcolor{darkerGreen}{3.96e-63} &  \textcolor{darkerGreen}{2.57e-31} &  \textcolor{darkerGreen}{1.74e-65} \\
B  &  \textcolor{darkerGreen}{3.76e-10} &                                 -- &  \textcolor{darkerGreen}{6.86e-20} &  \textcolor{darkerGreen}{5.14e-03} &  \textcolor{darkerGreen}{1.85e-18} &  \textcolor{darkerGreen}{4.08e-38} &  \textcolor{darkerGreen}{1.18e-08} &  \textcolor{darkerGreen}{3.09e-27} \\
R  &  \textcolor{darkerGreen}{1.80e-45} &  \textcolor{darkerGreen}{6.86e-20} &                                 -- &  \textcolor{darkerGreen}{4.27e-10} &      \textcolor{crimson}{7.51e-02} &  \textcolor{darkerGreen}{1.12e-05} &      \textcolor{crimson}{2.73e-02} &  \textcolor{darkerGreen}{4.78e-03} \\
TR &  \textcolor{darkerGreen}{8.22e-14} &  \textcolor{darkerGreen}{5.14e-03} &  \textcolor{darkerGreen}{4.27e-10} &                                 -- &  \textcolor{darkerGreen}{1.97e-05} &  \textcolor{darkerGreen}{3.44e-13} &  \textcolor{darkerGreen}{2.88e-03} &  \textcolor{darkerGreen}{1.06e-12} \\
BR &  \textcolor{darkerGreen}{1.15e-36} &  \textcolor{darkerGreen}{1.85e-18} &      \textcolor{crimson}{7.51e-02} &  \textcolor{darkerGreen}{1.97e-05} &                                 -- &  \textcolor{darkerGreen}{3.26e-06} &      \textcolor{crimson}{1.62e-01} &  \textcolor{darkerGreen}{8.91e-05} \\
L  &  \textcolor{darkerGreen}{3.96e-63} &  \textcolor{darkerGreen}{4.08e-38} &  \textcolor{darkerGreen}{1.12e-05} &  \textcolor{darkerGreen}{3.44e-13} &  \textcolor{darkerGreen}{3.26e-06} &                                 -- &  \textcolor{darkerGreen}{1.73e-14} &  \textcolor{darkerGreen}{9.46e-03} \\
TL &  \textcolor{darkerGreen}{2.57e-31} &  \textcolor{darkerGreen}{1.18e-08} &      \textcolor{crimson}{2.73e-02} &  \textcolor{darkerGreen}{2.88e-03} &      \textcolor{crimson}{1.62e-01} &  \textcolor{darkerGreen}{1.73e-14} &                                 -- &  \textcolor{darkerGreen}{1.82e-08} \\
BL &  \textcolor{darkerGreen}{1.74e-65} &  \textcolor{darkerGreen}{3.09e-27} &  \textcolor{darkerGreen}{4.78e-03} &  \textcolor{darkerGreen}{1.06e-12} &  \textcolor{darkerGreen}{8.91e-05} &  \textcolor{darkerGreen}{9.46e-03} &  \textcolor{darkerGreen}{1.82e-08} &                                 -- \\
\bottomrule
\end{tabular}
\end{adjustbox}
\caption{\textbf{User Study -- \texttt{PerceptPPO}:} Detailed $p$ values of Cluster 2 for all position pairs. Green cells indicate statistically significant differences, or red otherwise. We use a Two-Tailed test with $\alpha=0.05$; therefore, $p$ has to be less than $0.025$ to be significant.}
\label{tab:ppo-results_pvals_c2}
\end{table}

\begin{table}[H]
\centering
\setlength{\tabcolsep}{3pt}
\begin{adjustbox}{width=\columnwidth}
\begin{tabular}{c|cccccccc}
\toprule
{} &                                  T &                                  B &                                  R &                                 TR &                                 BR &                                  L &                                 TL &                                 BL \\
\midrule
T  &                                 -- &      \textcolor{crimson}{6.50e-02} &      \textcolor{crimson}{2.30e-01} &      \textcolor{crimson}{4.07e-02} &      \textcolor{crimson}{2.41e-01} &  \textcolor{darkerGreen}{9.99e-08} &  \textcolor{darkerGreen}{1.71e-10} &  \textcolor{darkerGreen}{9.62e-06} \\
B  &      \textcolor{crimson}{6.50e-02} &                                 -- &      \textcolor{crimson}{2.78e-01} &  \textcolor{darkerGreen}{1.09e-02} &      \textcolor{crimson}{2.29e-01} &  \textcolor{darkerGreen}{3.09e-05} &  \textcolor{darkerGreen}{6.09e-05} &  \textcolor{darkerGreen}{4.62e-04} \\
R  &      \textcolor{crimson}{2.30e-01} &      \textcolor{crimson}{2.78e-01} &                                 -- &  \textcolor{darkerGreen}{1.71e-02} &      \textcolor{crimson}{4.51e-01} &  \textcolor{darkerGreen}{1.16e-04} &  \textcolor{darkerGreen}{1.11e-04} &  \textcolor{darkerGreen}{4.26e-03} \\
TR &      \textcolor{crimson}{4.07e-02} &  \textcolor{darkerGreen}{1.09e-02} &  \textcolor{darkerGreen}{1.71e-02} &                                 -- &  \textcolor{darkerGreen}{3.42e-03} &  \textcolor{darkerGreen}{5.87e-06} &  \textcolor{darkerGreen}{1.55e-08} &  \textcolor{darkerGreen}{1.50e-05} \\
BR &      \textcolor{crimson}{2.41e-01} &      \textcolor{crimson}{2.29e-01} &      \textcolor{crimson}{4.51e-01} &  \textcolor{darkerGreen}{3.42e-03} &                                 -- &  \textcolor{darkerGreen}{1.40e-04} &  \textcolor{darkerGreen}{1.32e-05} &  \textcolor{darkerGreen}{9.08e-04} \\
L  &  \textcolor{darkerGreen}{9.99e-08} &  \textcolor{darkerGreen}{3.09e-05} &  \textcolor{darkerGreen}{1.16e-04} &  \textcolor{darkerGreen}{5.87e-06} &  \textcolor{darkerGreen}{1.40e-04} &                                 -- &      \textcolor{crimson}{4.59e-01} &      \textcolor{crimson}{1.99e-01} \\
TL &  \textcolor{darkerGreen}{1.71e-10} &  \textcolor{darkerGreen}{6.09e-05} &  \textcolor{darkerGreen}{1.11e-04} &  \textcolor{darkerGreen}{1.55e-08} &  \textcolor{darkerGreen}{1.32e-05} &      \textcolor{crimson}{4.59e-01} &                                 -- &      \textcolor{crimson}{1.75e-01} \\
BL &  \textcolor{darkerGreen}{9.62e-06} &  \textcolor{darkerGreen}{4.62e-04} &  \textcolor{darkerGreen}{4.26e-03} &  \textcolor{darkerGreen}{1.50e-05} &  \textcolor{darkerGreen}{9.08e-04} &      \textcolor{crimson}{1.99e-01} &      \textcolor{crimson}{1.75e-01} &                                 -- \\
\bottomrule
\end{tabular}
\end{adjustbox}
\caption{\textbf{User Study -- \texttt{PerceptPPO}:} Detailed $p$ values of Cluster 3 for all position pairs. Green cells indicate statistically significant differences, or red otherwise. We use a Two-Tailed test with $\alpha=0.05$; therefore, $p$ has to be less than $0.025$ to be significant.}
\label{tab:ppo-results_pvals_c3}
\end{table}

\section{Evaluation 1: Label Density}

In this evaluation, our objective was to understand the impact of label density on users' preferences. Recognizing that dense labeling can clutter a map and reduce its legibility, while too sparse a distribution might overlook important information, we sought to identify an optimal label density from a user-centered perspective.

\subsection{User Study Interface}
\begin{figure}[H]
    
    \includegraphics[width=\columnwidth]{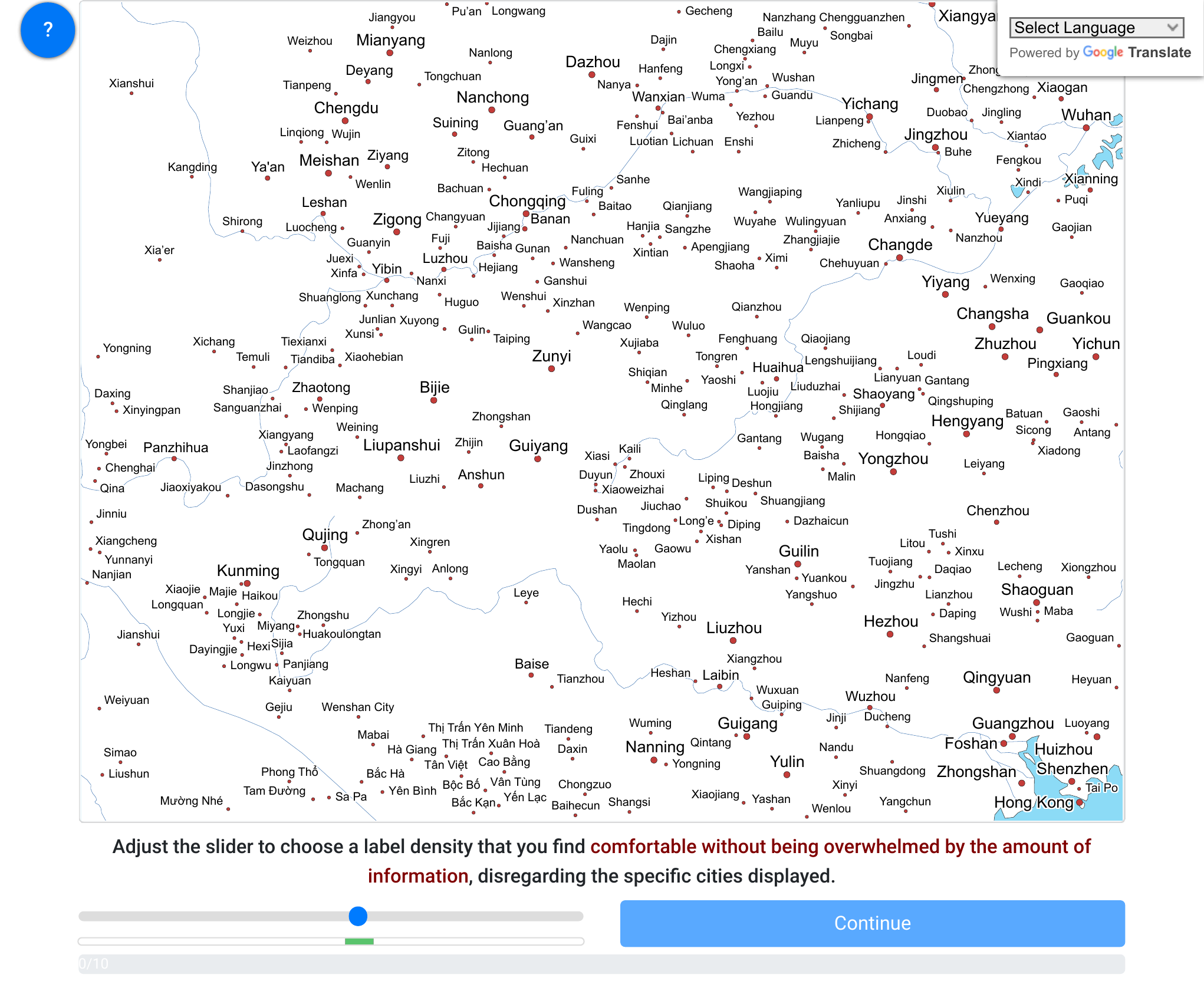}
    \caption{
        \textbf{Evaluation 1 -- Label Density:} Web interface for experiment. Initial position of slider is randomized in each trial.
    }
    \label{fig:web-density1}
\end{figure}

\begin{figure}[H]
    
    \includegraphics[width=\columnwidth]{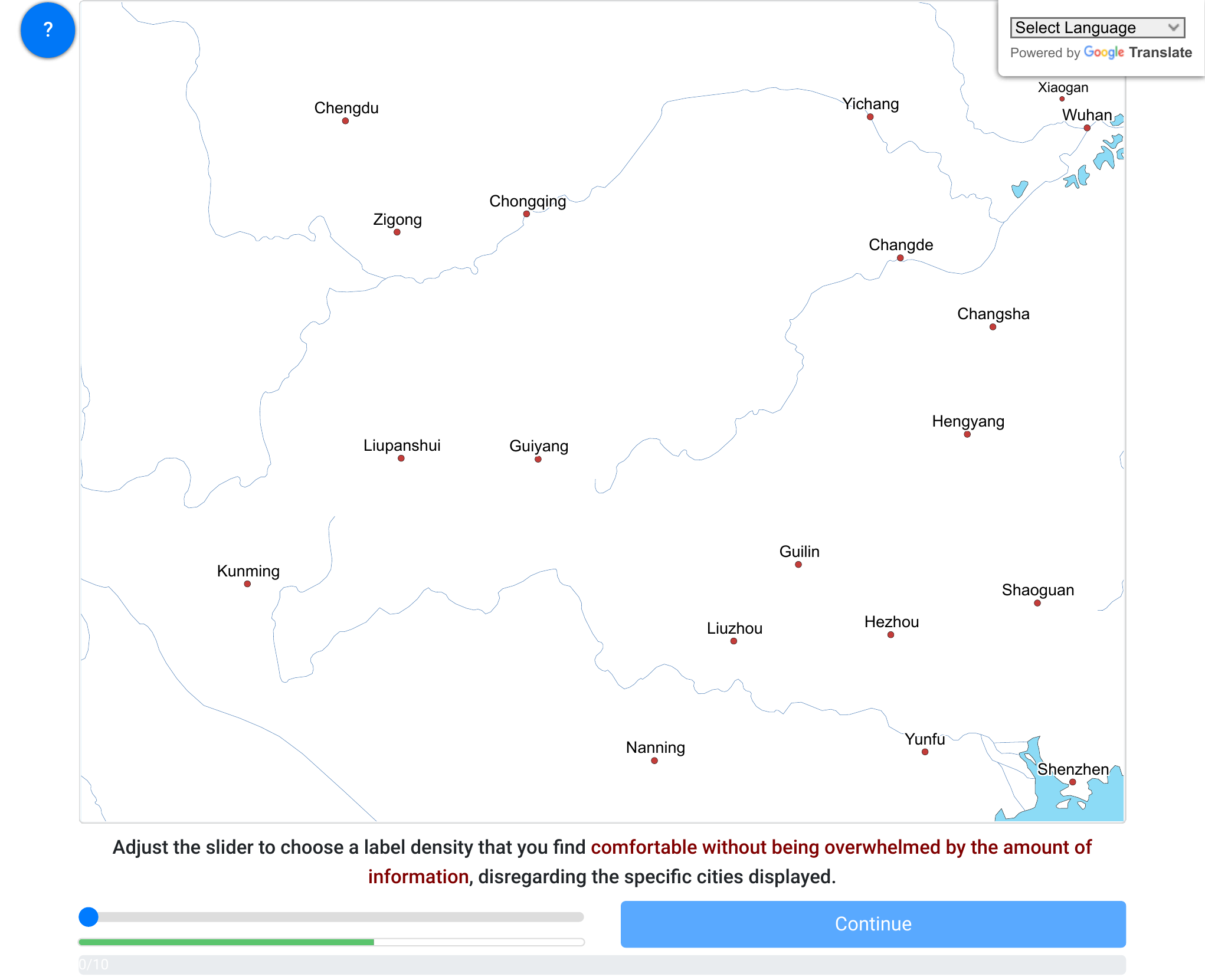}
    \caption{
        \textbf{Evaluation 1 -- Label Density:} Web interface for experiment. After setting the slider to the leftmost position while exploring the intermediate densities, as shown by the green bar below the slider.
    }
    \label{fig:web-density2}
\end{figure}

\begin{figure}[H]
        
        \includegraphics[width=\columnwidth]{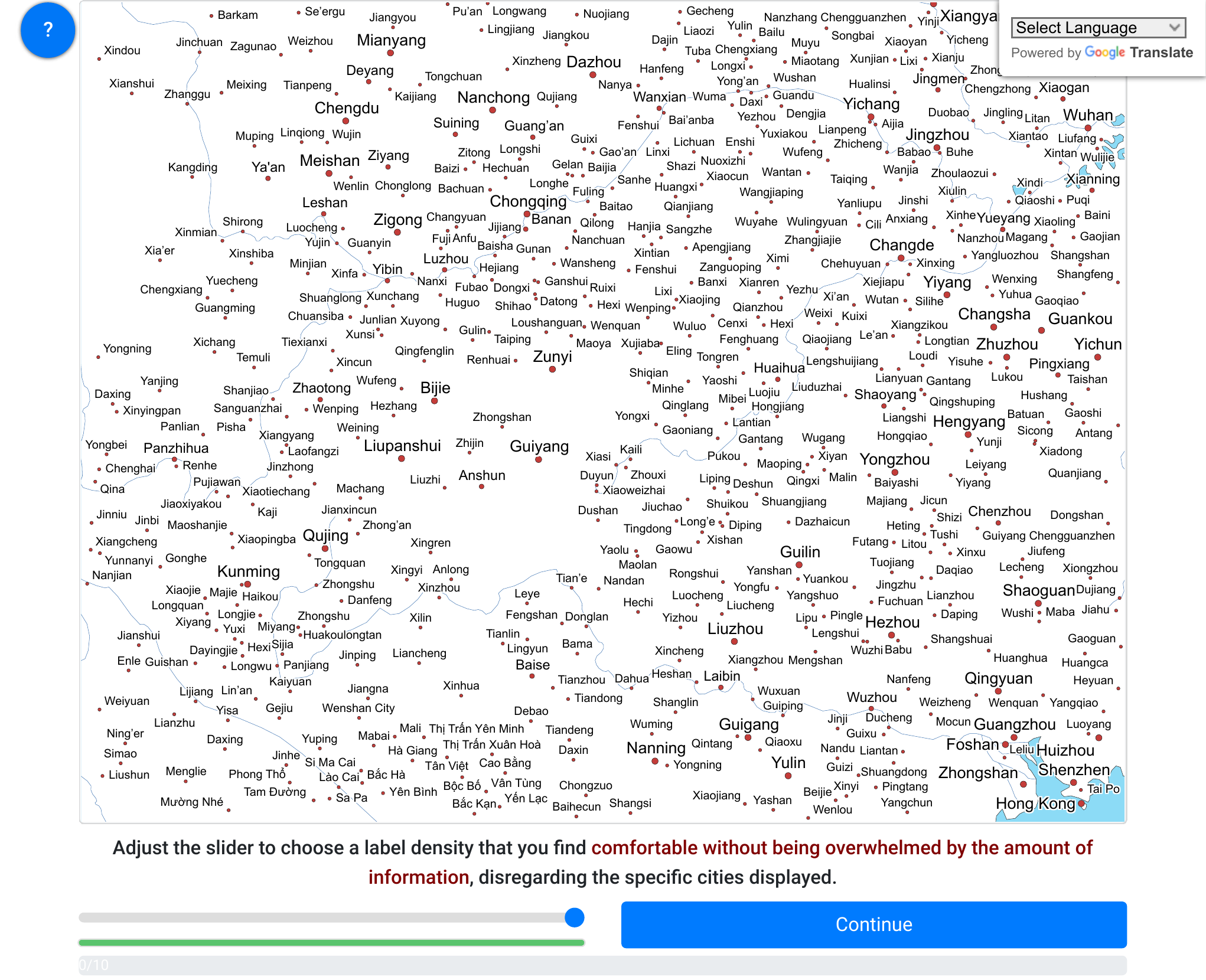}
    \caption{
        \textbf{Evaluation 1 -- Label Density:} Web interface for experiment. After setting the slider to the rightmost position while exploring the intermediate densities, as shown by the green bar below the slider.
    }
    \label{fig:web-density3}
\end{figure}

This section provides examples of user interface used to undercover preferred label density. The following paragraphs present textual information available to participants as seen in \autoref{fig:web-density1}, \autoref{fig:web-density2}, and \autoref{fig:web-density3}.

\textbf{Introduction} Welcome to our user study! We are excited to have you participate in this research to help us understand more about the perception of maps. Your feedback will play a crucial role in shaping future map-based products and services. 

\textbf{Task} You will see a map along with a slider positioned below it. The slider will alter the density (amount) of labels on the map. Your goal is to use the slider to set a label density on the map that feels comfortable and not overwhelming by the amount of information for you. Please do not focus on the specific cities or locations presented, but solely on the density of labels.

\textbf{Instructions} Move the slider with your mouse or use the left and right arrow keys to set a prefered label density. Press "Continue" to proceed to the next map. At the end of the assignment, you will have an opportunity to provide additional feedback. Afterward, press the "Submit" to submit your response.

\textbf{Note} Please note that opening multiple HITs simultaneously is not allowed, and only one submission per participant will be eligible for payment. Additionally, any attempt to tamper with the website's functionality is strictly prohibited and will result in disqualification from the study.

\textbf{Start Experiment} Ready to get started? Press the "Start Experiment" button below and follow the instructions on your screen. By proceeding, you consent to participate in this study. Provided data will be collected and used for research purposes only.

\textbf{Call For Action} Adjust the slider to choose a label density that you find comfortable without being overwhelmed by the amount of information, disregarding the specific cities displayed.

\subsection{Data}
\begin{figure*}[h!]
 
    \includegraphics[width=\textwidth]{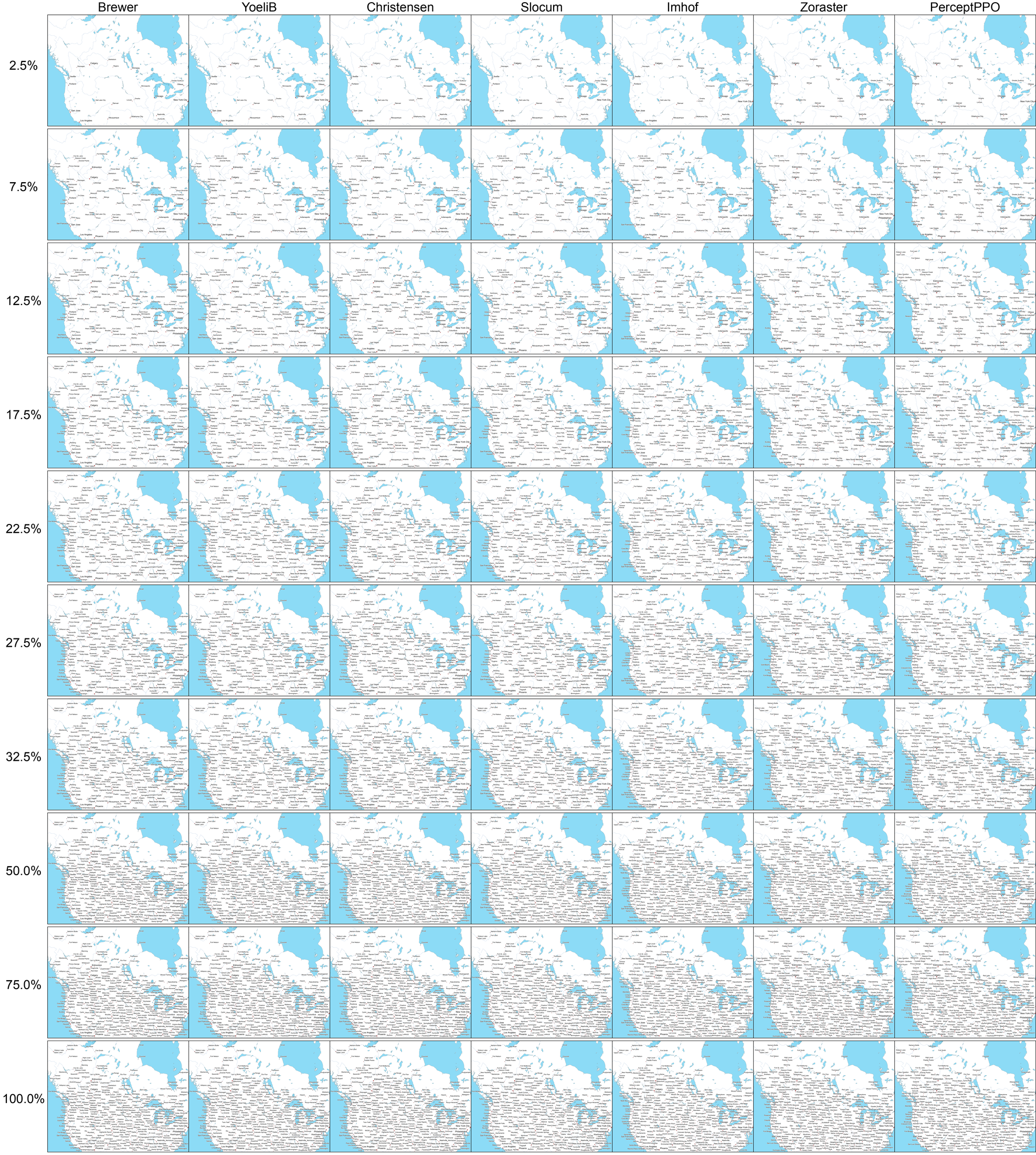}
    \caption{\textbf{Evaluation 1 -- Label Density Study:} Visualization of area 0 with various label density thresholds $LD_\mathrm{thr}$ used in the user study.}
    \label{fig:density_renders1}
\end{figure*}

\begin{figure*}[h!]
 
    \includegraphics[width=\textwidth]{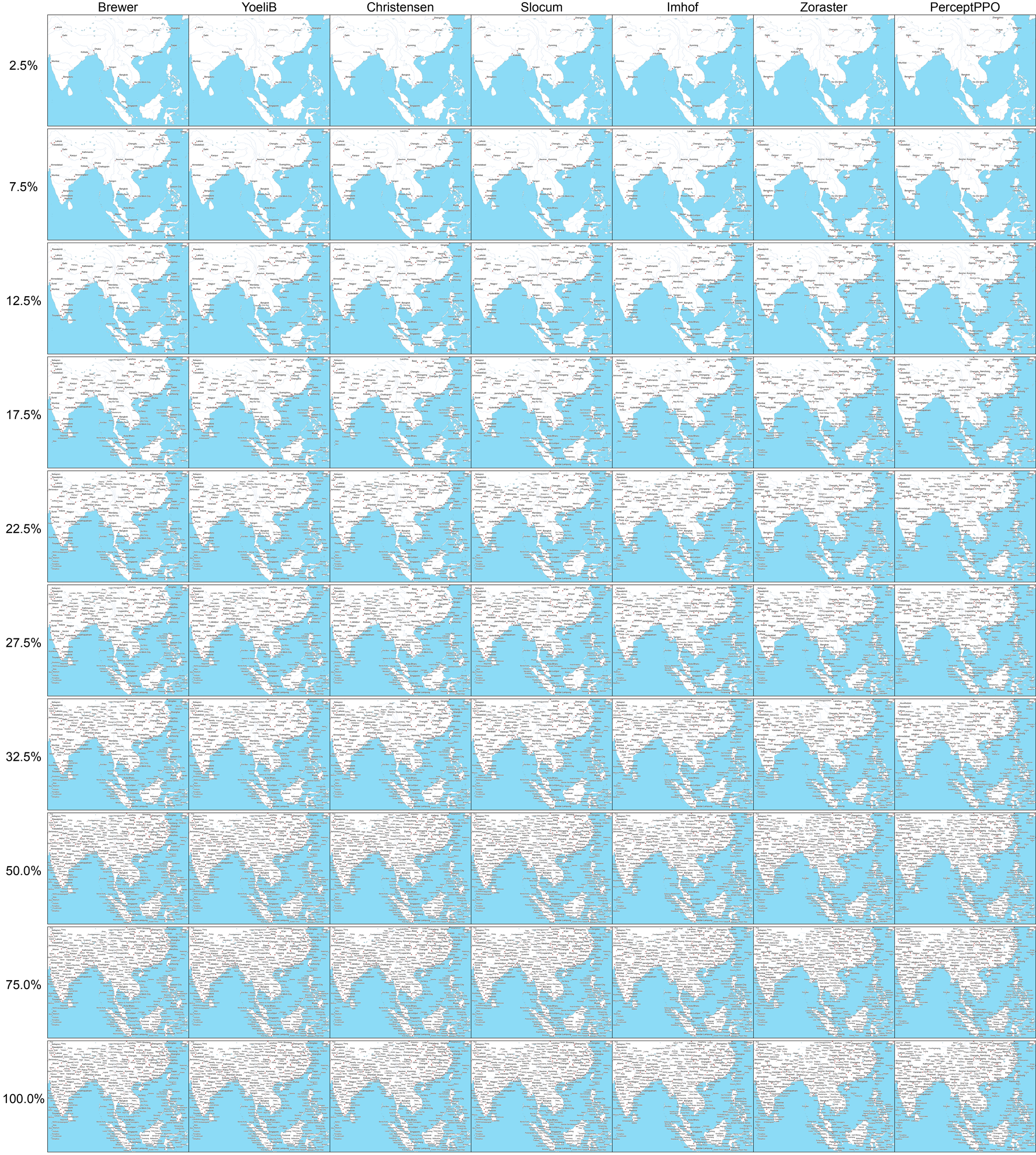}
    \caption{\textbf{Evaluation 1 -- Label Density Study:} Visualization of area 4 with various label density thresholds $LD_\mathrm{thr}$ used in the user study.}
    \label{fig:density_renders2}
\end{figure*}

\begin{figure*}[h!]
 
    \includegraphics[width=\textwidth]{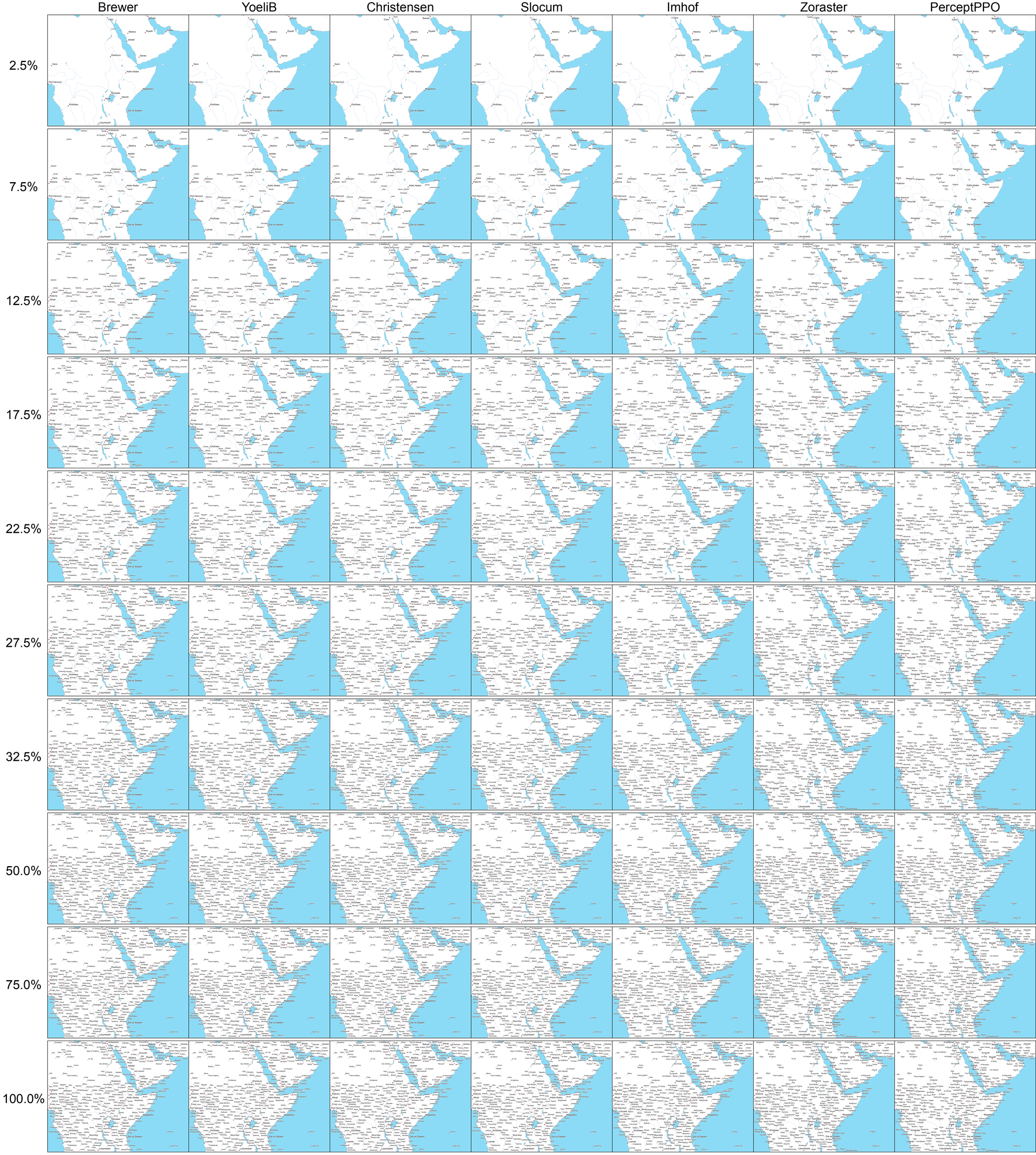}
    \caption{\textbf{Evaluation 1 -- Label Density Study:} Visualization of area 5 with various label density thresholds $LD_\mathrm{thr}$ used in the user study.}
    \label{fig:density_renders3}
\end{figure*}

\begin{figure*}[h!]
 
    \includegraphics[width=\textwidth]{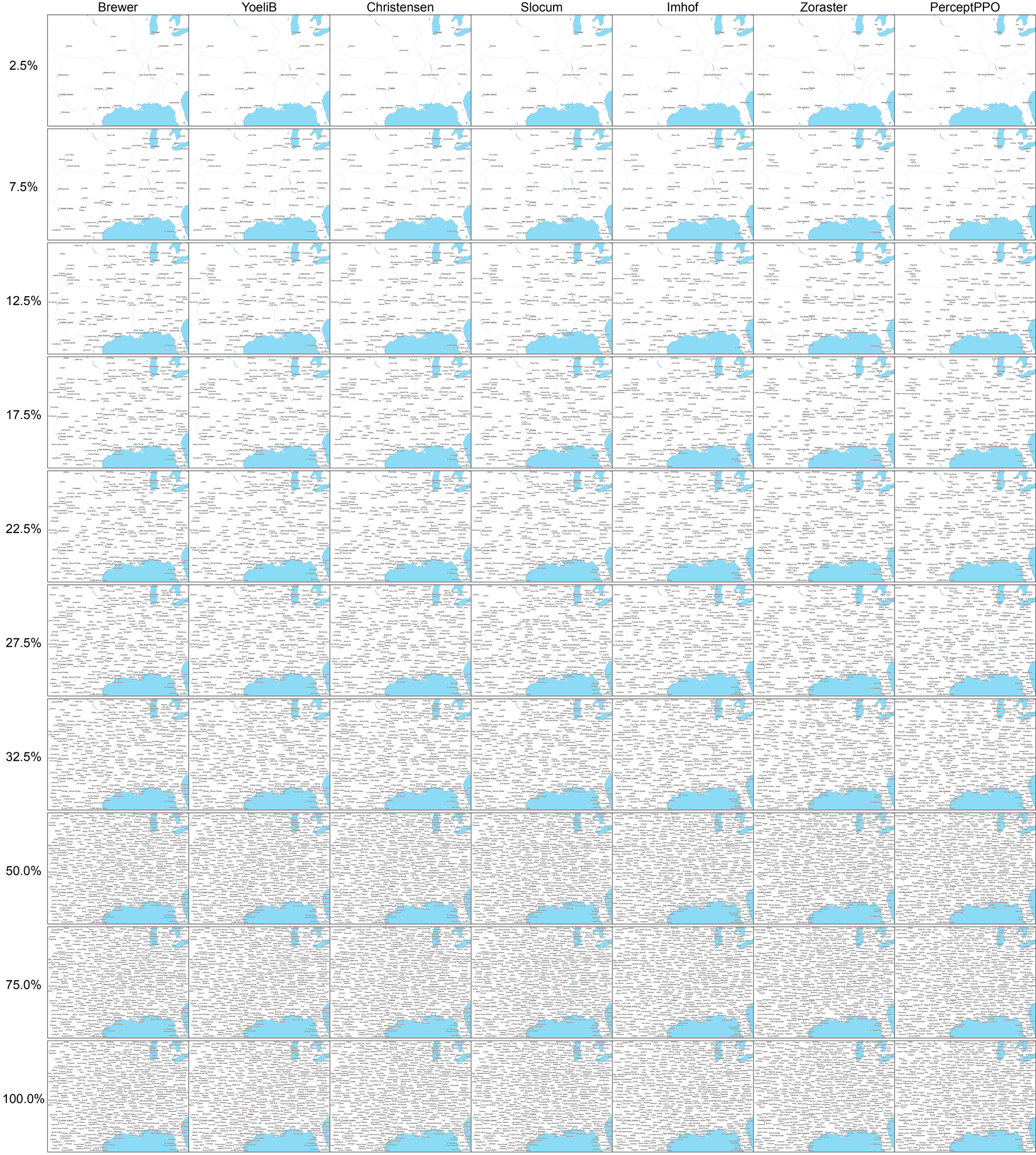}
    \caption{\textbf{Evaluation 1 -- Label Density Study:} Visualization of area 6 with various label density thresholds $LD_\mathrm{thr}$ used in the user study.}
    \label{fig:density_renders4}
\end{figure*}

\begin{figure*}[h!]
 
    \includegraphics[width=\textwidth]{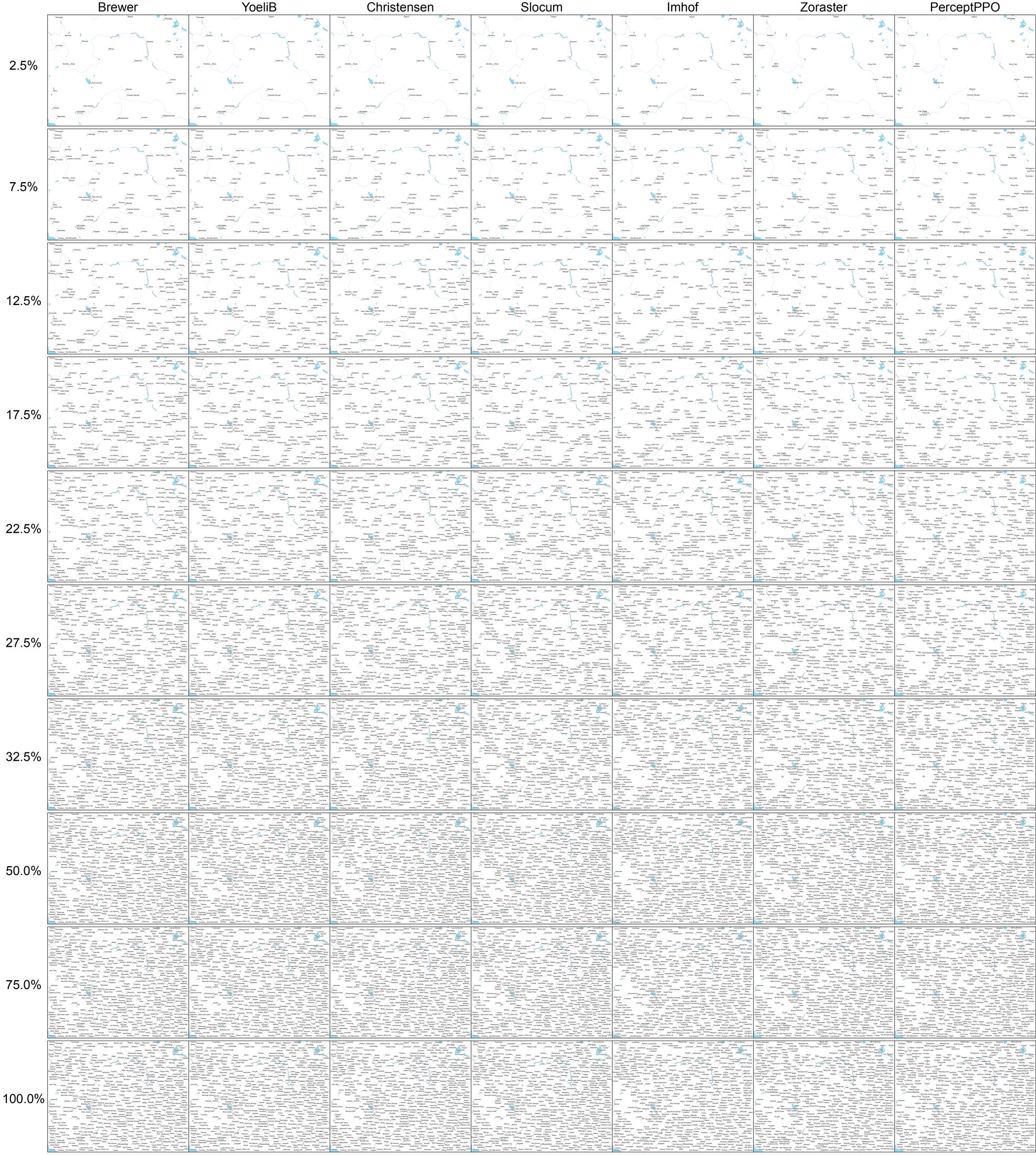}
    \caption{\textbf{Evaluation 1 -- Label Density Study:} Visualization of area 9 with various label density thresholds $LD_\mathrm{thr}$ used in the user study.}
    \label{fig:density_renders5}
\end{figure*}

\begin{figure*}[h!]
 
    \includegraphics[width=\textwidth]{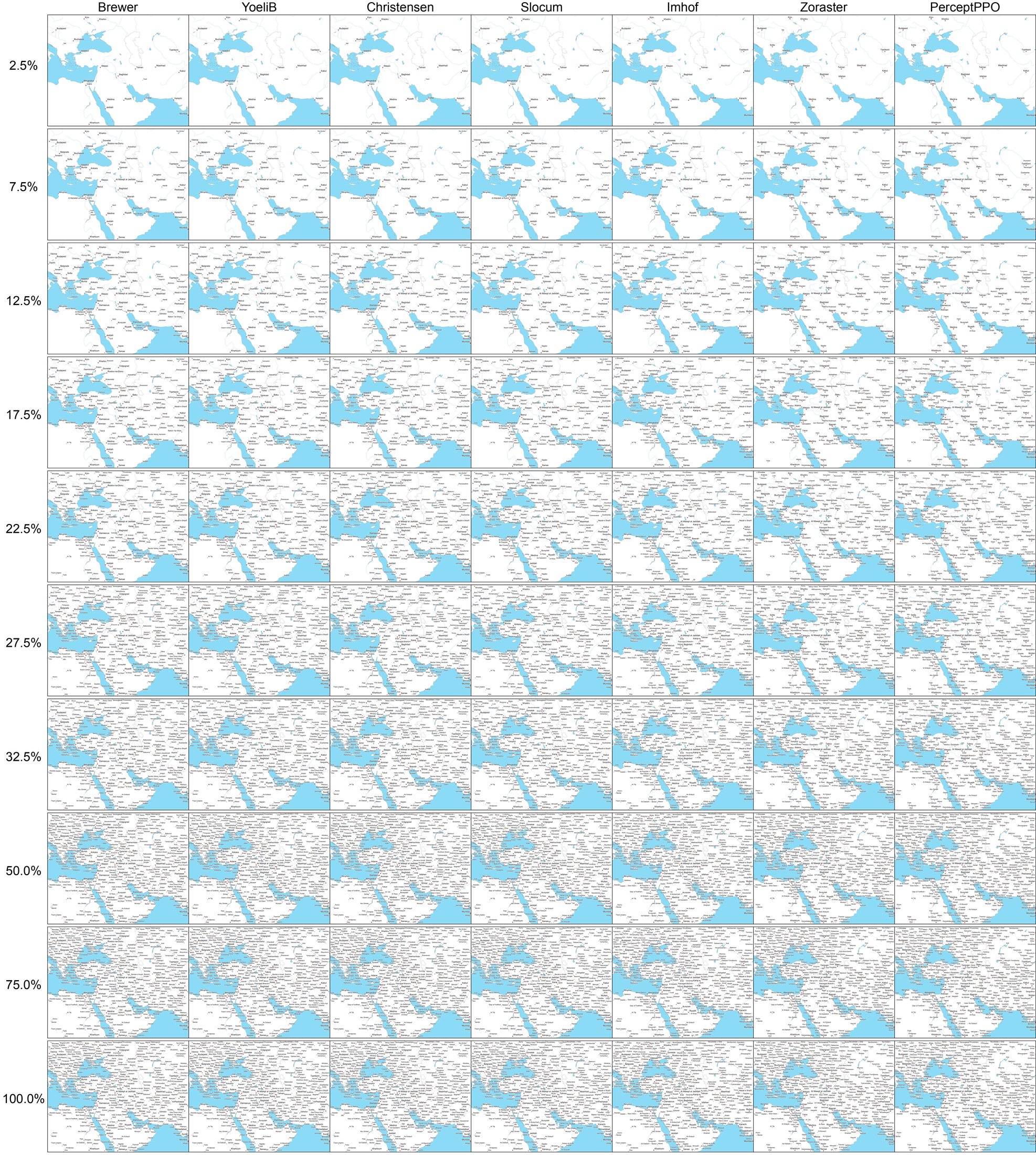}
    \caption{\textbf{Evaluation 1 -- Label Density Study:} Visualization of area 12 with various label density thresholds $LD_\mathrm{thr}$ used in the user study.}
    \label{fig:density_renders6}
\end{figure*}

\begin{figure*}[h!]
 
    \includegraphics[width=\textwidth]{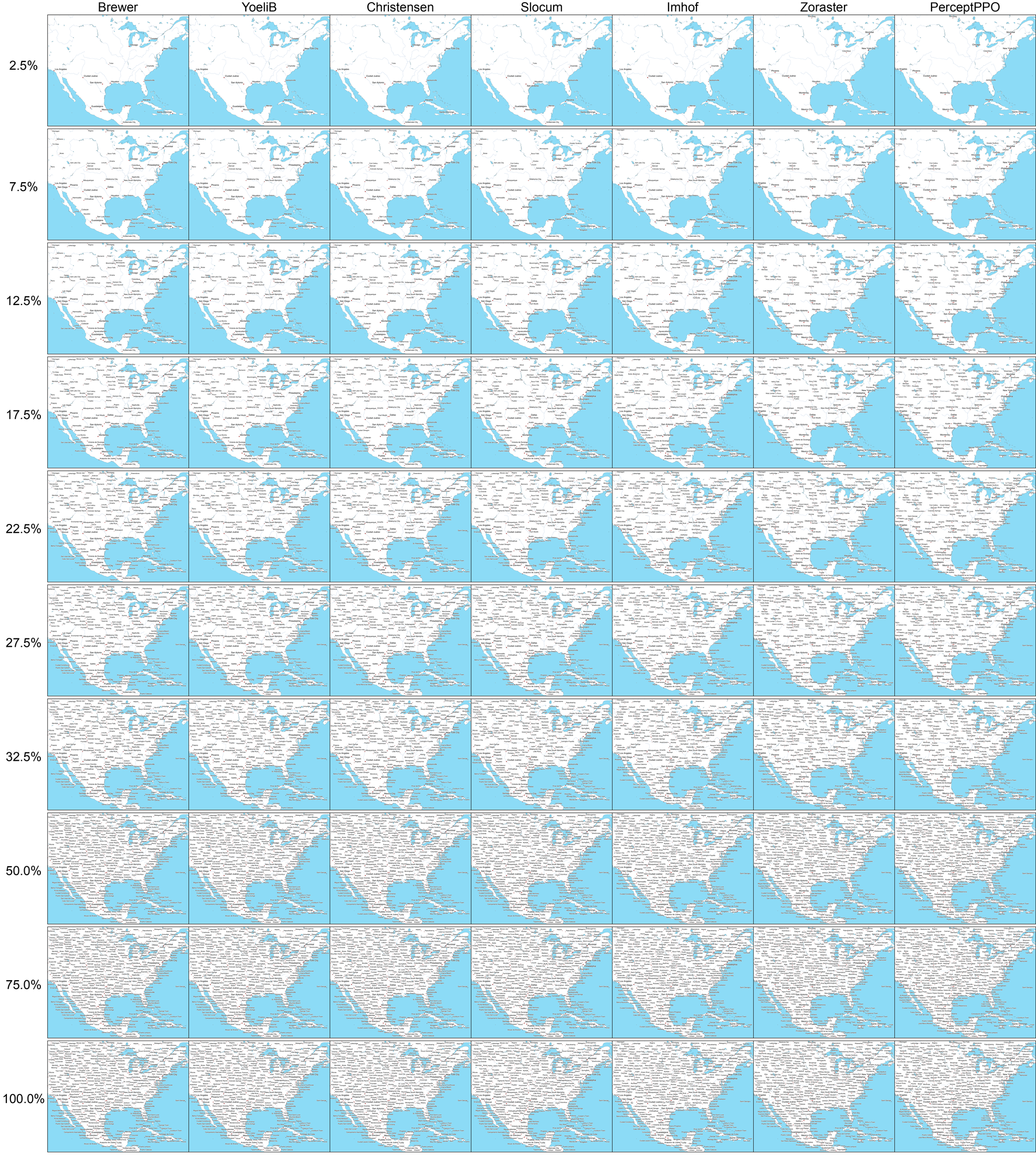}
    \caption{\textbf{Evaluation 1 -- Label Density Study:} Visualization of area 13 with various label density thresholds $LD_\mathrm{thr}$ used in the user study.}
    \label{fig:density_renders7}
\end{figure*}

\begin{figure*}[h!]
 
    \includegraphics[width=\textwidth]{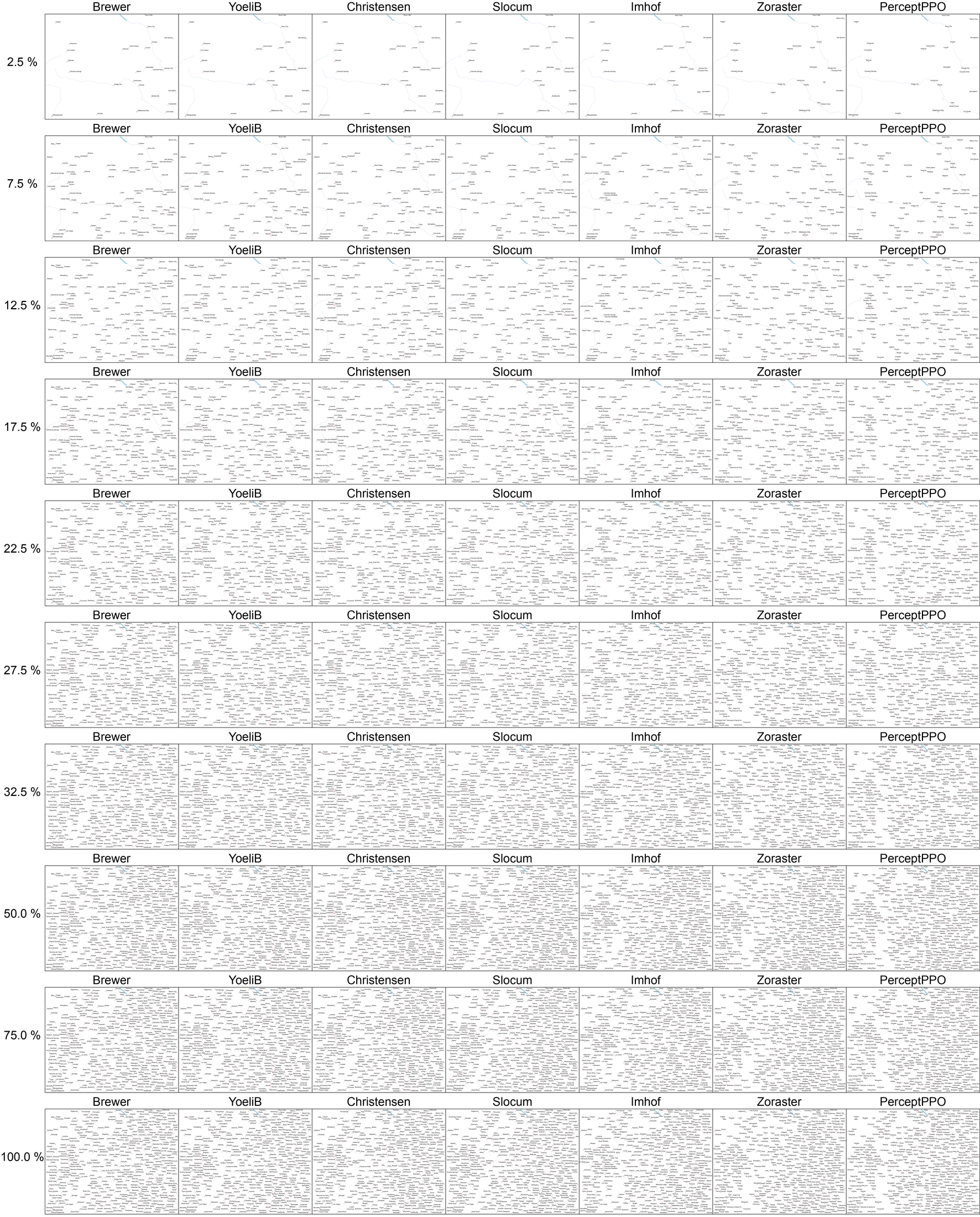}
    \caption{\textbf{Evaluation 1 -- Label Density Study:} Visualization of area 16 with various label density thresholds $LD_\mathrm{thr}$ used in the user study.}
    \label{fig:density_renders8}
\end{figure*}

\begin{figure*}[h!]
 
    \includegraphics[width=\textwidth]{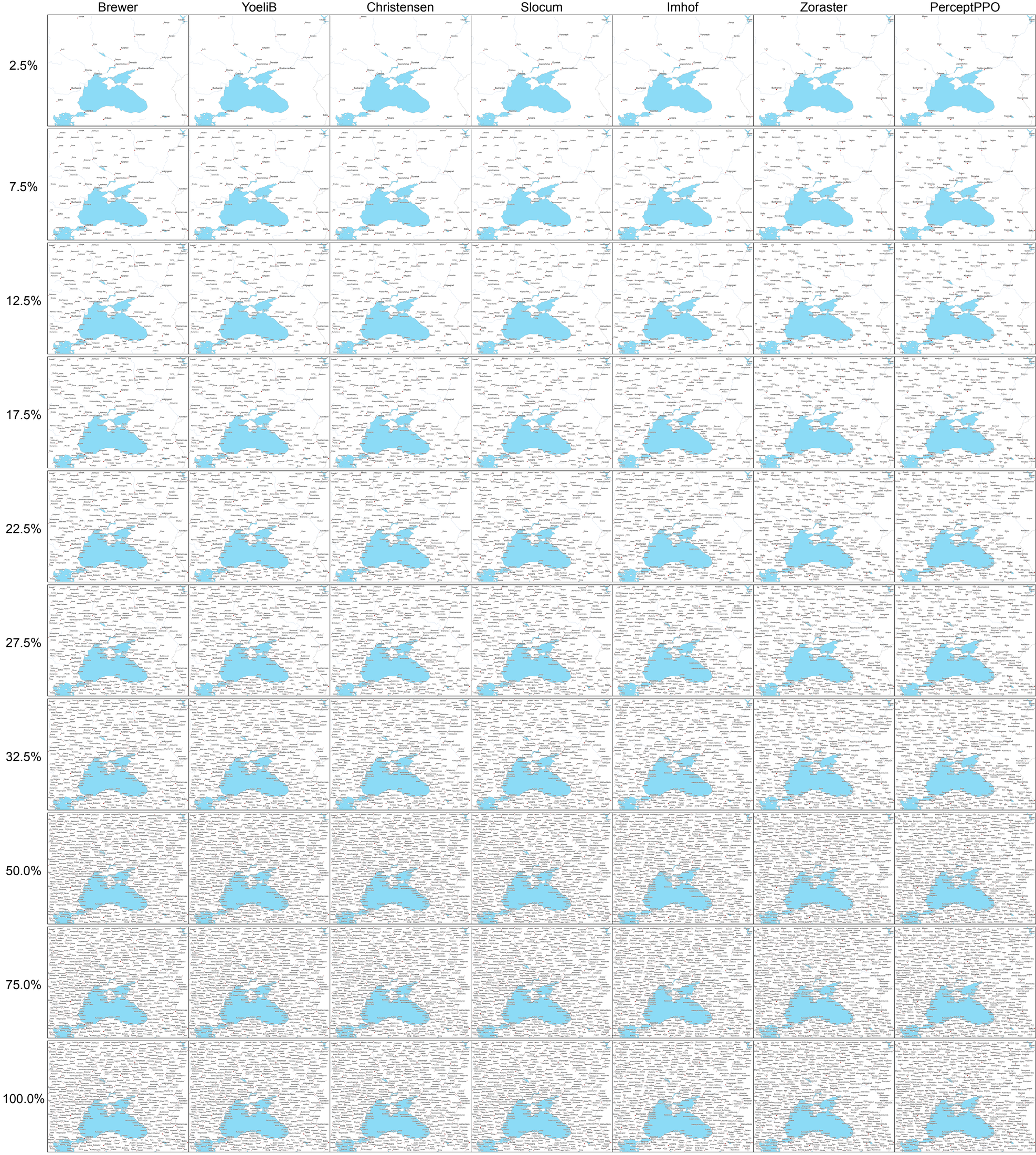}
    \caption{\textbf{Evaluation 1 -- Label Density Study:} Visualization of area 27 with various label density thresholds $LD_\mathrm{thr}$ used in the user study.}
    \label{fig:density_renders9}
\end{figure*}

\begin{figure*}[h!]
 
    \includegraphics[width=\textwidth]{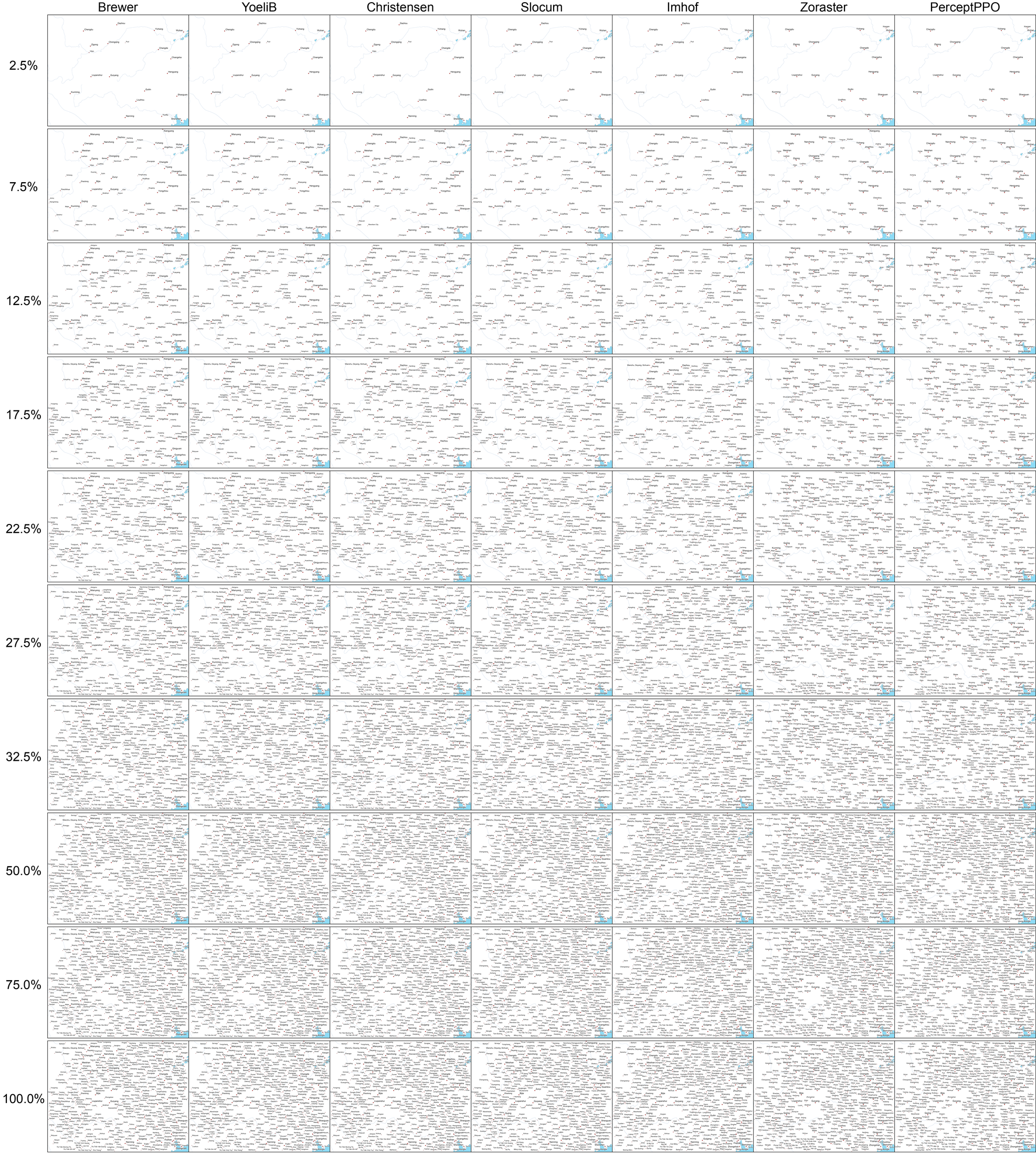}
    \caption{\textbf{Evaluation 1 -- Label Density Study:} Visualization of area 29 with various label density thresholds $LD_\mathrm{thr}$ used in the user study.}
    \label{fig:density_renders10}
\end{figure*}

\autoref{fig:density_renders1}, \autoref{fig:density_renders2}, \autoref{fig:density_renders3}, \autoref{fig:density_renders4}, \autoref{fig:density_renders5}, \autoref{fig:density_renders6}, \autoref{fig:density_renders7}, \autoref{fig:density_renders8}, \autoref{fig:density_renders9}, and \autoref{fig:density_renders10} display the map renders utilized in the label density user study, showcasing various label density thresholds $LD_\mathrm{thr}$ for all selected areas. We chose ten populated areas (0, 4, 5, 6, 9, 12, 13, 17, 27, 29) from the \texttt{PerceptPPO} study to cover a broad spectrum of label density scenarios, including geographical features like continent borders and bodies of water. This approach aimed to mirror maps encountered in everyday life while limiting extraneous influences on participant choices. We utilized Natural Earth data for physical features and maintained the same city data source as in previous sections, opting for a 1:50,000,000 scale to balance detail with manageable SVG image sizes.

\subsection{Results}

\begin{figure*}[!ht]
    \subfigure[] {
        
        \includegraphics[width=0.33\textwidth]{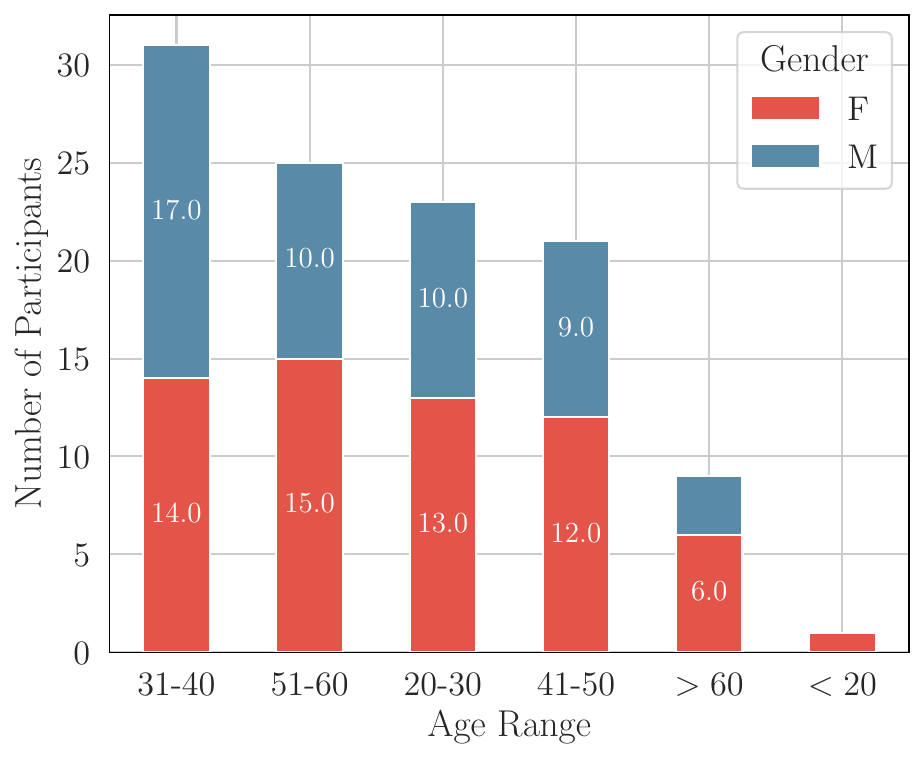}
    }
    \hspace{0em}
    \subfigure[] {
        
        \includegraphics[width=0.33\textwidth]{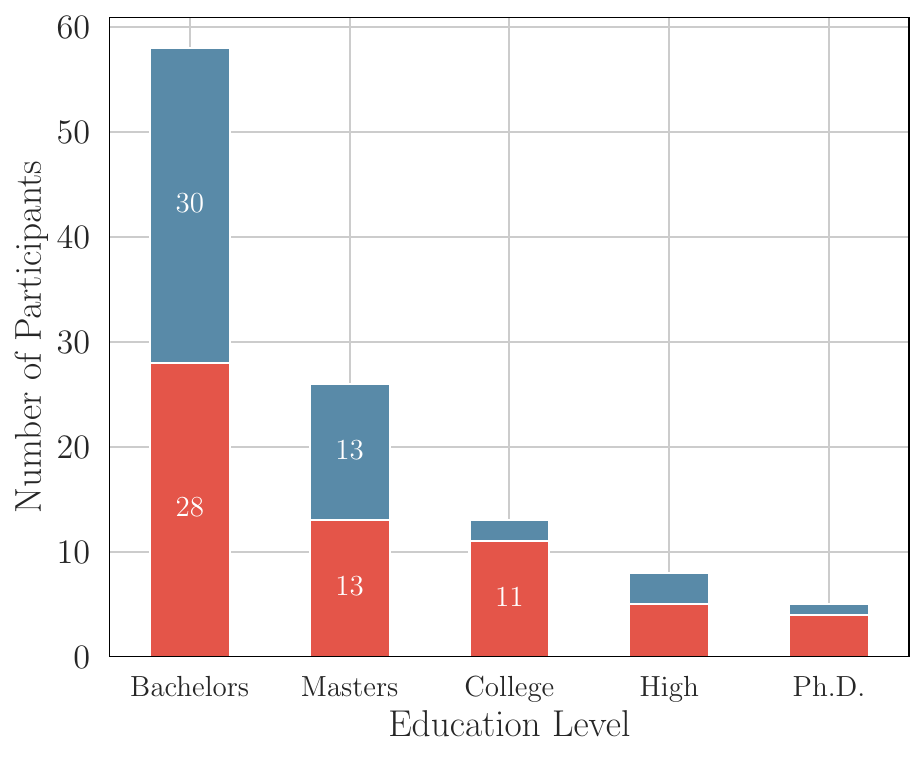}
    }
    \hspace{0em}
    \subfigure[] {
        
        \includegraphics[width=0.33\textwidth]{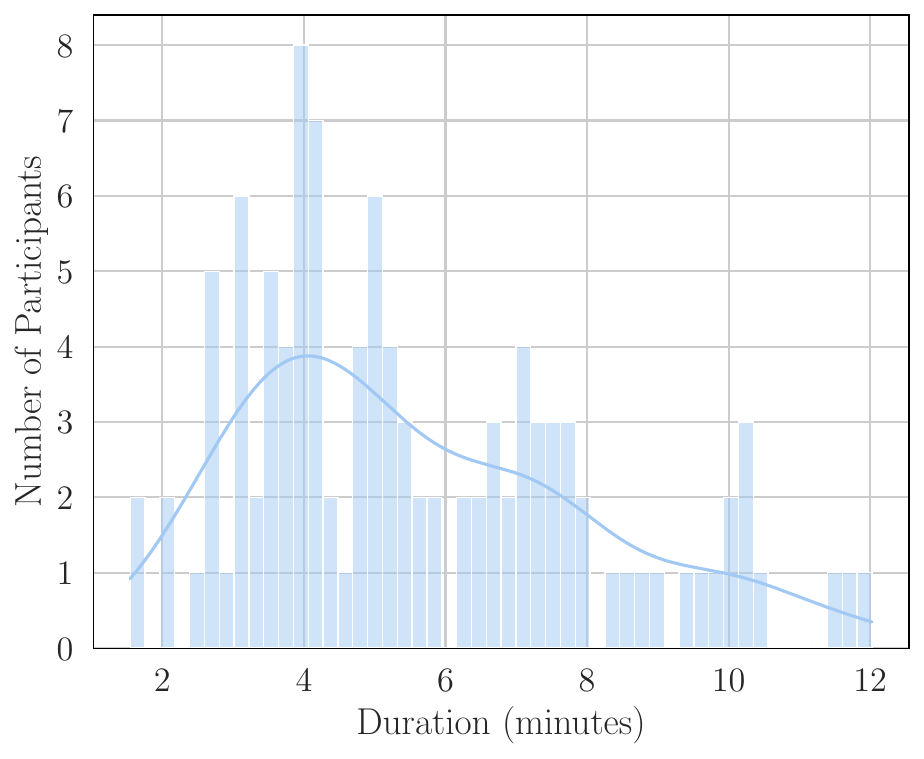}
    }
    \caption{
        \textbf{Evaluation 1 -- Label Density Study:} User statistics of age, education, and duration of study for all participants.
    }
    \label{fig:density_stats}
\end{figure*}

\begin{figure*}[!ht]
    \subfigure[] {
        
        \includegraphics[width=0.5\textwidth]{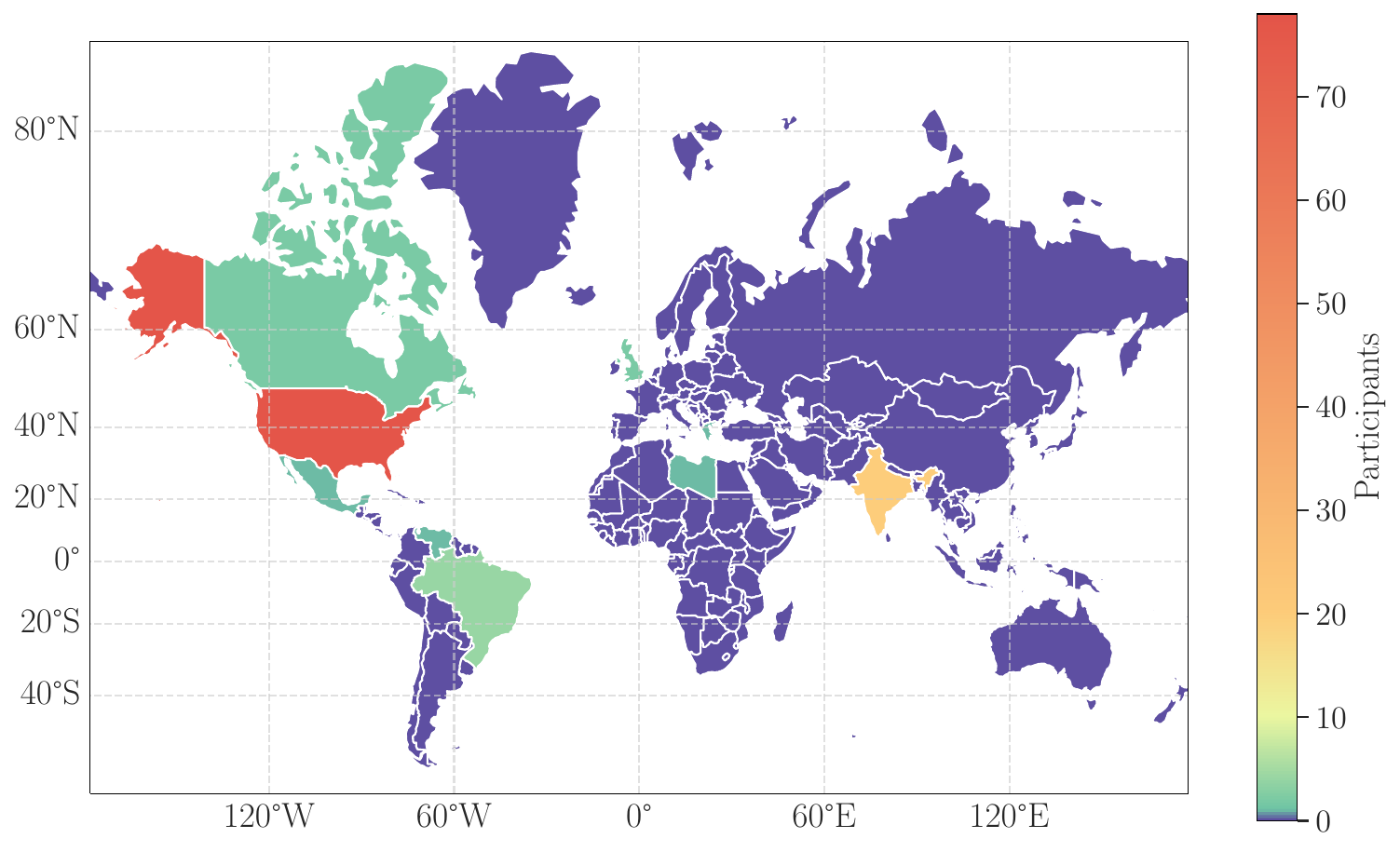}
    }
    \hspace{0em}
    \subfigure[] {
        
        \includegraphics[width=0.5\textwidth]{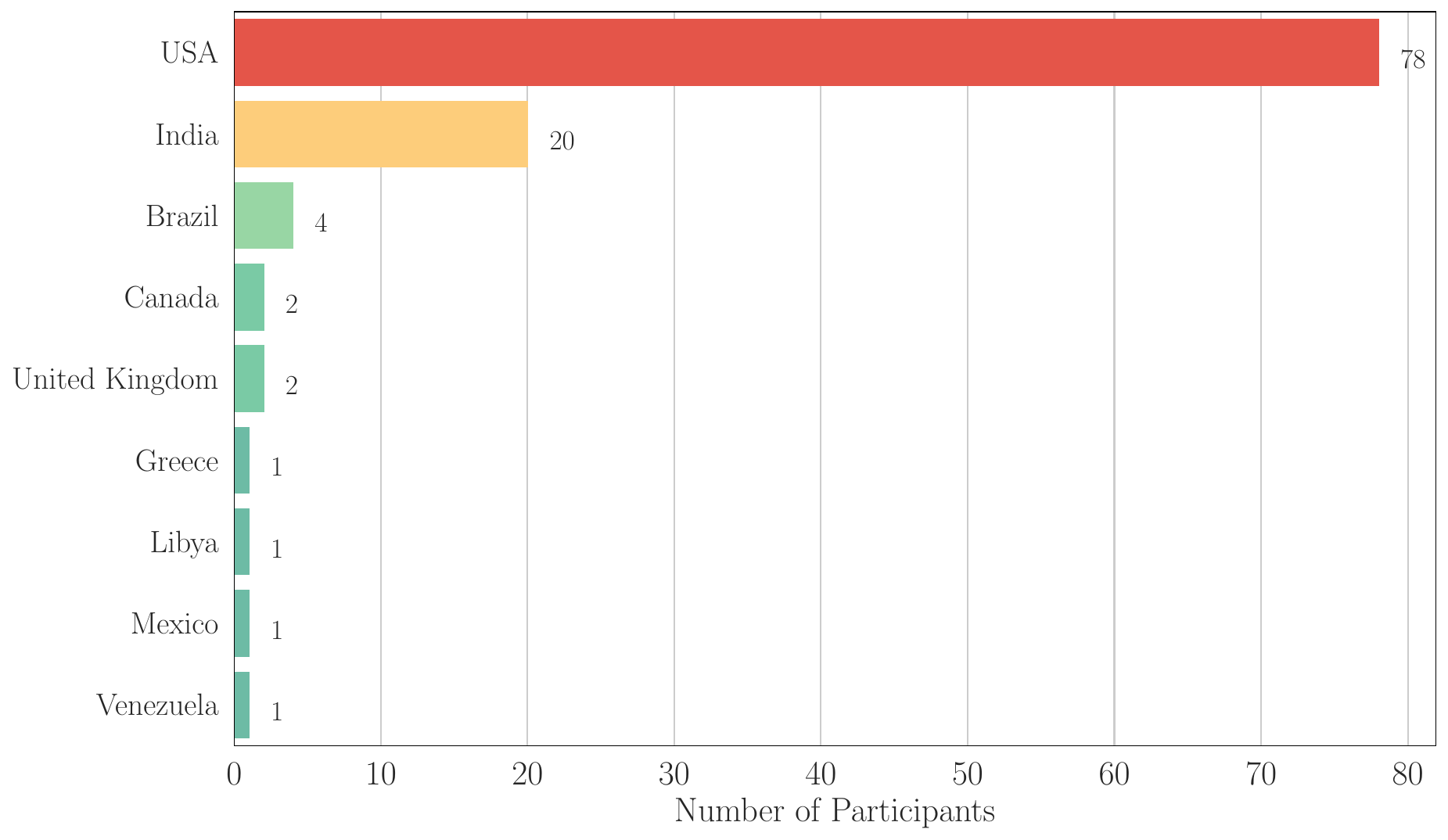}
    }
    \caption{
        \textbf{Evaluation 1 -- Label Density Study:} Country distribution for all participants.
    }
    \label{fig:density_countries}
\end{figure*}

\begin{figure*}[!ht]
    \subfigure[] {
        
        \includegraphics[width=0.33\textwidth, trim={0.5cm 0cm 0.5cm 0cm}, clip]{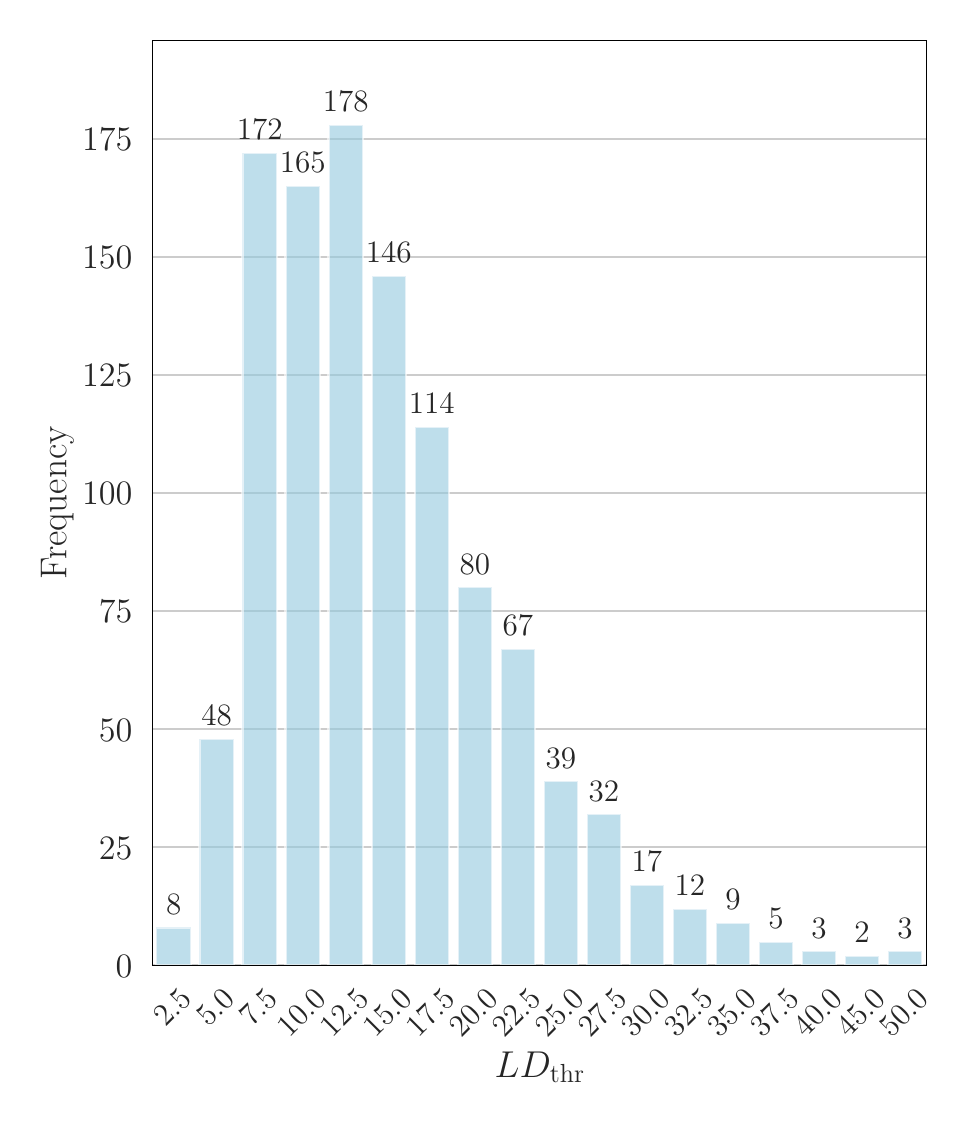}
    }
    \hspace{0em}
    \subfigure[] {
        
        \includegraphics[width=0.33\textwidth, trim={0.5cm 0cm 0.5cm 0cm}, clip]{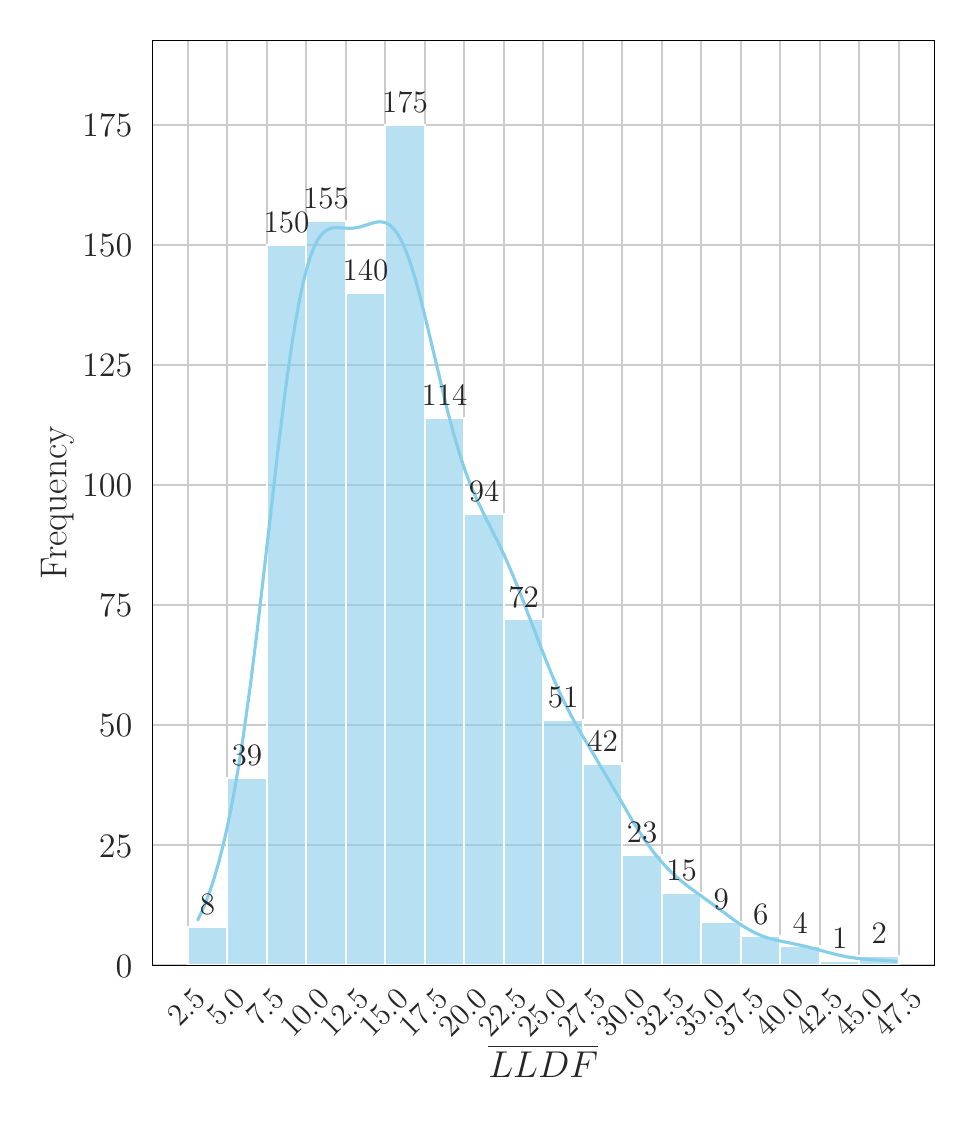}
    }
    \hspace{0em}
    \subfigure[] {
        
        \includegraphics[width=0.33\textwidth, trim={0.5cm 0cm 0.5cm 0cm}, clip]{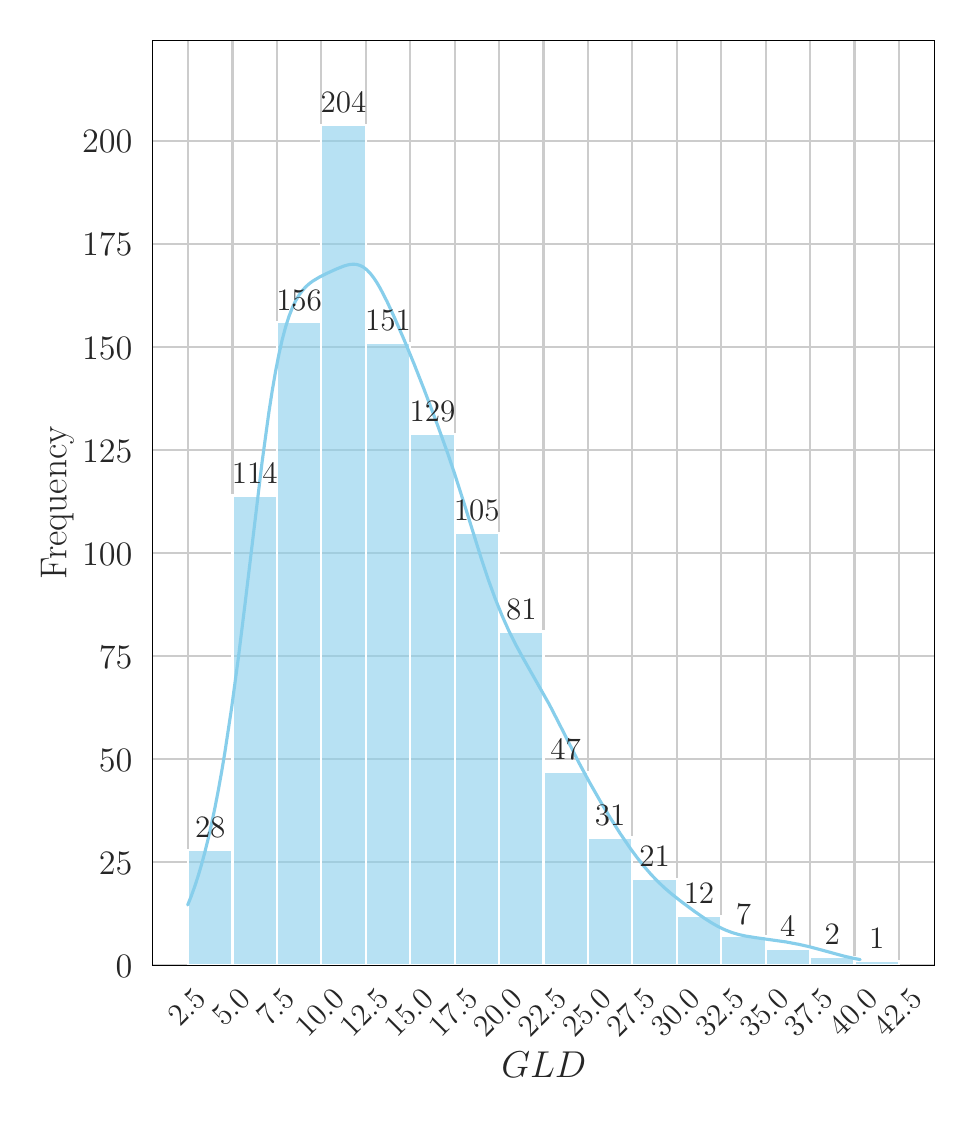}
    }
    \caption{
        \textbf{Evaluation 1 -- Label Density Study:} Frequency of  $LD_\mathrm{thr}$, $\overline{LLDF}$, and $GLD$ of study for all participants.
    }
    \label{fig:density_hits}
\end{figure*}

\begin{figure*}[h!]
    \centering
    \subfigure {
        
        \includegraphics[width=0.95\textwidth]{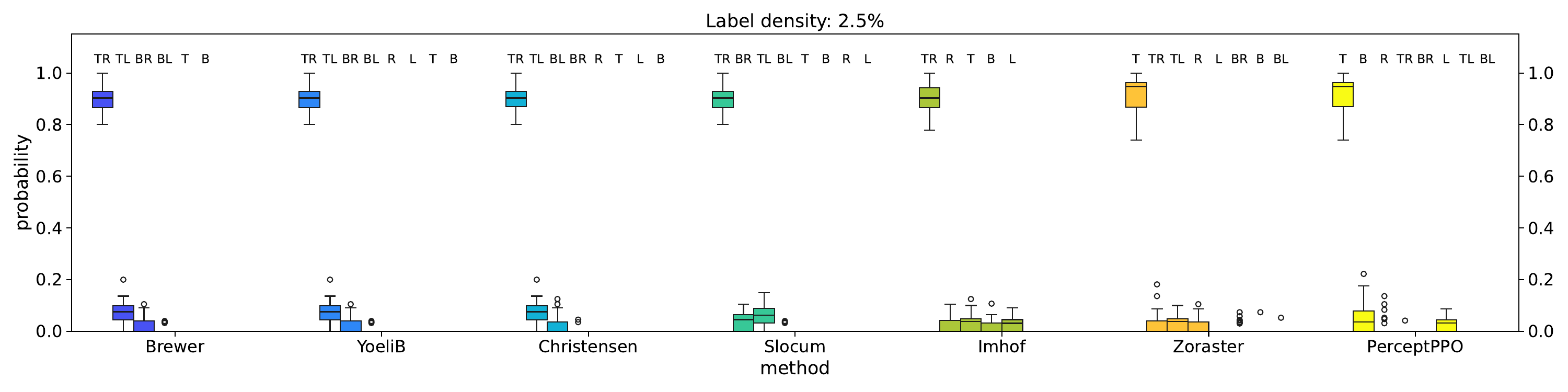}
    }\\
    \subfigure {
        
        \includegraphics[width=0.95\textwidth]{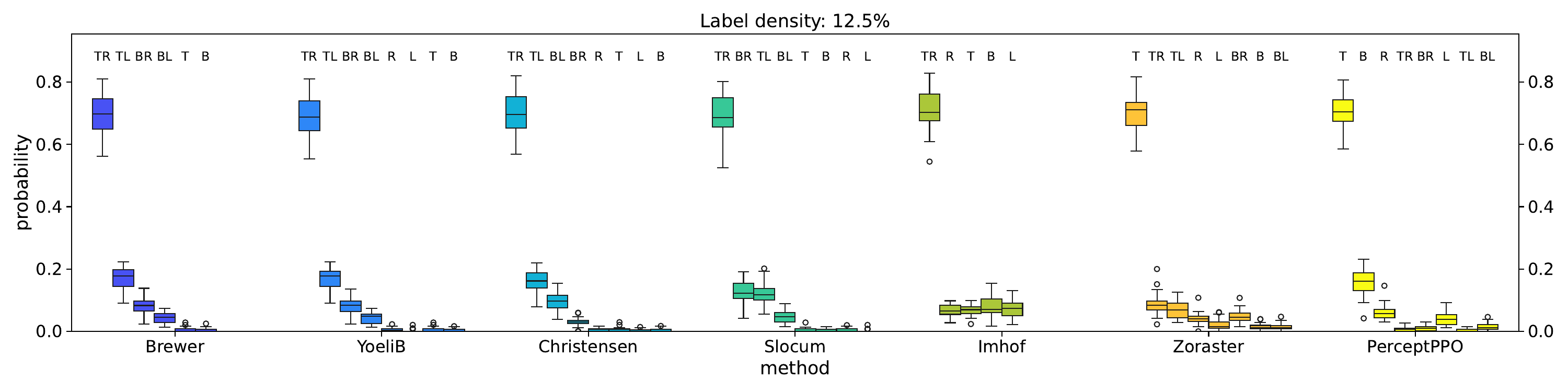}
    }\\
    \subfigure {
        
        \includegraphics[width=0.95\textwidth]{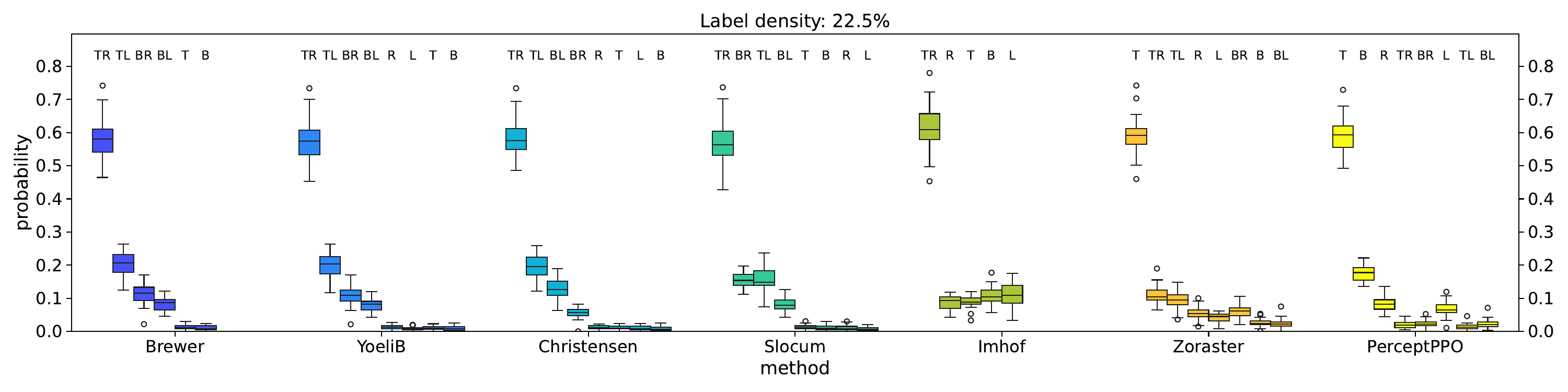}
    }\\
    \subfigure {
        
        \includegraphics[width=0.95\textwidth]{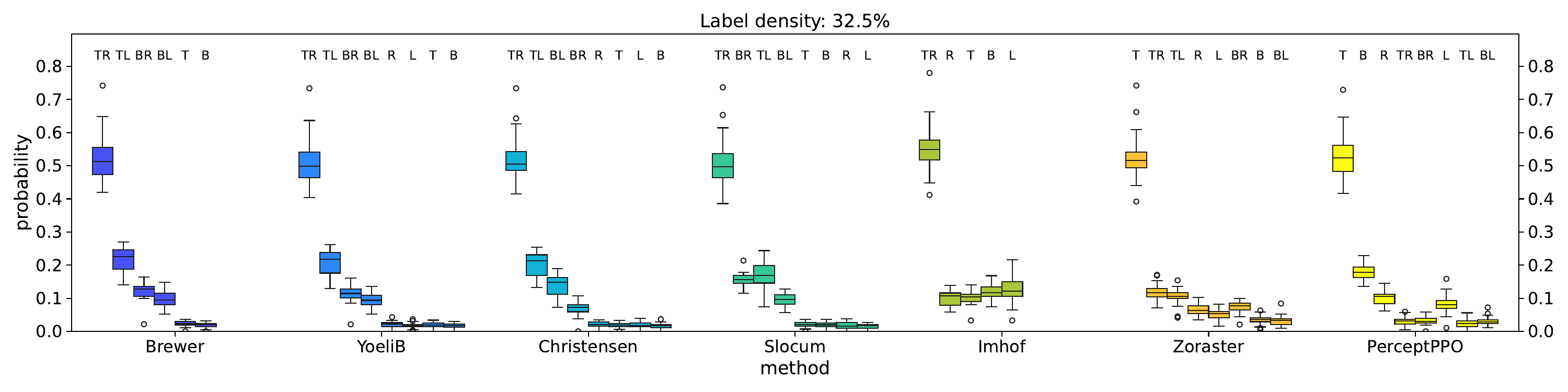}
    }\\
    \subfigure {
        
        \includegraphics[width=0.95\textwidth]{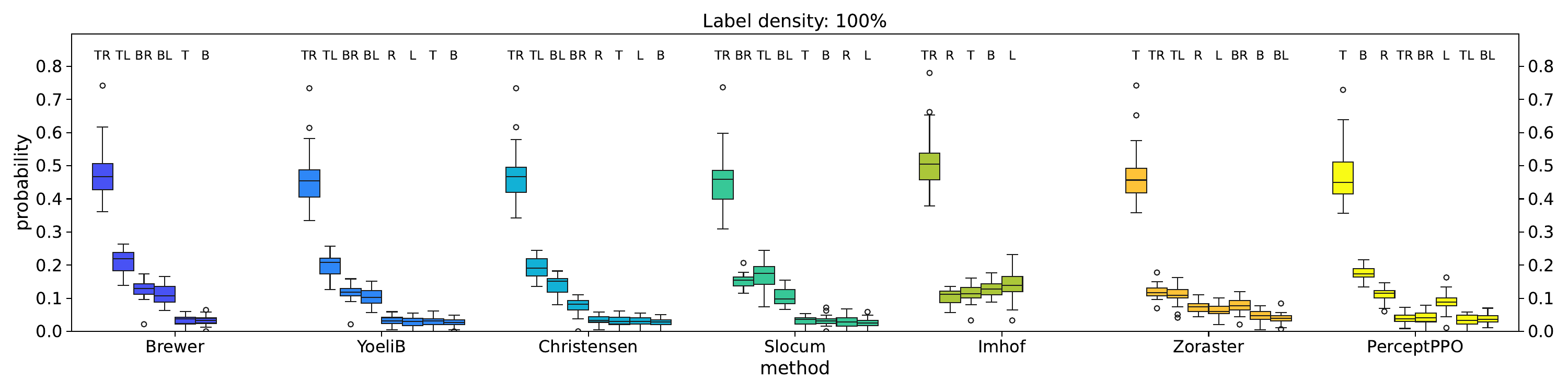}
    }
    \caption{\textbf{Evaluation 1 -- Label Density Study:} The probability that a label is placed at a specific position for each investigated method is dependent on label density. For each PPO, the positions (see the labels above the boxplots) are in ascending order with respect to their priorities from left to right. Label density threshold $LD_\mathrm{thr}$ increases from top to bottom.}
    \label{fig:probabilities_density_all_methods}
\end{figure*}

\autoref{fig:probabilities_density_all_methods} show probabilities that a label is placed at a specific position for each investigated method dependent on label density. \autoref{fig:density_hits} displays frequency of $LD_\mathrm{thr}$, $\overline{LLDF}$, and $GLD$ of study for all participants. \autoref{fig:density_stats} and \autoref{fig:density_countries} show the detailed statistics of engaged participants.

\section{Evaluation 2: Comparison of PPOs}

This segment of our research focuses on a comparative analysis of various Position Priority Orders (PPOs), as delineated in our study. We aim to ascertain the relative effectiveness and user preference across different PPO configurations, offering insights into how label placement strategies can be optimized for user satisfaction.

\subsection{User Study Interface}
\begin{figure}[H]
    \subfigure[Introduction] {
        
        \includegraphics[width=\columnwidth]{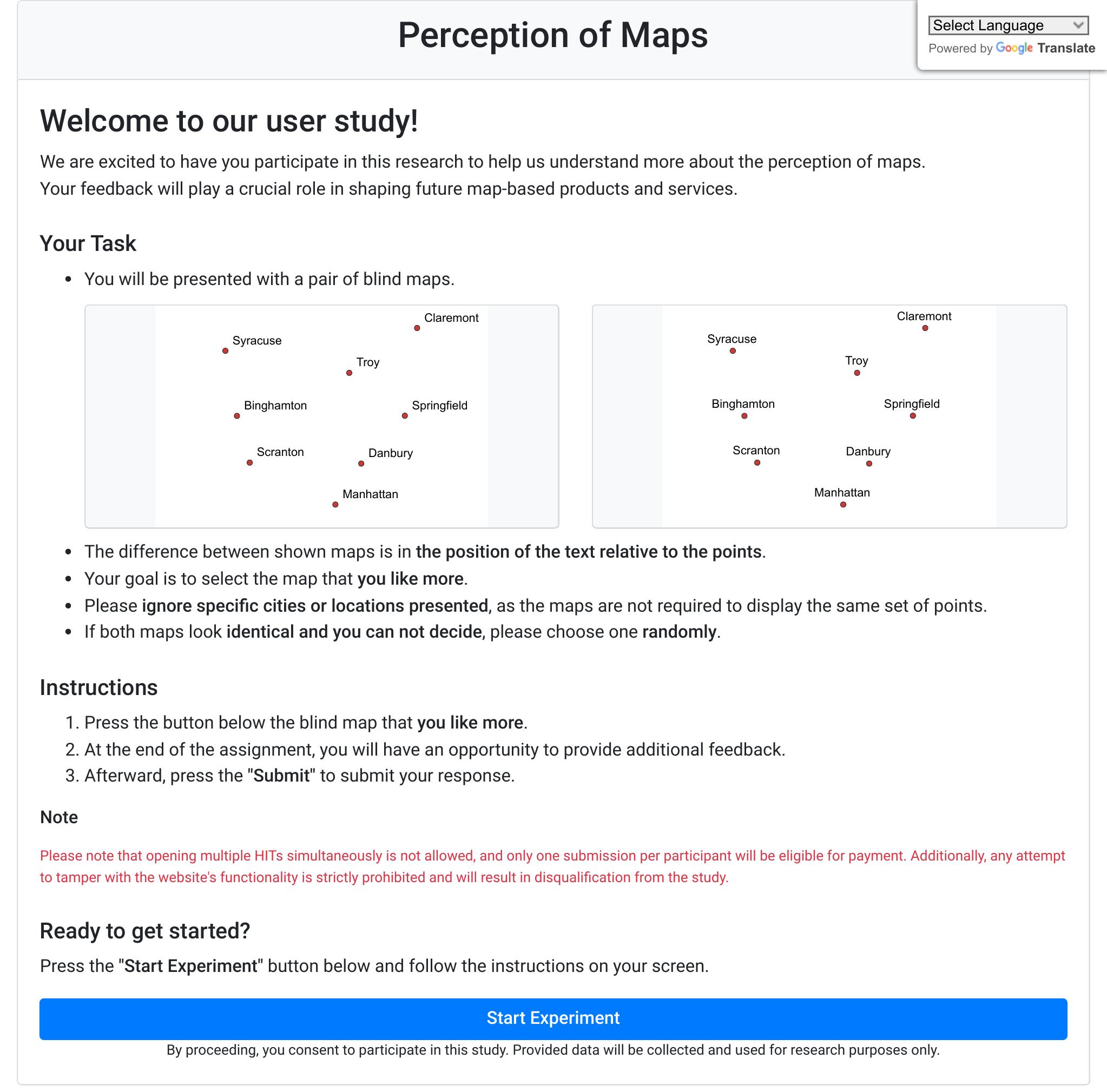}
    }
    \hspace{0em}
    \subfigure[Survey] {
        
        \includegraphics[width=\columnwidth]{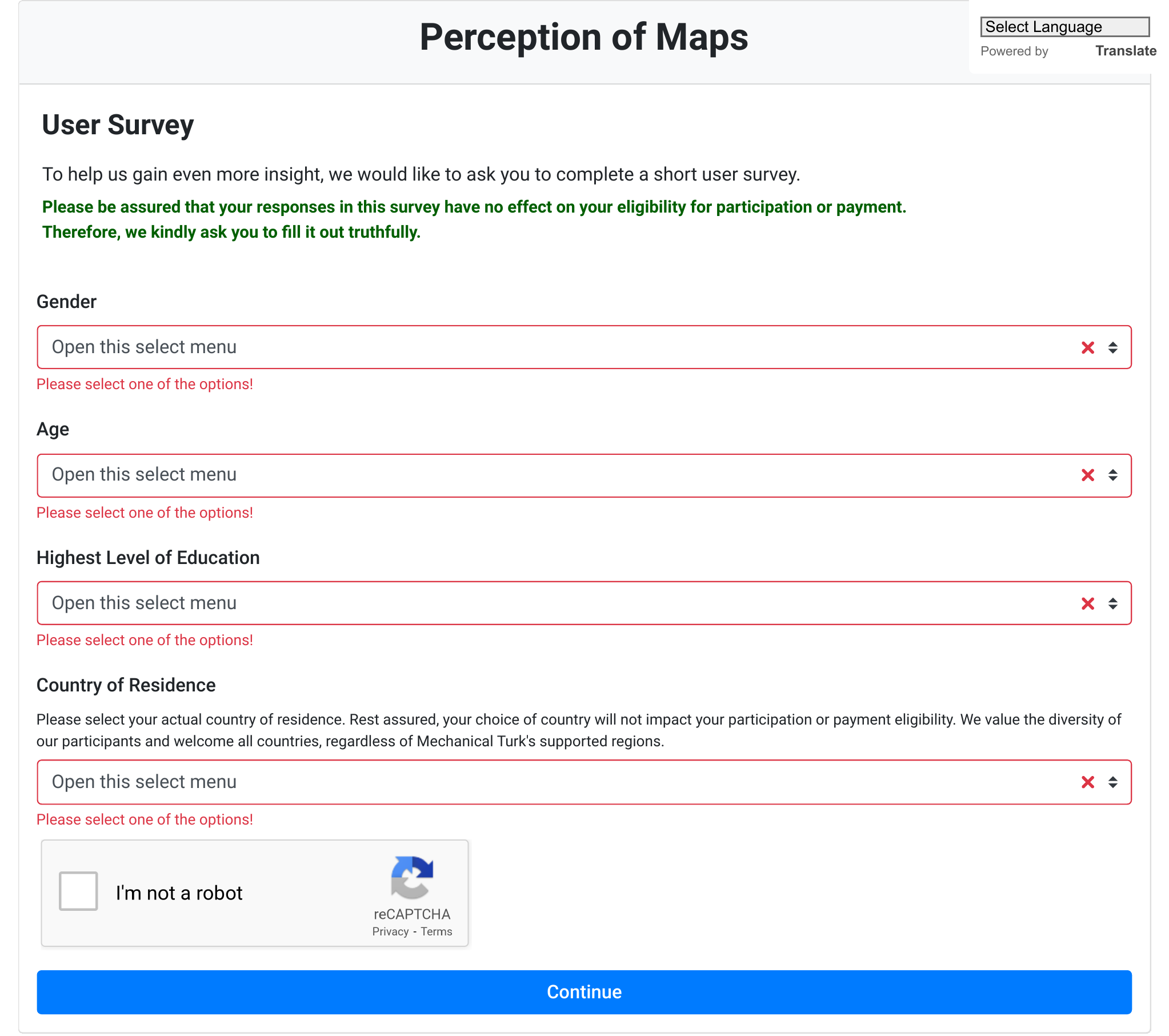}
    }
    \caption{
        \textbf{Evaluation 2 -- Comparison Study of PPOs:} Web interface for introduction and survey.
    }
    \label{fig:web-ppo_eval1}
\end{figure}

\begin{figure}[H]
    \subfigure[Experiment] {
        
        \includegraphics[width=\columnwidth]{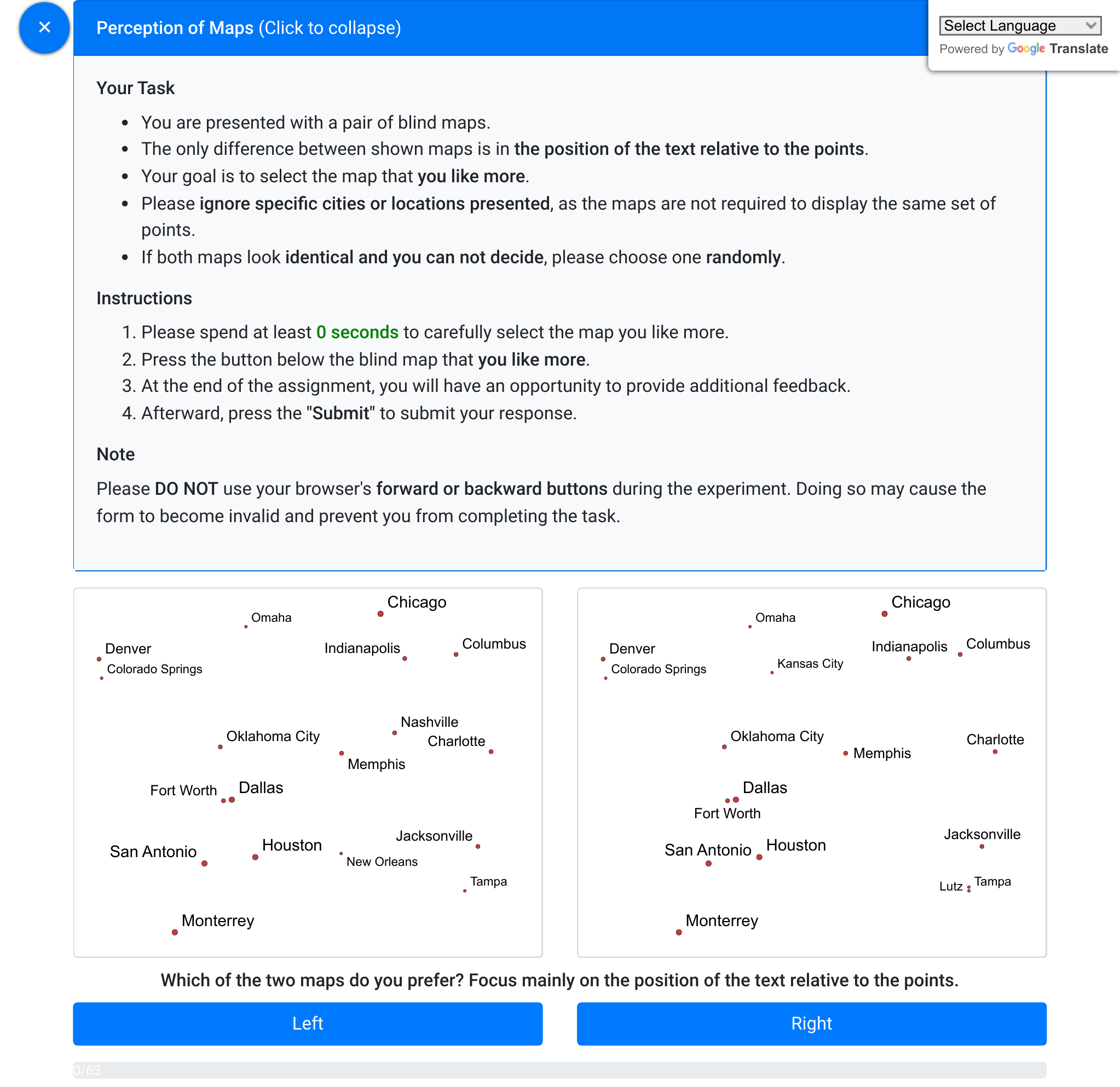}
    }
    \hspace{0em}
    \subfigure[Feedback] {
        
        \includegraphics[width=\columnwidth]{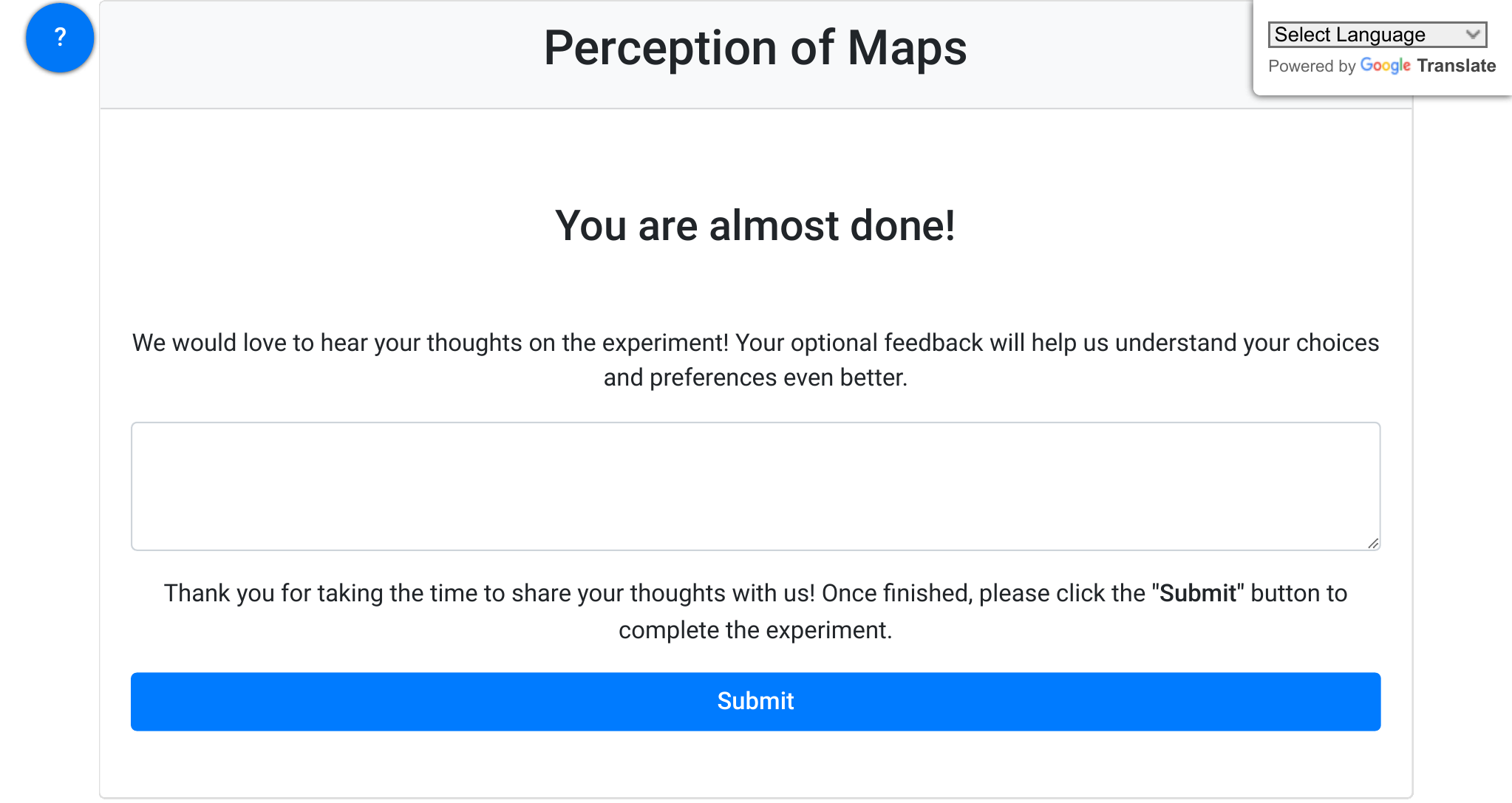}
    }
    \hspace{0em}
    \subfigure[Feedback translated to Japanese] {
        
        \includegraphics[width=\columnwidth]{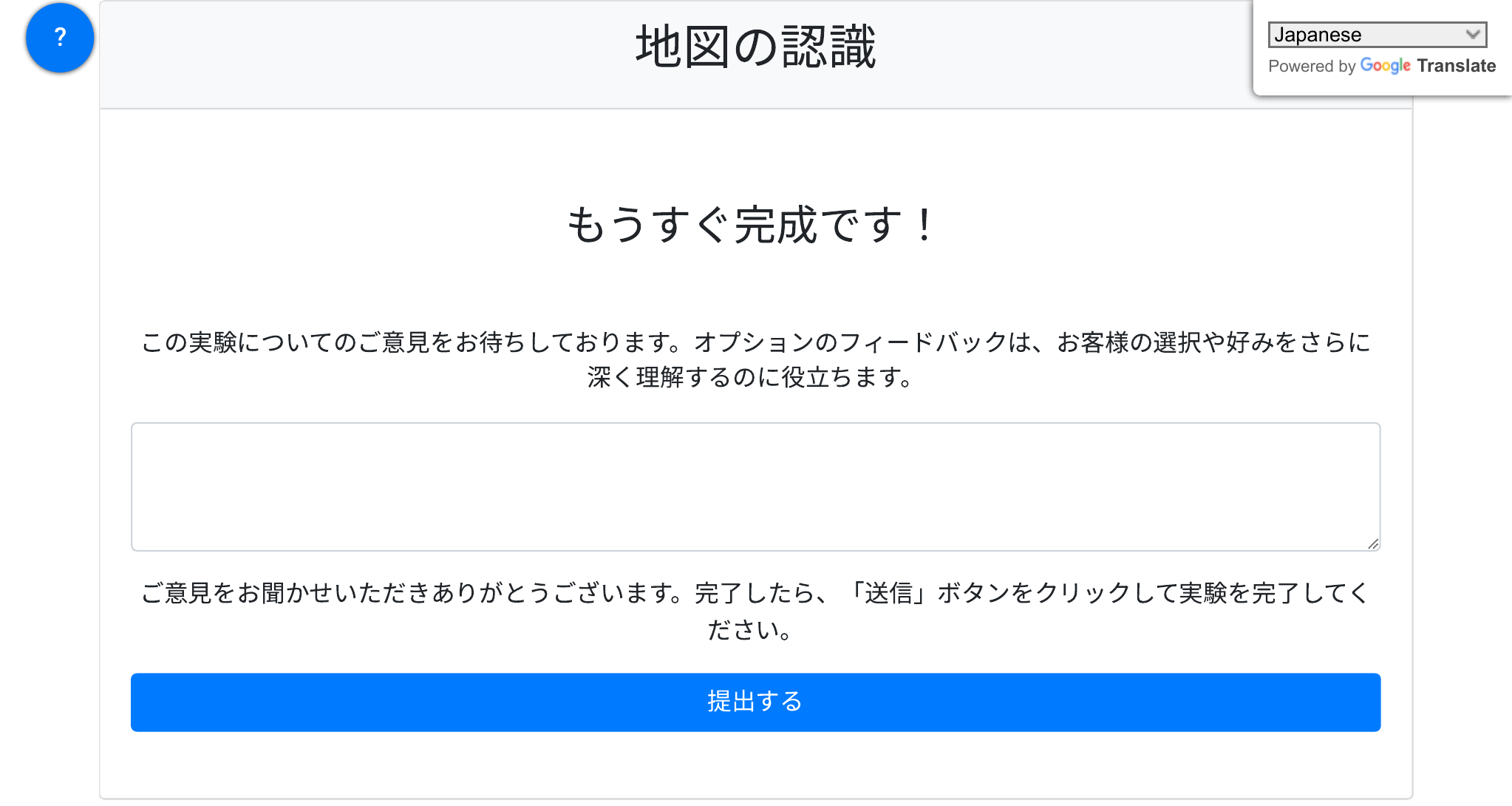}
    }
    \caption{
        \textbf{Evaluation 2 -- Comparison Study of PPOs:} Web interface for experiment and survey, with example of translation of instructions using Google Translate.
    }
  \label{fig:web-ppo_eval2}
\end{figure}

This section provides examples of user interface used to compare PPOs. The following paragraphs present textual information available to participants as seen in \autoref{fig:web-ppo_eval1} and \autoref{fig:web-ppo_eval2}.

\textbf{Introduction} Welcome to our user study! We are excited to have you participate in this research to help us understand more about the perception of maps. Your feedback will play a crucial role in shaping future map-based products and services. 

\textbf{Task} You will be presented with a pair of blind maps. The difference between shown maps is in the position of the text relative to the points. Your goal is to select the map that you like more. Please ignore specific cities or locations presented, as the maps are not required to display the same set of points. If both maps look identical and you can not decide, please choose one randomly.

\textbf{Instructions} Press the button below the blind map that you like more. At the end of the assignment, you will have an opportunity to provide additional feedback. Afterward, press the "Submit" to submit your response.

\textbf{Note} Please note that opening multiple HITs simultaneously is not allowed, and only one submission per participant will be eligible for payment. Additionally, any attempt to tamper with the website's functionality is strictly prohibited and will result in disqualification from the study.

\textbf{Start Experiment} Ready to get started? Press the "Start Experiment" button below and follow the instructions on your screen. By proceeding, you consent to participate in this study. Provided data will be collected and used for research purposes only.

\textbf{Survey} To help us gain even more insight, we would like to ask you to complete a short user survey. Please be assured that your responses in this survey have no effect on your eligibility for participation or payment. Therefore, we kindly ask you to fill it out truthfully.

\textbf{Call For Action} Which of the two maps do you prefer? Focus mainly on the position of the text relative to the points.

\subsection{Data}
\begin{figure*}[h!]
 
    \includegraphics[width=\textwidth]{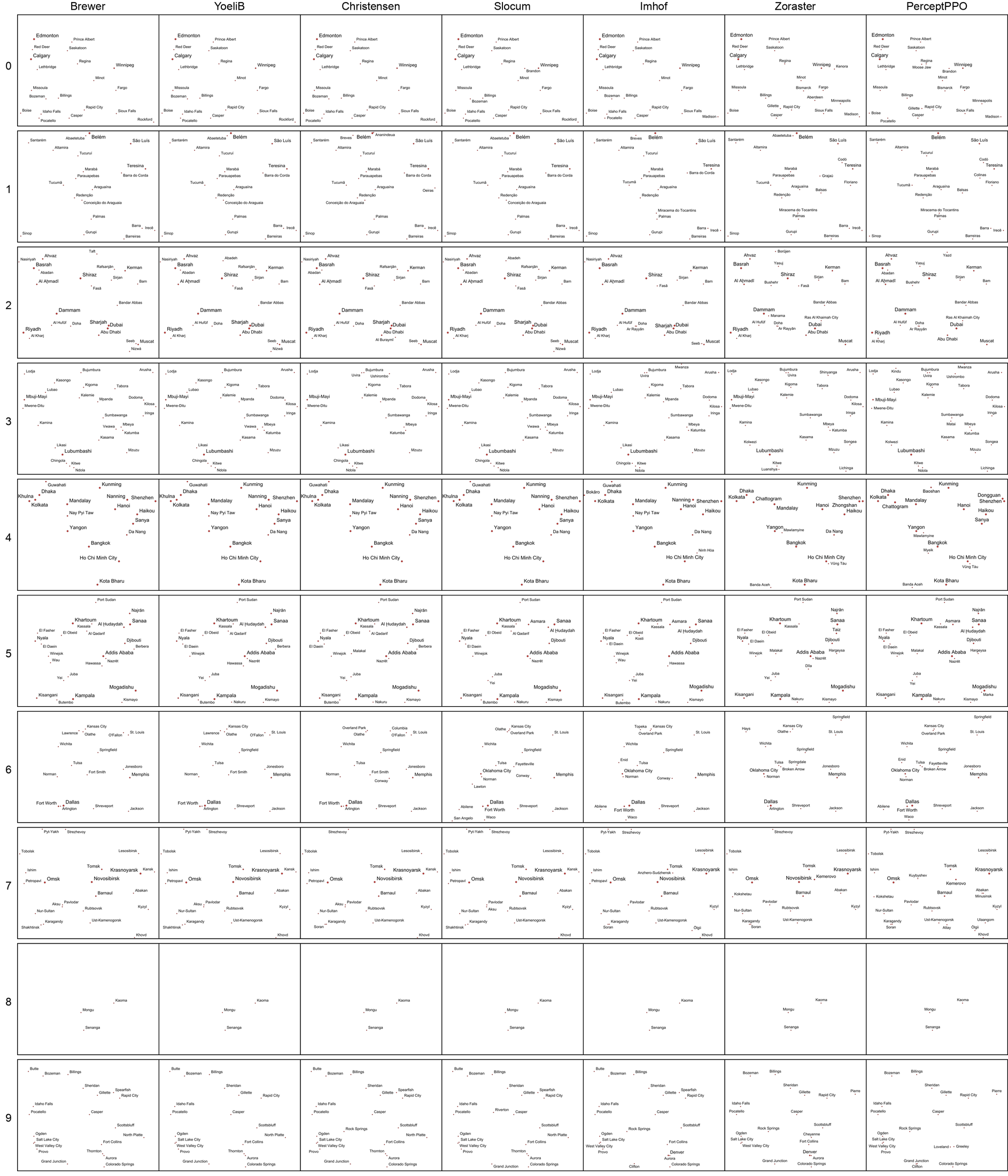}
    \caption{\textbf{Evaluation 2 -- Comparison Study of PPOs:} Visualizations of areas 0 through 9 used in the user study.}
    \label{fig:ppo_eval_renders1}
\end{figure*}

\begin{figure*}[h!]
 
    \includegraphics[width=\textwidth]{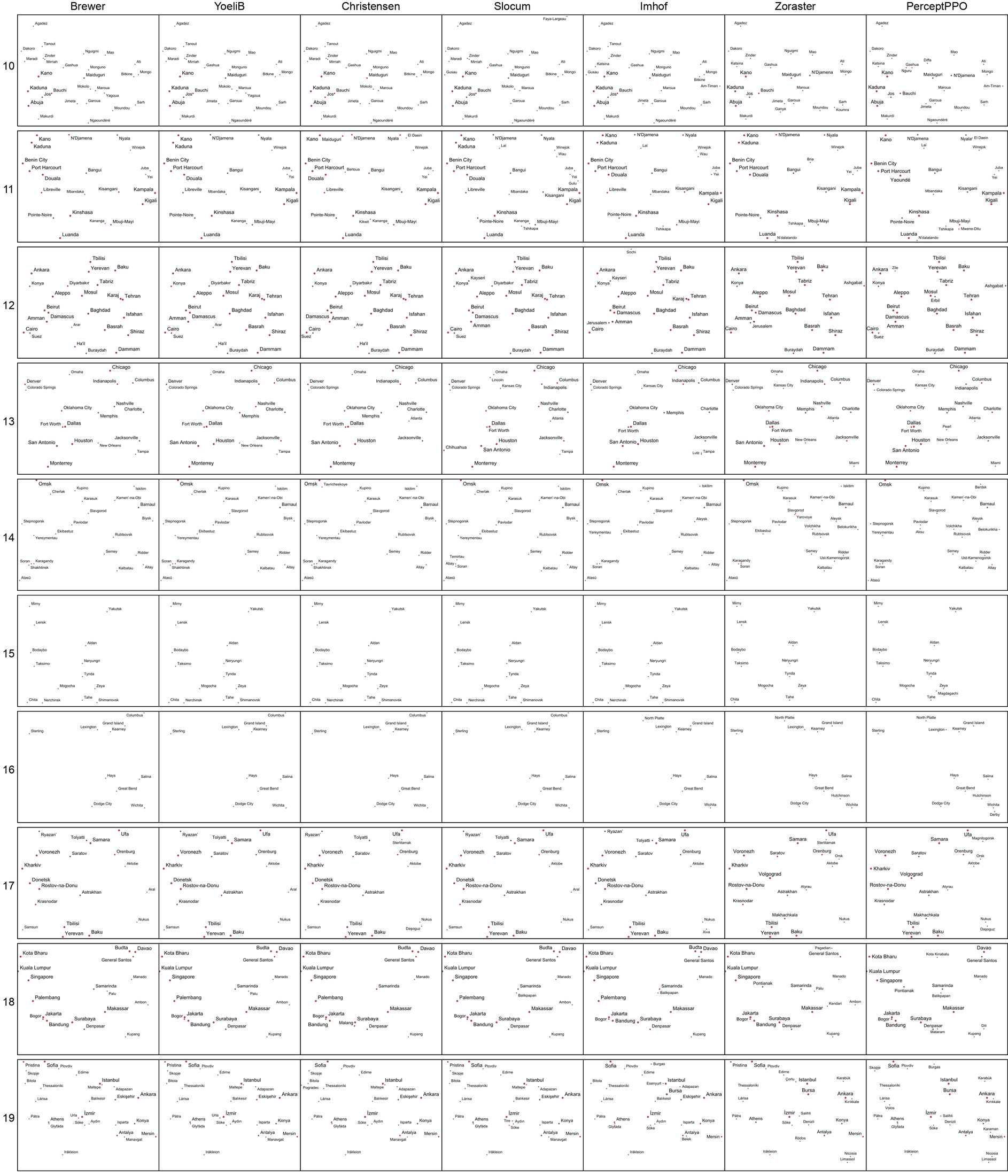}
    \caption{\textbf{Evaluation 2 -- Comparison Study of PPOs:} Visualizations of areas 10 through 19 used in the user study.}
    \label{fig:ppo_eval_renders2}
\end{figure*}

\begin{figure*}[h!]
 
    \includegraphics[width=\textwidth]{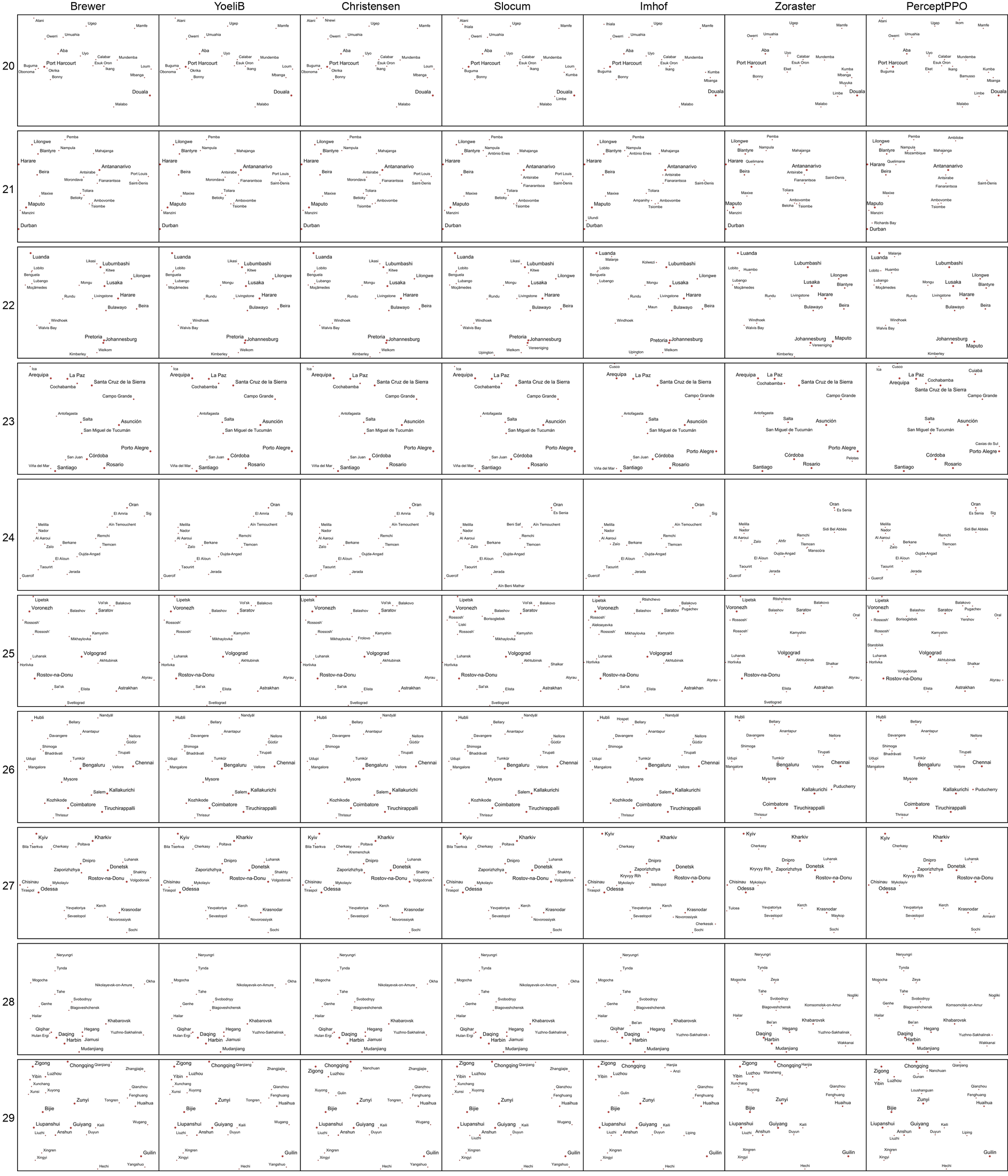}
    \caption{\textbf{Evaluation 2 -- Comparison Study of PPOs:} Visualizations of areas 20 through 29 used in the user study.}
    \label{fig:ppo_eval_renders3}
\end{figure*}

\autoref{fig:ppo_eval_renders1}, \autoref{fig:ppo_eval_renders2}, and \autoref{fig:ppo_eval_renders3} display the map renders utilized in the PPO comparison user study, showcasing all examined PPOs.

\subsection{Results}

\begin{figure*}[!ht]
    \subfigure[] {
        
        \includegraphics[width=0.33\textwidth]{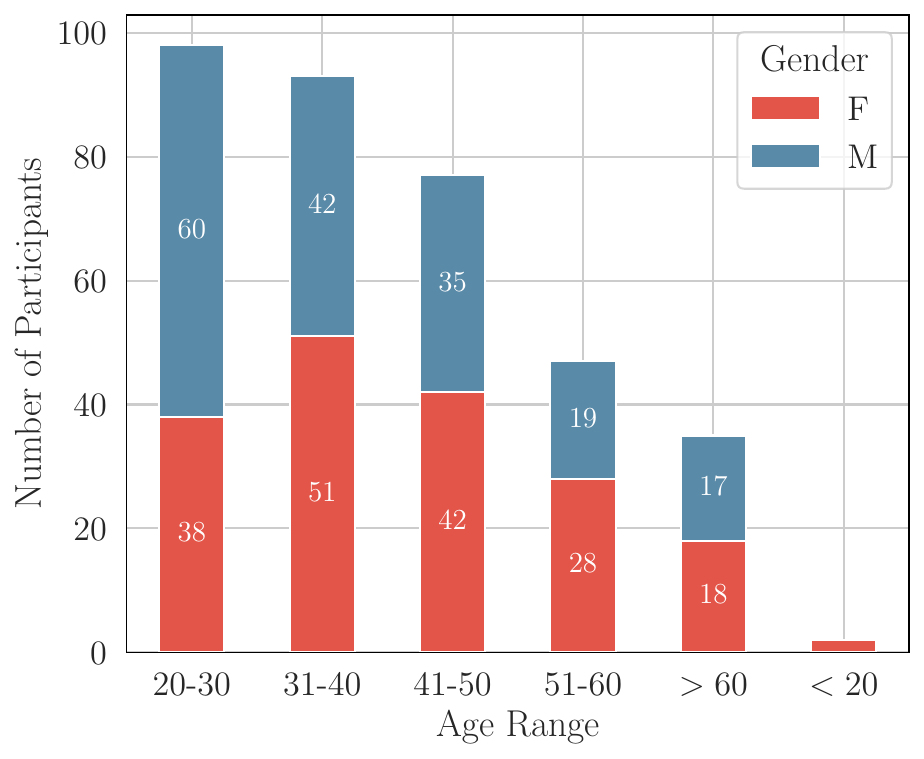}
    }
    \hspace{0em}
    \subfigure[] {
        
        \includegraphics[width=0.33\textwidth]{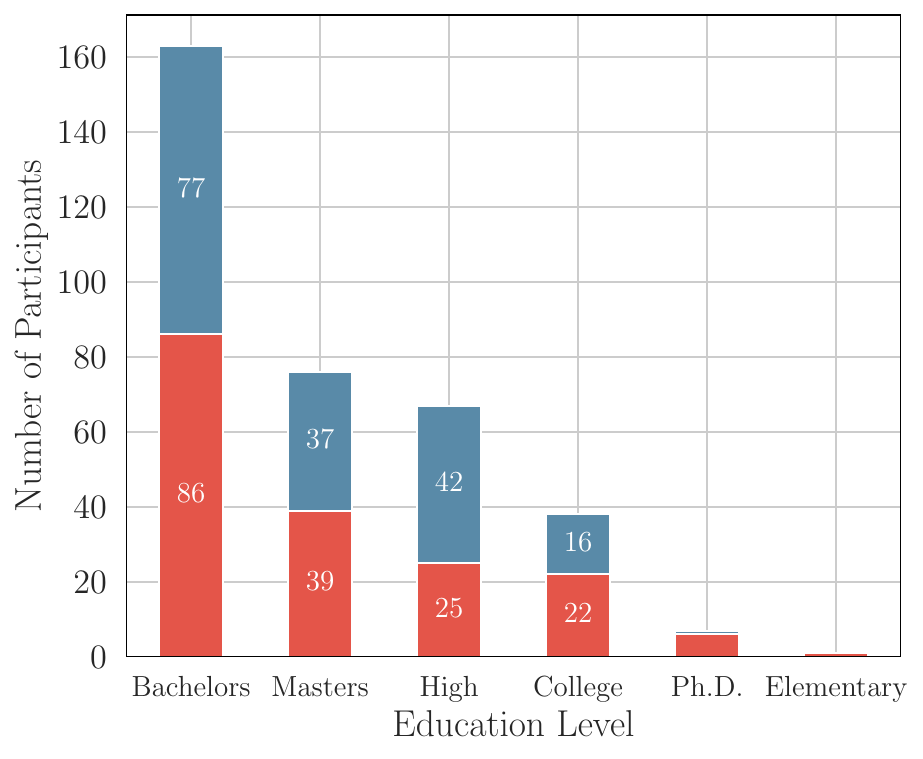}
    }
    \hspace{0em}
    \subfigure[] {
        
        \includegraphics[width=0.33\textwidth]{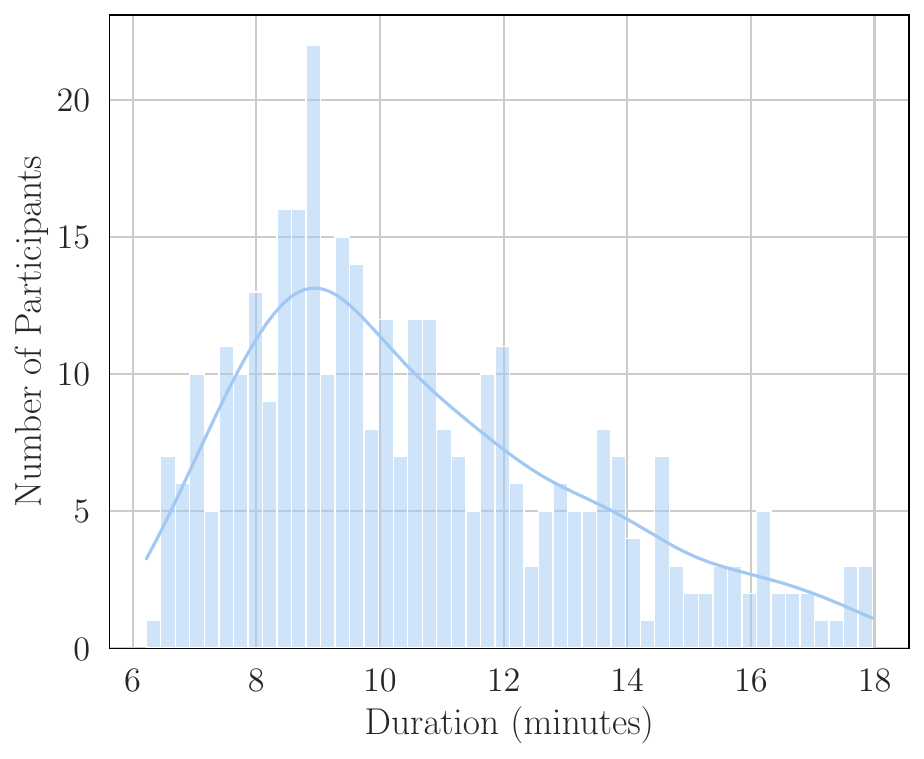}
    }
    \caption{
        \textbf{Evaluation 2 -- Comparison Study of PPOs:} User statistics of age, education, and duration of study for all participants.
    }
    \label{fig:ppo_eval_stats}
\end{figure*}

\begin{figure*}[!ht]
    \subfigure[] {
        
        \includegraphics[width=0.5\textwidth]{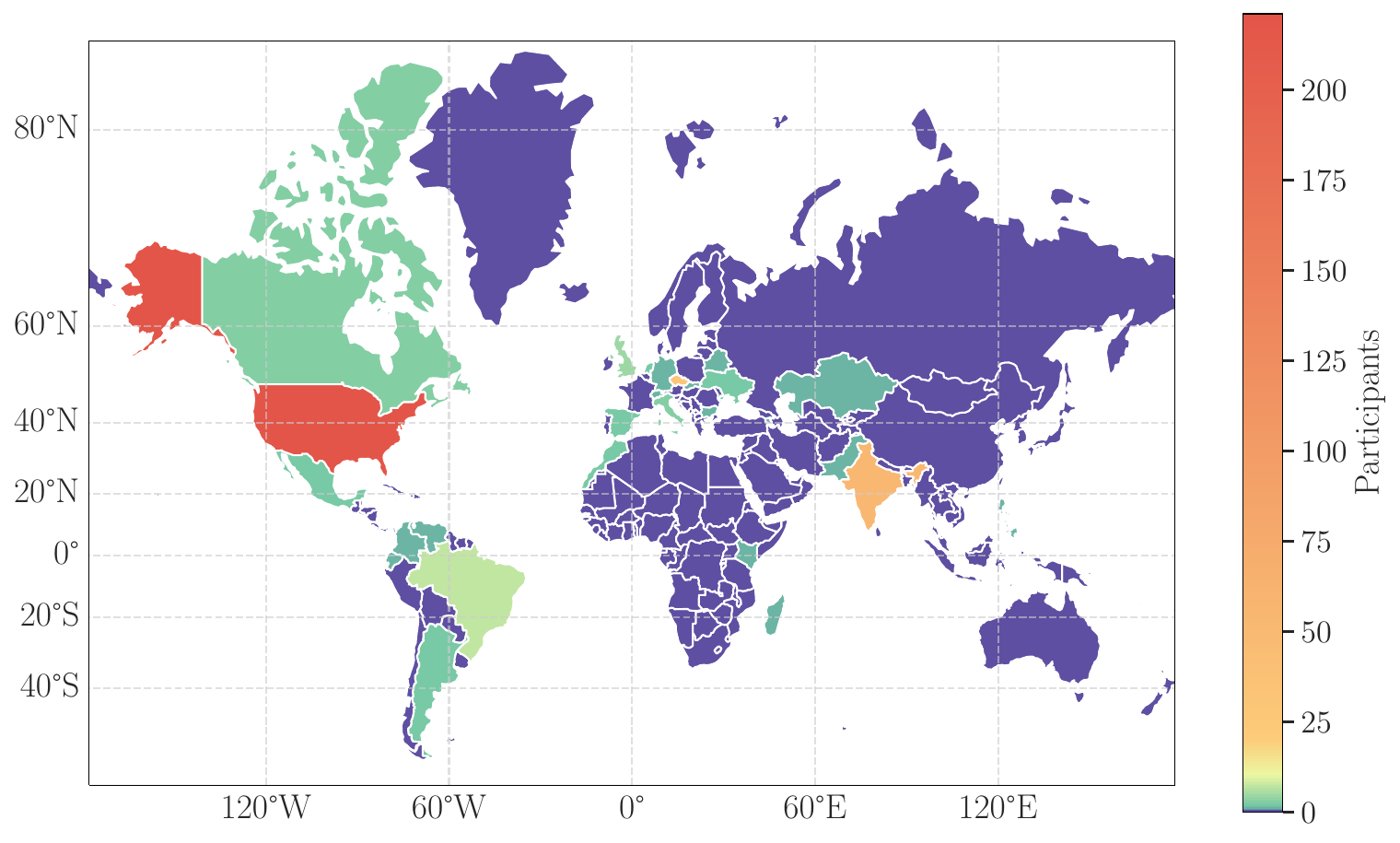}
    }
    \hspace{0em}
    \subfigure[] {
        
        \includegraphics[width=0.5\textwidth]{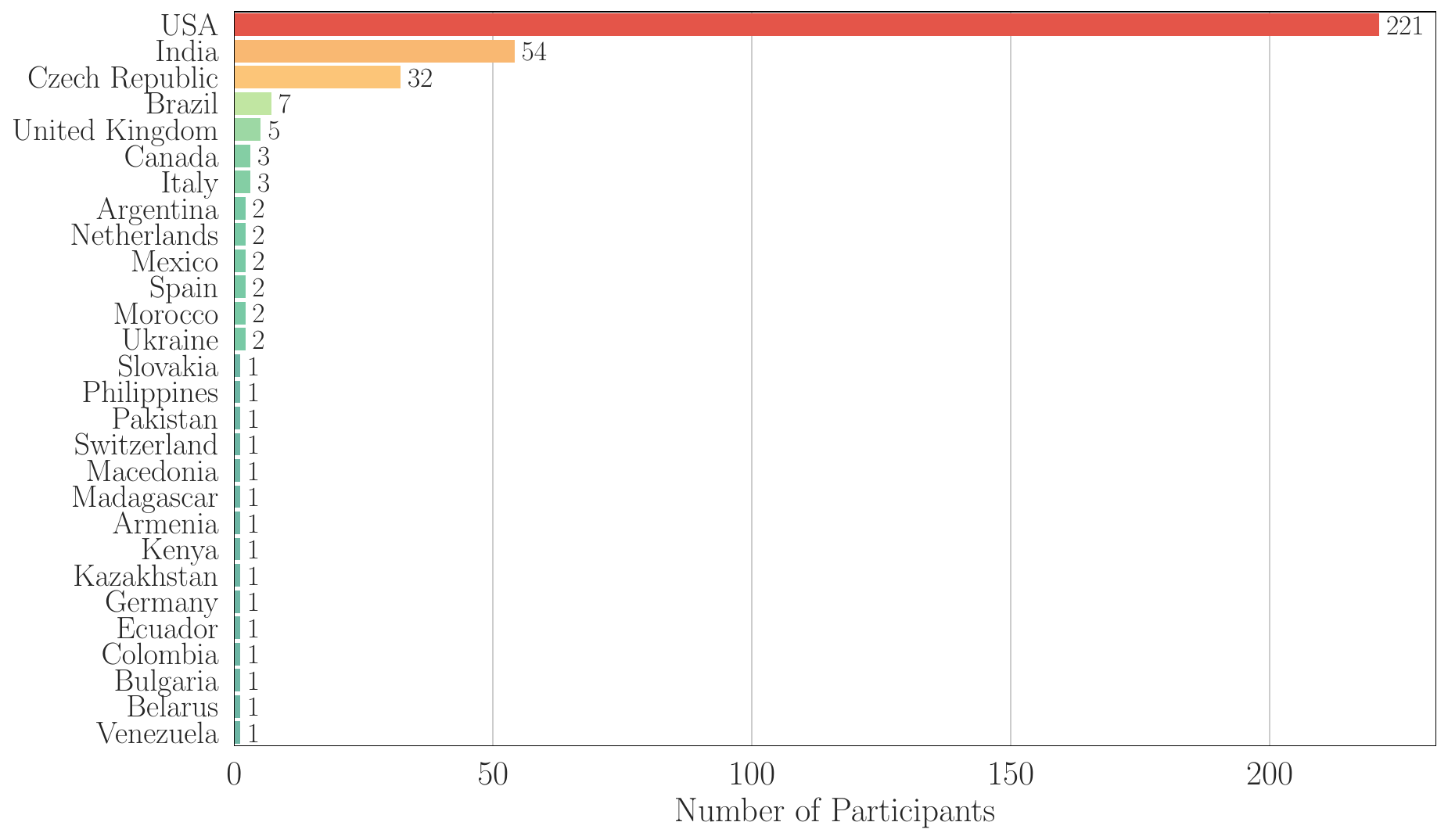}
    }
    \caption{
        \textbf{Evaluation 2 -- Comparison Study of PPOs:} Country distribution for all participants.
    }
    \label{fig:ppo_eval_countries}
\end{figure*}

\begin{figure*}[h!]
    \centering
    \includegraphics[width=0.95\textwidth]{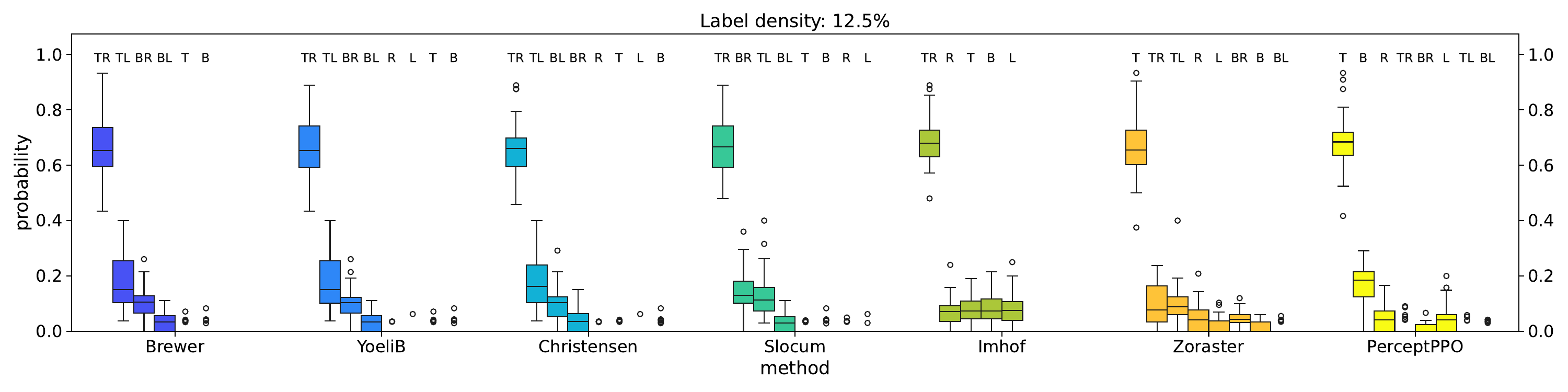}
    \caption{\textbf{Evaluation 2 -- Comparison Study of PPOs:} The probability that a label is placed at a specific position for each investigated method for the data used in Evaluation 2 ($652\times512$ size of renders and label density $LD_\mathrm{thr}=12.5\%$). For each PPO, the positions (see the labels above the boxplots) are in ascending order with respect to their priorities from left to right.}
    \label{fig:probabilities_density_all_methods_2}
\end{figure*}

\autoref{tab:ppo-eval_pvals} presents comprehensive $p$-values from pairwise comparisons of examined PPOs. Additionally, \autoref{tab:ppo-eval_pvals_c1}, \autoref{tab:ppo-eval_pvals_c2}, and \autoref{tab:ppo-eval_pvals_c3} detail the $p$-values for the pairwise comparisons within the identified clusters. \autoref{tab:ppo-eval-results} shows coefficient of consistency $\zeta$ and coefficient of agreement $u$ along with its corresponding values of \repA{$u_\mathrm{min}$}{min $u$}, $\chi^2$ and $p$-value for examined PPOs for each map area reported over all responses and clusters obtained by hierarchical clustering. \autoref{fig:ppo_eval_stats} and \autoref{fig:ppo_eval_countries} show the detailed statistics of engaged participants.     \autoref{fig:probabilities_density_all_methods_2} show probabilities that a label is placed at a specific position for each investigated method for the data used in the PPO comparison user study ($652\times512$ size of renders and label density $LD_\mathrm{thr}=12.5\%$).

\begin{table}[H]
\centering
\setlength{\tabcolsep}{3pt}
\begin{adjustbox}{width=\columnwidth}
\begin{tabular}{c|cccccccc}
\toprule
{} &                             Brewer &                             YoeliB &                        Christensen &                             Slocum &                              Imhof &                           Zoraster &                         PerceptPPO \\
\midrule
Brewer      &                                 -- &      \textcolor{crimson}{2.37e-01} &  \textcolor{darkerGreen}{1.80e-02} &      \textcolor{crimson}{4.45e-01} &      \textcolor{crimson}{6.72e-02} &  \textcolor{darkerGreen}{4.68e-11} &  \textcolor{darkerGreen}{8.35e-14} \\
YoeliB      &      \textcolor{crimson}{2.37e-01} &                                 -- &      \textcolor{crimson}{8.60e-02} &      \textcolor{crimson}{2.34e-01} &  \textcolor{darkerGreen}{2.42e-02} &  \textcolor{darkerGreen}{1.35e-11} &  \textcolor{darkerGreen}{2.09e-15} \\
Christensen &  \textcolor{darkerGreen}{1.80e-02} &      \textcolor{crimson}{8.60e-02} &                                 -- &  \textcolor{darkerGreen}{1.85e-02} &  \textcolor{darkerGreen}{3.38e-04} &  \textcolor{darkerGreen}{7.71e-14} &  \textcolor{darkerGreen}{7.07e-18} \\
Slocum      &      \textcolor{crimson}{4.45e-01} &      \textcolor{crimson}{2.34e-01} &  \textcolor{darkerGreen}{1.85e-02} &                                 -- &      \textcolor{crimson}{9.62e-02} &  \textcolor{darkerGreen}{1.79e-10} &  \textcolor{darkerGreen}{3.41e-13} \\
Imhof       &      \textcolor{crimson}{6.72e-02} &  \textcolor{darkerGreen}{2.42e-02} &  \textcolor{darkerGreen}{3.38e-04} &      \textcolor{crimson}{9.62e-02} &                                 -- &  \textcolor{darkerGreen}{5.29e-10} &  \textcolor{darkerGreen}{6.71e-13} \\
Zoraster    &  \textcolor{darkerGreen}{4.68e-11} &  \textcolor{darkerGreen}{1.35e-11} &  \textcolor{darkerGreen}{7.71e-14} &  \textcolor{darkerGreen}{1.79e-10} &  \textcolor{darkerGreen}{5.29e-10} &                                 -- &      \textcolor{crimson}{3.35e-02} \\
PerceptPPO  &  \textcolor{darkerGreen}{8.35e-14} &  \textcolor{darkerGreen}{2.09e-15} &  \textcolor{darkerGreen}{7.07e-18} &  \textcolor{darkerGreen}{3.41e-13} &  \textcolor{darkerGreen}{6.71e-13} &      \textcolor{crimson}{3.35e-02} &                                 -- \\
\bottomrule
\end{tabular}
\end{adjustbox}
\caption{\textbf{Evaluation 2 -- Comparison Study of PPOs:} Detailed overall $p$ values disregard identified clusters for the examined PPOs. Green cells indicate statistically significant differences, or red otherwise. We use a Two-Tailed test with $\alpha=0.05$; therefore, $p$ has to be less than $0.025$ to be significant.}
\label{tab:ppo-eval_pvals}
\end{table}

\begin{table}[H]
\centering
\setlength{\tabcolsep}{3pt}
\begin{adjustbox}{width=\columnwidth}
\begin{tabular}{c|cccccccc}
\toprule
{} &                             Brewer &                             YoeliB &                         Christensen &                             Slocum &                              Imhof &                            Zoraster &                         PerceptPPO \\
\midrule
Brewer      &                                 -- &  \textcolor{darkerGreen}{1.08e-03} &   \textcolor{darkerGreen}{2.62e-06} &  \textcolor{darkerGreen}{8.13e-03} &  \textcolor{darkerGreen}{1.43e-03} &   \textcolor{darkerGreen}{1.97e-57} &  \textcolor{darkerGreen}{1.73e-49} \\
YoeliB      &  \textcolor{darkerGreen}{1.08e-03} &                                 -- &       \textcolor{crimson}{8.01e-02} &      \textcolor{crimson}{2.08e-01} &  \textcolor{darkerGreen}{3.75e-07} &   \textcolor{darkerGreen}{7.23e-98} &  \textcolor{darkerGreen}{1.09e-80} \\
Christensen &  \textcolor{darkerGreen}{2.62e-06} &      \textcolor{crimson}{8.01e-02} &                                  -- &  \textcolor{darkerGreen}{1.51e-02} &  \textcolor{darkerGreen}{6.79e-12} &  \textcolor{darkerGreen}{2.42e-105} &  \textcolor{darkerGreen}{9.26e-88} \\
Slocum      &  \textcolor{darkerGreen}{8.13e-03} &      \textcolor{crimson}{2.08e-01} &   \textcolor{darkerGreen}{1.51e-02} &                                 -- &  \textcolor{darkerGreen}{4.39e-06} &   \textcolor{darkerGreen}{1.26e-72} &  \textcolor{darkerGreen}{1.43e-66} \\
Imhof       &  \textcolor{darkerGreen}{1.43e-03} &  \textcolor{darkerGreen}{3.75e-07} &   \textcolor{darkerGreen}{6.79e-12} &  \textcolor{darkerGreen}{4.39e-06} &                                 -- &   \textcolor{darkerGreen}{1.30e-50} &  \textcolor{darkerGreen}{1.07e-41} \\
Zoraster    &  \textcolor{darkerGreen}{1.97e-57} &  \textcolor{darkerGreen}{7.23e-98} &  \textcolor{darkerGreen}{2.42e-105} &  \textcolor{darkerGreen}{1.26e-72} &  \textcolor{darkerGreen}{1.30e-50} &                                  -- &      \textcolor{crimson}{4.24e-01} \\
PerceptPPO  &  \textcolor{darkerGreen}{1.73e-49} &  \textcolor{darkerGreen}{1.09e-80} &   \textcolor{darkerGreen}{9.26e-88} &  \textcolor{darkerGreen}{1.43e-66} &  \textcolor{darkerGreen}{1.07e-41} &       \textcolor{crimson}{4.24e-01} &                                 -- \\
\bottomrule
\end{tabular}
\end{adjustbox}
\caption{\textbf{Evaluation 2 -- Comparison Study of PPOs:} Detailed $p$ values of Cluster 1 for the examined PPOs. Green cells indicate statistically significant differences, or red otherwise. We use a Two-Tailed test with $\alpha=0.05$; therefore, $p$ has to be less than $0.025$ to be significant.}
\label{tab:ppo-eval_pvals_c1}
\end{table}

\begin{table}[H]
\centering
\setlength{\tabcolsep}{3pt}
\begin{adjustbox}{width=\columnwidth}
\begin{tabular}{c|cccccccc}
\toprule
{} &                             Brewer &                             YoeliB &                        Christensen &                             Slocum &                              Imhof &                           Zoraster &                         PerceptPPO \\
\midrule
Brewer      &                                 -- &  \textcolor{darkerGreen}{5.45e-03} &  \textcolor{darkerGreen}{9.59e-03} &      \textcolor{crimson}{1.71e-01} &      \textcolor{crimson}{1.12e-01} &  \textcolor{darkerGreen}{8.18e-38} &  \textcolor{darkerGreen}{3.94e-29} \\
YoeliB      &  \textcolor{darkerGreen}{5.45e-03} &                                 -- &      \textcolor{crimson}{4.76e-01} &  \textcolor{darkerGreen}{5.57e-04} &      \textcolor{crimson}{2.77e-01} &  \textcolor{darkerGreen}{9.61e-22} &  \textcolor{darkerGreen}{3.83e-21} \\
Christensen &  \textcolor{darkerGreen}{9.59e-03} &      \textcolor{crimson}{4.76e-01} &                                 -- &  \textcolor{darkerGreen}{1.47e-03} &      \textcolor{crimson}{3.08e-01} &  \textcolor{darkerGreen}{2.08e-26} &  \textcolor{darkerGreen}{3.26e-24} \\
Slocum      &      \textcolor{crimson}{1.71e-01} &  \textcolor{darkerGreen}{5.57e-04} &  \textcolor{darkerGreen}{1.47e-03} &                                 -- &  \textcolor{darkerGreen}{2.41e-02} &  \textcolor{darkerGreen}{7.05e-39} &  \textcolor{darkerGreen}{2.14e-33} \\
Imhof       &      \textcolor{crimson}{1.12e-01} &      \textcolor{crimson}{2.77e-01} &      \textcolor{crimson}{3.08e-01} &  \textcolor{darkerGreen}{2.41e-02} &                                 -- &  \textcolor{darkerGreen}{1.59e-27} &  \textcolor{darkerGreen}{2.30e-33} \\
Zoraster    &  \textcolor{darkerGreen}{8.18e-38} &  \textcolor{darkerGreen}{9.61e-22} &  \textcolor{darkerGreen}{2.08e-26} &  \textcolor{darkerGreen}{7.05e-39} &  \textcolor{darkerGreen}{1.59e-27} &                                 -- &      \textcolor{crimson}{1.56e-01} \\
PerceptPPO  &  \textcolor{darkerGreen}{3.94e-29} &  \textcolor{darkerGreen}{3.83e-21} &  \textcolor{darkerGreen}{3.26e-24} &  \textcolor{darkerGreen}{2.14e-33} &  \textcolor{darkerGreen}{2.30e-33} &      \textcolor{crimson}{1.56e-01} &                                 -- \\
\bottomrule
\end{tabular}
\end{adjustbox}
\caption{\textbf{Evaluation 2 -- Comparison Study of PPOs:} Detailed $p$ values of Cluster 2 for the examined PPOs. Green cells indicate statistically significant differences, or red otherwise. We use a Two-Tailed test with $\alpha=0.05$; therefore, $p$ has to be less than $0.025$ to be significant.}
\label{tab:ppo-eval_pvals_c2}
\end{table}

\begin{table}[H]
\centering
\setlength{\tabcolsep}{3pt}
\begin{adjustbox}{width=\columnwidth}
\begin{tabular}{c|cccccccc}
\toprule
{} &                             Brewer &                             YoeliB &                        Christensen &                             Slocum &                              Imhof &                           Zoraster &                         PerceptPPO \\
\midrule
Brewer      &                                 -- &  \textcolor{darkerGreen}{4.52e-04} &      \textcolor{crimson}{2.58e-02} &      \textcolor{crimson}{5.20e-02} &      \textcolor{crimson}{4.72e-01} &  \textcolor{darkerGreen}{2.77e-03} &  \textcolor{darkerGreen}{3.58e-08} \\
YoeliB      &  \textcolor{darkerGreen}{4.52e-04} &                                 -- &      \textcolor{crimson}{1.65e-01} &      \textcolor{crimson}{1.63e-01} &  \textcolor{darkerGreen}{3.80e-04} &      \textcolor{crimson}{4.55e-01} &  \textcolor{darkerGreen}{2.92e-03} \\
Christensen &      \textcolor{crimson}{2.58e-02} &      \textcolor{crimson}{1.65e-01} &                                 -- &      \textcolor{crimson}{4.56e-01} &  \textcolor{darkerGreen}{1.16e-02} &      \textcolor{crimson}{1.36e-01} &  \textcolor{darkerGreen}{5.91e-05} \\
Slocum      &      \textcolor{crimson}{5.20e-02} &      \textcolor{crimson}{1.63e-01} &      \textcolor{crimson}{4.56e-01} &                                 -- &      \textcolor{crimson}{3.25e-02} &      \textcolor{crimson}{1.10e-01} &  \textcolor{darkerGreen}{5.80e-05} \\
Imhof       &      \textcolor{crimson}{4.72e-01} &  \textcolor{darkerGreen}{3.80e-04} &  \textcolor{darkerGreen}{1.16e-02} &      \textcolor{crimson}{3.25e-02} &                                 -- &  \textcolor{darkerGreen}{2.41e-04} &  \textcolor{darkerGreen}{5.84e-13} \\
Zoraster    &  \textcolor{darkerGreen}{2.77e-03} &      \textcolor{crimson}{4.55e-01} &      \textcolor{crimson}{1.36e-01} &      \textcolor{crimson}{1.10e-01} &  \textcolor{darkerGreen}{2.41e-04} &                                 -- &  \textcolor{darkerGreen}{4.53e-03} \\
PerceptPPO  &  \textcolor{darkerGreen}{3.58e-08} &  \textcolor{darkerGreen}{2.92e-03} &  \textcolor{darkerGreen}{5.91e-05} &  \textcolor{darkerGreen}{5.80e-05} &  \textcolor{darkerGreen}{5.84e-13} &  \textcolor{darkerGreen}{4.53e-03} &                                 -- \\
\bottomrule
\end{tabular}
\end{adjustbox}
\caption{\textbf{Evaluation 2 -- Comparison Study of PPOs:} $p$ values of Cluster 3 for the examined PPOs. Green cells indicate statistically significant differences, or red otherwise. We use a Two-Tailed test with $\alpha=0.05$; therefore, $p$ has to be less than $0.025$ to be significant.}
\label{tab:ppo-eval_pvals_c3}
\end{table}


\begin{table*}[t!]
\centering
\setlength{\tabcolsep}{3pt}
\begin{adjustbox}{width=\textwidth}
\begin{tabular}{c|ccccc|ccccc|ccccc|ccccc}
\toprule
Area & \multicolumn{5}{c}{Overall} & \multicolumn{5}{c}{Cluster 1} & \multicolumn{5}{c}{Cluster 2} & \multicolumn{5}{c}{Cluster 3} \\
         & mean $\zeta$ &   $u$ & \repA{$u_\mathrm{min}$}{min $u$} & $\chi^2$ &                        $p$-value & mean $\zeta$ &   $u$ & \repA{$u_\mathrm{min}$}{min $u$} & $\chi^2$ &                        $p$-value & mean $\zeta$ &   $u$ & \repA{$u_\mathrm{min}$}{min $u$} & $\chi^2$ &                        $p$-value & mean $\zeta$ &    $u$ & \repA{$u_\mathrm{min}$}{min $u$} & $\chi^2$ &                        $p$-value \\
\midrule
       0 &        0.720 & 0.142 &  -0.043 &  127.937 & \textcolor{darkerGreen}{2.2e-13} &        0.900 & 0.310 &  -0.091 &  150.960 & \textcolor{darkerGreen}{1.0e-15} &        0.750 & 0.286 &  -0.333 &  212.000 &   \textcolor{darkerGreen}{0.012} &        0.438 &  0.013 &  -0.143 &   46.222 &       \textcolor{crimson}{0.363} \\
       1 &        0.619 & 0.102 &  -0.059 &   87.969 & \textcolor{darkerGreen}{8.7e-07} &        0.821 & 0.293 &  -0.143 &  115.040 & \textcolor{darkerGreen}{1.3e-07} &        0.767 & 0.381 &  -0.333 &  228.000 &   \textcolor{darkerGreen}{0.001} &        0.388 & -0.018 &  -0.143 &   38.222 &       \textcolor{crimson}{0.700} \\
       2 &        0.679 & 0.152 &  -0.048 &  126.681 & \textcolor{darkerGreen}{4.9e-13} &        0.754 & 0.316 &  -0.091 &  153.360 & \textcolor{darkerGreen}{4.0e-16} &        0.812 & 0.250 &  -0.333 &  124.000 &   \textcolor{darkerGreen}{0.003} &        0.390 & -0.071 &  -0.200 &   47.556 &       \textcolor{crimson}{0.915} \\
       3 &        0.652 & 0.102 &  -0.040 &  106.019 & \textcolor{darkerGreen}{6.3e-10} &        0.739 & 0.210 &  -0.111 &  101.143 & \textcolor{darkerGreen}{5.8e-07} &        0.879 & 0.367 &  -0.143 &  132.640 & \textcolor{darkerGreen}{4.4e-10} &        0.389 &  0.008 &  -0.111 &   42.857 &       \textcolor{crimson}{0.398} \\
       4 &        0.625 & 0.102 &  -0.048 &   98.340 & \textcolor{darkerGreen}{1.4e-08} &        0.819 & 0.311 &  -0.143 &  124.222 & \textcolor{darkerGreen}{1.1e-09} &        0.733 & 0.095 &  -0.333 &  180.000 &       \textcolor{crimson}{0.250} &        0.455 &  0.070 &  -0.091 &   61.580 &   \textcolor{darkerGreen}{0.009} \\
       5 &        0.474 & 0.044 &  -0.040 &   63.932 & \textcolor{darkerGreen}{6.1e-04} &        0.731 & 0.186 &  -0.143 &   91.556 & \textcolor{darkerGreen}{2.9e-05} &        0.867 & 0.286 &  -0.333 &  212.000 &   \textcolor{darkerGreen}{0.012} &        0.243 &  0.010 &  -0.077 &   39.389 &       \textcolor{crimson}{0.296} \\
       6 &        0.688 & 0.099 &  -0.048 &   93.839 & \textcolor{darkerGreen}{7.6e-08} &        0.894 & 0.480 &  -0.143 &  168.222 & \textcolor{darkerGreen}{1.5e-16} &        0.925 & 0.488 &  -0.333 &  164.000 & \textcolor{darkerGreen}{4.2e-07} &        0.400 &  0.012 &  -0.111 &   44.000 &       \textcolor{crimson}{0.352} \\
       7 &        0.700 & 0.148 &  -0.048 &  124.155 & \textcolor{darkerGreen}{1.3e-12} &        0.861 & 0.448 &  -0.111 &  169.714 & \textcolor{darkerGreen}{1.5e-17} &        0.975 & 0.571 &  -1.000 &      inf &       \textcolor{darkerGreen}{0} &        0.500 &  0.019 &  -0.111 &   44.875 &       \textcolor{crimson}{0.252} \\
       8 &        0.774 & 0.116 &  -0.037 &  122.170 & \textcolor{darkerGreen}{1.2e-12} &        0.896 & 0.319 &  -0.077 &  162.281 & \textcolor{darkerGreen}{6.3e-18} &        0.810 & 0.371 &  -0.200 &  130.222 & \textcolor{darkerGreen}{1.0e-06} &        0.578 & -0.028 &  -0.111 &   32.571 &       \textcolor{crimson}{0.828} \\
       9 &        0.647 & 0.127 &  -0.059 &   98.116 & \textcolor{darkerGreen}{3.6e-08} &        0.825 & 0.343 &  -0.200 &  123.500 & \textcolor{darkerGreen}{1.2e-07} &        0.850 & 0.143 &  -0.333 &  188.000 &       \textcolor{crimson}{0.139} &        0.438 &  0.074 &  -0.143 &   62.222 &   \textcolor{darkerGreen}{0.033} \\
      10 &        0.594 & 0.094 &  -0.059 &   83.719 & \textcolor{darkerGreen}{3.4e-06} &        0.780 & 0.257 &  -0.200 &  108.889 & \textcolor{darkerGreen}{2.3e-04} &        0.850 & 0.250 &  -0.333 &  124.000 &   \textcolor{darkerGreen}{0.003} &        0.378 &  0.044 &  -0.111 &   53.143 &       \textcolor{crimson}{0.099} \\
      11 &        0.664 & 0.114 &  -0.048 &  105.940 & \textcolor{darkerGreen}{9.4e-10} &        0.744 & 0.226 &  -0.111 &  105.714 & \textcolor{darkerGreen}{1.4e-07} &        0.825 & 0.352 &  -0.200 &  125.500 & \textcolor{darkerGreen}{6.5e-08} &        0.421 &  0.034 &  -0.143 &   54.240 &       \textcolor{crimson}{0.219} \\
      12 &        0.615 & 0.167 &  -0.043 &  144.698 & \textcolor{darkerGreen}{3.1e-16} &        0.722 & 0.317 &  -0.111 &  132.000 & \textcolor{darkerGreen}{1.9e-11} &        0.790 & 0.457 &  -0.200 &  146.222 & \textcolor{darkerGreen}{1.0e-08} &        0.411 &  0.111 &  -0.111 &   72.571 &   \textcolor{darkerGreen}{0.002} \\
      13 &        0.714 & 0.136 &  -0.048 &  119.940 & \textcolor{darkerGreen}{5.4e-12} &        0.905 & 0.431 &  -0.091 &  185.136 & \textcolor{darkerGreen}{3.1e-21} &        0.933 & 0.333 &  -0.333 &  220.000 &   \textcolor{darkerGreen}{0.004} &        0.369 & -0.010 &  -0.143 &   40.222 &       \textcolor{crimson}{0.616} \\
      14 &        0.640 & 0.165 &  -0.034 &  170.299 & \textcolor{darkerGreen}{4.3e-21} &        0.800 & 0.357 &  -0.077 &  177.554 & \textcolor{darkerGreen}{1.4e-20} &        0.900 & 0.524 &  -0.333 &  252.000 & \textcolor{darkerGreen}{2.9e-05} &        0.419 &  0.046 &  -0.077 &   53.917 &   \textcolor{darkerGreen}{0.029} \\
      15 &        0.717 & 0.127 &  -0.043 &  117.651 & \textcolor{darkerGreen}{1.1e-11} &        0.911 & 0.480 &  -0.111 &  178.857 & \textcolor{darkerGreen}{4.3e-19} &        0.770 & 0.400 &  -0.200 &  135.556 & \textcolor{darkerGreen}{2.3e-07} &        0.494 &  0.032 &  -0.111 &   49.714 &       \textcolor{crimson}{0.169} \\
      16 &        0.661 & 0.114 &  -0.043 &  108.889 & \textcolor{darkerGreen}{2.8e-10} &        0.764 & 0.361 &  -0.091 &  161.136 & \textcolor{darkerGreen}{4.3e-17} &        0.900 & 0.429 &  -0.333 &  236.000 & \textcolor{darkerGreen}{4.2e-04} &        0.456 &  0.012 &  -0.111 &   44.000 &       \textcolor{crimson}{0.352} \\
      17 &        0.674 & 0.093 &  -0.059 &   81.049 & \textcolor{darkerGreen}{9.6e-06} &        0.925 & 0.438 &  -0.200 &  143.500 & \textcolor{darkerGreen}{2.2e-10} &        0.800 & 0.429 &  -0.333 &  236.000 & \textcolor{darkerGreen}{4.2e-04} &        0.438 & -0.028 &  -0.143 &   35.556 &       \textcolor{crimson}{0.801} \\
      18 &        0.773 & 0.179 &  -0.040 &  166.931 & \textcolor{darkerGreen}{2.5e-20} &        0.850 & 0.286 &  -0.111 &  128.875 & \textcolor{darkerGreen}{1.9e-11} &        0.925 & 0.548 &  -0.143 &  186.222 & \textcolor{darkerGreen}{1.5e-19} &        0.525 &  0.122 &  -0.143 &   74.889 &   \textcolor{darkerGreen}{0.002} \\
      19 &        0.673 & 0.105 &  -0.040 &  111.264 & \textcolor{darkerGreen}{8.3e-11} &        0.810 & 0.295 &  -0.111 &  131.875 & \textcolor{darkerGreen}{6.6e-12} &        0.880 & 0.314 &  -0.200 &  119.556 & \textcolor{darkerGreen}{1.7e-05} &        0.455 &  0.016 &  -0.091 &   42.914 &       \textcolor{crimson}{0.270} \\
      20 &        0.656 & 0.094 &  -0.040 &  100.454 & \textcolor{darkerGreen}{4.7e-09} &        0.855 & 0.444 &  -0.111 &  178.875 & \textcolor{darkerGreen}{1.1e-19} &        0.670 & 0.257 &  -0.200 &  108.889 & \textcolor{darkerGreen}{2.3e-04} &        0.450 &  0.041 &  -0.111 &   51.875 &       \textcolor{crimson}{0.088} \\
      21 &        0.725 & 0.134 &  -0.059 &  104.969 & \textcolor{darkerGreen}{2.6e-09} &        0.939 & 0.294 &  -0.111 &  125.143 & \textcolor{darkerGreen}{2.1e-10} &        0.767 & 0.429 &  -0.333 &  236.000 & \textcolor{darkerGreen}{4.2e-04} &        0.383 & -0.043 &  -0.200 &   42.500 &       \textcolor{crimson}{0.836} \\
      22 &        0.636 & 0.040 &  -0.048 &   57.740 &   \textcolor{darkerGreen}{0.004} &        0.867 & 0.224 &  -0.200 &   98.500 & \textcolor{darkerGreen}{1.3e-04} &        0.925 & 0.179 &  -0.333 &  112.000 &   \textcolor{darkerGreen}{0.022} &        0.425 & -0.015 &  -0.091 &   30.960 &       \textcolor{crimson}{0.746} \\
      23 &        0.758 & 0.168 &  -0.043 &  149.570 & \textcolor{darkerGreen}{3.8e-17} &        0.910 & 0.256 &  -0.111 &  119.375 & \textcolor{darkerGreen}{5.5e-10} &        0.808 & 0.405 &  -0.200 &  136.500 & \textcolor{darkerGreen}{2.1e-09} &        0.531 &  0.122 &  -0.143 &   74.889 &   \textcolor{darkerGreen}{0.002} \\
      24 &        0.681 & 0.076 &  -0.048 &   79.102 & \textcolor{darkerGreen}{9.6e-06} &        0.875 & 0.390 &  -0.143 &  144.889 & \textcolor{darkerGreen}{8.2e-13} &        0.875 & 0.357 &  -1.000 &      inf &       \textcolor{darkerGreen}{0} &        0.505 & -0.013 &  -0.091 &   33.136 &       \textcolor{crimson}{0.695} \\
      25 &        0.626 & 0.125 &  -0.043 &  116.127 & \textcolor{darkerGreen}{1.9e-11} &        0.840 & 0.310 &  -0.111 &  136.375 & \textcolor{darkerGreen}{1.3e-12} &        0.900 & 0.143 &  -1.000 &      inf &       \textcolor{darkerGreen}{0} &        0.382 &  0.026 &  -0.091 &   46.469 &       \textcolor{crimson}{0.164} \\
      26 &        0.715 & 0.136 &  -0.043 &  123.365 & \textcolor{darkerGreen}{1.3e-12} &        0.791 & 0.286 &  -0.091 &  135.358 & \textcolor{darkerGreen}{7.7e-13} &        0.863 & 0.476 &  -0.333 &  162.000 & \textcolor{darkerGreen}{6.9e-07} &        0.537 &  0.112 &  -0.143 &   72.222 &   \textcolor{darkerGreen}{0.004} \\
      27 &        0.722 & 0.150 &  -0.053 &  121.506 & \textcolor{darkerGreen}{4.2e-12} &        0.837 & 0.339 &  -0.143 &  131.556 & \textcolor{darkerGreen}{9.1e-11} &        0.833 & 0.348 &  -0.200 &  124.500 & \textcolor{darkerGreen}{8.8e-08} &        0.458 &  0.105 &  -0.200 &   73.500 &   \textcolor{darkerGreen}{0.029} \\
      28 &        0.740 & 0.130 &  -0.040 &  126.193 & \textcolor{darkerGreen}{3.3e-13} &        0.875 & 0.281 &  -0.111 &  127.375 & \textcolor{darkerGreen}{3.3e-11} &        0.850 & 0.305 &  -0.200 &  115.500 & \textcolor{darkerGreen}{1.2e-06} &        0.517 &  0.147 &  -0.111 &   82.857 & \textcolor{darkerGreen}{1.3e-04} \\
      29 &        0.636 & 0.132 &  -0.040 &  128.280 & \textcolor{darkerGreen}{1.5e-13} &        0.767 & 0.292 &  -0.091 &  144.560 & \textcolor{darkerGreen}{1.2e-14} &        0.983 & 0.714 &  -0.333 &  284.000 & \textcolor{darkerGreen}{5.6e-08} &        0.375 &  0.003 &  -0.111 &   39.875 &       \textcolor{crimson}{0.448} \\
\bottomrule
\end{tabular}
\end{adjustbox}

\caption{\textbf{\texttt{PerceptPPO} Study:} Coefficient of consistency $\zeta$ and coefficient of agreement $u$ along with its corresponding values of \ch{$u_\mathrm{min}$}
, $\chi^2$ and $p$-value for positions of \texttt{PerceptPPO} for each map area reported over all responses and clusters obtained by hierarchical clustering.}
\label{tab:ppo-results}
\end{table*}

\begin{table*}[ht!]
\centering
\setlength{\tabcolsep}{3pt}
\begin{adjustbox}{width=\textwidth}
\begin{tabular}{c|ccccc|ccccc|ccccc|ccccc}
\toprule
Area & \multicolumn{5}{c}{Overall} & \multicolumn{5}{c}{Cluster 1} & \multicolumn{5}{c}{Cluster 2} & \multicolumn{5}{c}{Cluster 3} \\
         & mean $\zeta$ &      $u$ & \repA{$u_\mathrm{min}$}{min $u$} & $\chi^2$ &                        $p$-value & mean $\zeta$ &   $u$ & \repA{$u_\mathrm{min}$}{min $u$} & $\chi^2$ &                        $p$-value & mean $\zeta$ &   $u$ & \repA{$u_\mathrm{min}$}{min $u$} & $\chi^2$ &                        $p$-value & mean $\zeta$ &     $u$ & \repA{$u_\mathrm{min}$}{min $u$} & $\chi^2$ &                      $p$-value \\
\midrule
       0 &        0.639 &    0.001 &  -0.030 &   23.947 &       \textcolor{crimson}{0.407} &        0.776 & 0.167 &  -0.067 &   82.556 & \textcolor{darkerGreen}{8.9e-08} &        0.735 & 0.356 &  -0.143 &   97.280 & \textcolor{darkerGreen}{1.0e-07} &        0.411 &   0.032 &  -0.091 &   36.120 &     \textcolor{crimson}{0.132} \\
       1 &        0.599 &    0.055 &  -0.029 &   65.478 & \textcolor{darkerGreen}{5.6e-06} &        0.684 & 0.263 &  -0.053 &  135.557 & \textcolor{darkerGreen}{3.2e-17} &        0.536 & 0.067 &  -0.200 &   48.875 &       \textcolor{crimson}{0.143} &        0.487 &   0.075 &  -0.091 &   47.407 & \textcolor{darkerGreen}{0.015} \\
       2 &        0.550 &    0.038 &  -0.030 &   50.572 & \textcolor{darkerGreen}{7.8e-04} &        0.652 & 0.253 &  -0.067 &  111.479 & \textcolor{darkerGreen}{1.6e-12} &        0.633 & 0.048 &  -0.143 &   42.880 &       \textcolor{crimson}{0.177} &        0.375 &  -0.007 &  -0.091 &   25.320 &     \textcolor{crimson}{0.596} \\
       3 &        0.608 &    0.002 &  -0.027 &   23.977 &       \textcolor{crimson}{0.396} &        0.675 & 0.243 &  -0.111 &   82.857 & \textcolor{darkerGreen}{1.2e-06} &        0.643 & 0.165 &  -0.091 &   70.519 & \textcolor{darkerGreen}{2.0e-05} &        0.550 &   0.013 &  -0.059 &   29.920 &     \textcolor{crimson}{0.244} \\
       4 &        0.560 &   -0.006 &  -0.029 &   18.419 &       \textcolor{crimson}{0.729} &        0.595 & 0.134 &  -0.091 &   64.520 & \textcolor{darkerGreen}{9.2e-05} &        0.587 & 0.079 &  -0.111 &   47.429 &   \textcolor{darkerGreen}{0.029} &        0.514 &  -0.021 &  -0.067 &   18.556 &     \textcolor{crimson}{0.858} \\
       5 &        0.567 &    0.010 &  -0.029 &   30.523 &       \textcolor{crimson}{0.133} &        0.638 & 0.233 &  -0.067 &  104.710 & \textcolor{darkerGreen}{2.2e-11} &        0.541 & 0.084 &  -0.143 &   49.280 &       \textcolor{crimson}{0.059} &        0.500 &  -0.028 &  -0.077 &   18.347 &     \textcolor{crimson}{0.895} \\
       6 &        0.599 &    0.020 &  -0.029 &   38.183 &   \textcolor{darkerGreen}{0.023} &        0.611 & 0.114 &  -0.059 &   70.727 & \textcolor{darkerGreen}{3.2e-06} &        0.643 & 0.048 &  -0.143 &   41.333 &       \textcolor{crimson}{0.142} &        0.543 &   0.024 &  -0.111 &   34.781 &     \textcolor{crimson}{0.232} \\
       7 &        0.582 &    0.045 &  -0.030 &   54.947 & \textcolor{darkerGreen}{2.1e-04} &        0.670 & 0.199 &  -0.077 &   85.983 & \textcolor{darkerGreen}{4.8e-08} &        0.690 & 0.048 &  -0.200 &   45.875 &       \textcolor{crimson}{0.221} &        0.454 &   0.007 &  -0.077 &   28.375 &     \textcolor{crimson}{0.368} \\
       8 &        0.601 &    0.041 &  -0.029 &   54.536 & \textcolor{darkerGreen}{2.1e-04} &        0.714 & 0.298 &  -0.077 &  121.042 & \textcolor{darkerGreen}{5.0e-14} &        0.667 & 0.384 &  -0.200 &   98.875 & \textcolor{darkerGreen}{5.0e-07} &        0.478 &   0.029 &  -0.067 &   36.000 &     \textcolor{crimson}{0.086} \\
       9 &        0.573 &    0.018 &  -0.029 &   36.099 &   \textcolor{darkerGreen}{0.040} &        0.607 & 0.111 &  -0.111 &   55.281 &   \textcolor{darkerGreen}{0.003} &        0.686 & 0.314 &  -0.200 &   89.333 & \textcolor{darkerGreen}{1.7e-04} &        0.529 &   0.024 &  -0.053 &   35.074 &     \textcolor{crimson}{0.079} \\
      10 &        0.589 &    0.039 &  -0.032 &   49.947 & \textcolor{darkerGreen}{1.0e-03} &        0.729 & 0.306 &  -0.111 &  101.281 & \textcolor{darkerGreen}{8.6e-10} &        0.540 & 0.069 &  -0.111 &   45.143 &   \textcolor{darkerGreen}{0.047} &        0.516 &   0.060 &  -0.077 &   44.529 & \textcolor{darkerGreen}{0.019} \\
      11 &        0.547 &    0.029 &  -0.029 &   44.826 &   \textcolor{darkerGreen}{0.004} &        0.656 & 0.110 &  -0.091 &   56.296 &   \textcolor{darkerGreen}{0.001} &        0.589 & 0.224 &  -0.143 &   76.000 & \textcolor{darkerGreen}{2.6e-05} &        0.451 &   0.025 &  -0.067 &   34.571 &     \textcolor{crimson}{0.114} \\
      12 &        0.532 &    0.015 &  -0.029 &   34.183 &       \textcolor{crimson}{0.061} &        0.669 & 0.193 &  -0.091 &   77.630 & \textcolor{darkerGreen}{2.0e-06} &        0.500 & 0.072 &  -0.077 &   48.165 &   \textcolor{darkerGreen}{0.008} &        0.440 &  -0.007 &  -0.091 &   25.320 &     \textcolor{crimson}{0.596} \\
      13 &        0.675 &    0.072 &  -0.029 &   79.007 & \textcolor{darkerGreen}{4.3e-08} &        0.795 & 0.356 &  -0.067 &  146.556 & \textcolor{darkerGreen}{9.7e-19} &        0.679 & 0.276 &  -0.200 &   81.875 & \textcolor{darkerGreen}{8.2e-05} &        0.552 &   0.015 &  -0.067 &   30.864 &     \textcolor{crimson}{0.237} \\
      14 &        0.589 &    0.030 &  -0.029 &   45.948 &   \textcolor{darkerGreen}{0.003} &        0.590 & 0.164 &  -0.067 &   81.325 & \textcolor{darkerGreen}{1.4e-07} &        0.673 & 0.156 &  -0.143 &   62.080 &   \textcolor{darkerGreen}{0.004} &        0.546 &  -0.002 &  -0.077 &   25.708 &     \textcolor{crimson}{0.509} \\
      15 &        0.577 &    0.017 &  -0.029 &   35.830 &   \textcolor{darkerGreen}{0.042} &        0.675 & 0.244 &  -0.059 &  122.727 & \textcolor{darkerGreen}{7.9e-15} &        0.520 & 0.111 &  -0.143 &   54.080 &   \textcolor{darkerGreen}{0.022} &        0.455 &  -0.004 &  -0.091 &   26.963 &     \textcolor{crimson}{0.548} \\
      16 &        0.604 &    0.046 &  -0.027 &   59.634 & \textcolor{darkerGreen}{4.0e-05} &        0.729 & 0.320 &  -0.067 &  134.249 & \textcolor{darkerGreen}{1.6e-16} &        0.774 & 0.359 &  -0.200 &   94.875 & \textcolor{darkerGreen}{1.8e-06} &        0.424 &  -0.006 &  -0.067 &   23.429 &     \textcolor{crimson}{0.593} \\
      17 &        0.608 &    0.006 &  -0.029 &   27.614 &       \textcolor{crimson}{0.228} &        0.665 & 0.326 &  -0.077 &  123.802 & \textcolor{darkerGreen}{2.5e-14} &        0.550 & 0.007 &  -0.111 &   30.781 &       \textcolor{crimson}{0.402} &        0.595 &  -0.016 &  -0.091 &   22.920 &     \textcolor{crimson}{0.724} \\
      18 &        0.504 &    0.005 &  -0.029 &   26.301 &       \textcolor{crimson}{0.281} &        0.533 & 0.092 &  -0.077 &   53.983 &   \textcolor{darkerGreen}{0.002} &        0.476 & 0.003 &  -0.200 &   38.875 &       \textcolor{crimson}{0.493} &        0.492 &   0.048 &  -0.059 &   43.253 & \textcolor{darkerGreen}{0.015} \\
      19 &        0.623 &    0.041 &  -0.029 &   55.007 & \textcolor{darkerGreen}{1.8e-04} &        0.743 & 0.282 &  -0.067 &  121.325 & \textcolor{darkerGreen}{3.1e-14} &        0.531 & 0.011 &  -0.143 &   36.480 &       \textcolor{crimson}{0.413} &        0.541 &   0.014 &  -0.077 &   30.708 &     \textcolor{crimson}{0.263} \\
      20 &        0.591 &        0 &  -0.029 &   22.772 &       \textcolor{crimson}{0.468} &        0.770 & 0.285 &  -0.077 &  117.042 & \textcolor{darkerGreen}{2.5e-13} &        0.487 & 0.120 &  -0.091 &   58.963 & \textcolor{darkerGreen}{6.8e-04} &        0.468 &   0.027 &  -0.091 &   34.963 &     \textcolor{crimson}{0.189} \\
      21 &        0.552 &   -0.006 &  -0.027 &   18.034 &       \textcolor{crimson}{0.747} &        0.539 & 0.238 &  -0.091 &   89.185 & \textcolor{darkerGreen}{3.6e-08} &        0.667 & 0.240 &  -0.091 &   93.720 & \textcolor{darkerGreen}{4.3e-09} &        0.464 &   0.015 &  -0.077 &   31.042 &     \textcolor{crimson}{0.249} \\
      22 &        0.553 &    0.003 &  -0.029 &   25.433 &       \textcolor{crimson}{0.326} &        0.634 & 0.224 &  -0.143 &   76.000 & \textcolor{darkerGreen}{2.6e-05} &        0.684 & 0.011 &  -0.143 &   36.480 &       \textcolor{crimson}{0.413} &        0.475 &   0.011 &  -0.053 &   29.296 &     \textcolor{crimson}{0.236} \\
      23 &        0.580 &    0.013 &  -0.029 &   32.463 &       \textcolor{crimson}{0.090} &        0.625 & 0.245 &  -0.091 &   95.320 & \textcolor{darkerGreen}{2.4e-09} &        0.643 & 0.143 &  -0.111 &   62.781 & \textcolor{darkerGreen}{3.4e-04} &        0.489 &  -0.018 &  -0.077 &   21.256 &     \textcolor{crimson}{0.777} \\
      24 &        0.584 &    0.044 &  -0.029 &   56.463 & \textcolor{darkerGreen}{1.2e-04} &        0.589 & 0.196 &  -0.067 &   96.000 & \textcolor{darkerGreen}{4.8e-10} &        0.464 & 0.143 &  -0.200 &   60.875 &   \textcolor{darkerGreen}{0.016} &        0.632 &   0.065 &  -0.077 &   45.983 & \textcolor{darkerGreen}{0.013} \\
      25 &        0.635 &    0.011 &  -0.029 &   31.242 &       \textcolor{crimson}{0.114} &        0.612 & 0.161 &  -0.067 &   83.429 & \textcolor{darkerGreen}{5.1e-08} &        0.736 & 0.081 &  -0.111 &   48.281 &   \textcolor{darkerGreen}{0.016} &        0.571 &  -0.007 &  -0.111 &   27.281 &     \textcolor{crimson}{0.584} \\
      26 &        0.557 &  8.8e-04 &  -0.029 &   23.493 &       \textcolor{crimson}{0.429} &        0.571 & 0.233 &  -0.077 &   96.165 & \textcolor{darkerGreen}{1.1e-09} &        0.565 & 0.201 &  -0.091 &   82.920 & \textcolor{darkerGreen}{2.1e-07} &        0.529 &   0.029 &  -0.111 &   35.781 &     \textcolor{crimson}{0.198} \\
      27 &        0.627 & -4.0e-04 &  -0.029 &   22.523 &       \textcolor{crimson}{0.486} &        0.708 & 0.352 &  -0.091 &  118.519 & \textcolor{darkerGreen}{6.1e-13} &        0.706 & 0.201 &  -0.111 &   73.714 & \textcolor{darkerGreen}{2.3e-05} &        0.519 & 4.5e-04 &  -0.067 &   25.941 &     \textcolor{crimson}{0.472} \\
      28 &        0.612 &    0.018 &  -0.032 &   35.680 &   \textcolor{darkerGreen}{0.046} &        0.643 & 0.231 &  -0.091 &   87.407 & \textcolor{darkerGreen}{6.8e-08} &        0.619 & 0.212 &  -0.111 &   76.000 & \textcolor{darkerGreen}{1.1e-05} &        0.577 &   0.053 &  -0.091 &   42.120 & \textcolor{darkerGreen}{0.039} \\
      29 &        0.605 &    0.018 &  -0.029 &   36.889 &   \textcolor{darkerGreen}{0.032} &        0.633 & 0.214 &  -0.077 &   94.375 & \textcolor{darkerGreen}{1.6e-09} &        0.698 & 0.132 &  -0.111 &   58.857 &   \textcolor{darkerGreen}{0.002} &        0.511 &   0.009 &  -0.077 &   29.256 &     \textcolor{crimson}{0.352} \\
\bottomrule
\end{tabular}
\end{adjustbox}
\caption{\textbf{Evaluation 2 -- Comparison Study of PPOs:} Coefficient of consistency $\zeta$ and coefficient of agreement $u$ along with its corresponding values of \ch{$u_\mathrm{min}$}
, $\chi^2$ and $p$-value for examined PPOs for each map area reported over all responses and clusters obtained by hierarchical clustering.}
\label{tab:ppo-eval-results}
\end{table*}

\bibliographystyle{abbrv-doi}
